\documentclass[12pt]{article}
\usepackage[T1]{fontenc}
\usepackage[utf8]{inputenc}
\usepackage{microtype}
\usepackage[a4paper,margin=1in]{geometry}
\usepackage{setspace}
\usepackage[sc]{mathpazo}
\usepackage{comment}
\usepackage{ragged2e}
\usepackage{etoc}
\usepackage{amsmath,amsthm,mathrsfs,mathtools,bm,amssymb}
\usepackage{dsfont}
\usepackage{booktabs,array,tabularx,longtable}
\usepackage{enumitem}
\usepackage{graphicx}
\usepackage{tikz}
\usepackage{pgfplots}
\pgfplotsset{compat=1.18}
\usetikzlibrary{arrows.meta,patterns,positioning,calc,decorations.pathreplacing}

\usepackage{natbib}
\usepackage{xcolor}
\usepackage{hyperref}
\definecolor{linkblue}{RGB}{36,73,113}
\definecolor{warmdarkred}{RGB}{120, 20, 30}
\definecolor{myblue}{RGB}{0,30,180}
\hypersetup{
    colorlinks=true,
    linkcolor=warmdarkred,
    urlcolor=warmdarkred,
    citecolor=warmdarkred
}
\usepackage[nameinlink,noabbrev]{cleveref}
\usepackage[colorinlistoftodos,textsize=tiny]{todonotes}
\newtheoremstyle{econplain}
  {0.8em}{0.8em}{\itshape}{}{\bfseries}{.}{0.55em}{}
\theoremstyle{econplain}

\newtheorem{proposition}{Proposition}
\newtheorem{lemma}{Lemma}
\newtheorem{corollary}{Corollary}
\newtheorem{theorem}{Theorem}

\theoremstyle{definition}
\newtheorem{assumption}{Assumption}

\newtheoremstyle{econremark}
  {0.8em}{0.8em}{}{}{\bfseries}{.}{0.55em}{}
\theoremstyle{econremark}

\crefname{proposition}{Proposition}{Propositions}
\Crefname{proposition}{Proposition}{Propositions}
\crefname{lemma}{Lemma}{Lemmas}
\Crefname{lemma}{Lemma}{Lemmas}
\crefname{corollary}{Corollary}{Corollaries}
\Crefname{corollary}{Corollary}{Corollaries}
\crefname{theorem}{Theorem}{Theorems}
\Crefname{theorem}{Theorem}{Theorems}
\crefname{assumption}{Assumption}{Assumptions}
\Crefname{assumption}{Assumption}{Assumptions}
\crefname{definition}{Definition}{Definitions}
\Crefname{definition}{Definition}{Definitions}
\crefname{remark}{Remark}{Remarks}
\Crefname{remark}{Remark}{Remarks}
\crefname{example}{Example}{Examples}
\Crefname{example}{Example}{Examples}
\crefname{appendix}{Appendix}{Appendices}
\Crefname{appendix}{Appendix}{Appendices}
\crefname{supplement}{Supplement}{Supplements}
\Crefname{supplement}{Supplement}{Supplements}

\allowdisplaybreaks
\setlist[itemize]{topsep=0.35em,itemsep=0.2em,parsep=0pt,leftmargin=1.5em}
\setlist[enumerate]{topsep=0.35em,itemsep=0.2em,parsep=0pt,leftmargin=1.7em}

\newcommand{\thL}{\theta_{\ell}}
\newcommand{\thH}{\theta_{h}}
\newcommand{\Th}{\Theta}
\newcommand{\prior}{\pi}
\newcommand{\DeltaT}{\Delta}
\newcommand{\thbar}{\bar\theta}
\newcommand{\dpool}{d_{P}}

\newcommand{\hardsig}{\tau}
\newcommand{\sigvec}{\mathbf{s}}

\newcommand{\hardQ}{Q}

\newcommand{\rhoHard}{\rho_{\tau}}

\newcommand{\belief}{\mu}
\newcommand{\action}{a}
\newcommand{\receiverweight}{\lambda}

\newcommand{\costL}{c_{\ell}}
\newcommand{\costH}{c_{h}}

\newcommand{\detval}{\Psi}
\newcommand{\rileycost}{\psi}
\newcommand{\rileyone}{\psi^{1}}

\newcommand{\capcost}{\Gamma}
\newcommand{\welfare}{\mathcal W}
\newcommand{\profit}{\Pi}

\newcommand{\E}{\mathbb E}

\newcommand{\Var}{\operatorname{Var}}

\newcommand{\ind}{\mathds{1}}

\usepackage{titlesec}

\titleformat{\section}
{\normalfont\large\bfseries\centering\sc} 
{\thesection.}                          
{1em}                                   
{}                                      

\titleformat{\subsection}
{\normalfont\normalsize\bfseries\centering\sc}
{\thesubsection.}
{1em}
{}

\title{\bf The Market for AI-Assisted Signaling}

\author{Federico Vaccari\\
        {\small University of Bergamo}\\
        {\small \texttt{vaccari.econ@gmail.com}}}
\date{}

\begin{document}

\maketitle


\begin{abstract}
This paper studies a market in which AI providers choose technologies and prices for assistance used to signal ability. Individual adoption is unobserved. Profitable assistance can lower welfare relative to signaling without AI, even when ability remains fully revealed. Easier imitation increases the output required to demonstrate ability, and the extra signaling can outweigh production savings. Providers can profit from improving assistance for lower-ability senders because this increases higher-ability senders' purchases. Such improvements can lower welfare relative to retaining existing tools. Competition among providers offering the same technology supports full information, but development costs can sustain monopoly provision that conceals ability. Greater verification capacity can also reduce information by changing which customers providers serve.
\end{abstract}

\bigskip


\thispagestyle{empty}


\clearpage
\etocdepthtag.toc{main}
\begingroup
\hypersetup{linkcolor=black}
\renewcommand{\contentsname}{Contents}
\etocsettagdepth{main}{subsection}
\etocsettagdepth{appendix}{none}
\etocsettagdepth{supplement}{none}
\tableofcontents
\endgroup
\clearpage


\section{Introduction}
\label{sec:introduction}

An applicant preparing a cover letter can write it unaided or use artificial intelligence. The employer observes the letter but may not observe how it was produced. The letter's informational value depends on how difficult it is for a weaker applicant to imitate a stronger one. Artificial intelligence (AI) makes imitation easier, and narrows differences among applicants. The price of AI assistance also plays a role. A membership fee determines when assistance becomes worth buying, while a usage price changes the cost of producing a more impressive letter. The provider's offer can change both the letters applicants submit and what the employer learns from them.

This paper studies how AI providers choose technology and prices when their customers use assistance to signal ability. It asks when these choices preserve information and whether AI provision improves welfare. The model has senders with one of two ability levels, and a receiver who values accurate ability assessment. Senders can demonstrate ability through observable output, such as a written application, and through evidence that AI is less able to substitute, such as a supervised work trial. I call the output an \emph{artifact}, and the second form of evidence a \emph{hard signal}. Institutional capacity limits the amount of hard evidence each sender can provide.

The provider chooses how much editing assisted production requires from each ability group, and how much production the technology can assist. Then, it charges a membership fee and a price per unit of AI input. Senders know their ability when deciding whether to subscribe, how much assistance to buy, and which signals to produce. Prices and technology are public, but individual adoption and production methods are unobserved.\footnote{Observed AI use would make the production method an additional signal. Ability is fixed, and the artifact has no direct productive return in the welfare analysis.} Providers anticipate the signaling behavior induced by their offers.\footnote{I apply D1, which compares the ability groups' incentives to deviate to unexpected signals using costs minimized over production methods. The analysis uses finite observable support and a sender-favorable selection rule. \Cref{sec:model} states the equilibrium and selection criteria, whereas \Cref{app:refinement-details} gives the formal D1 test and selection ties.}

Lower-ability senders save more effort from AI and begin using it at a less demanding artifact than higher-ability senders. Over a range of output, producing a better artifact can be cheaper for an assisted lower-ability sender than for a higher-ability sender who still works manually. A small improvement in the artifact need not credibly identify higher ability. Distinguishing oneself can require a much larger, costly improvement. Pooling outcomes, in which both ability groups submit the same signals and receive the same assessment, survive when this expenditure exceeds the gain from recognition and lower-ability senders are willing to participate. Partial pooling can also arise. Some lower-ability senders join the shared signal, while the rest reveal their ability by submitting only the basic artifact.

Hard evidence offers another way to distinguish ability. The model assumes that higher-ability senders have a stronger cost advantage in producing this evidence than in producing the artifact. The separating outcome with the lowest private cost for these senders uses hard evidence first, adding artifact production when capacity alone cannot deter imitation. Pooling requires senders to exhaust this capacity. For otherwise, a small increase in hard evidence would credibly reveal higher ability. Increasing capacity can eliminate pooling without eliminating partial pooling, so the disappearance of a signal shared by everyone need not restore full information.

The provider's optimal offer depends on which ability group purchases assistance. When signals fully reveal ability and optimal operating profit is positive, the provider serves higher-ability senders and extracts their production savings. The threat of imitation by lower-ability senders limits how much assistance the provider can sell. In pooling and partial pooling, participating lower-ability senders buy assistance, and higher-ability senders produce manually. The provider charges for the manual effort its customers avoid, subject to participation and the cost of distinguishing oneself. Its technology choice compares these operating returns with development costs.

Membership and usage prices affect these incentives separately. Two offers can collect the same payment for the chosen artifact, yet impose different costs on someone producing a better artifact to demonstrate higher ability. Optimal usage prices can lie below compute cost, including free usage, with membership fees covering the resource bill. Such subsidies can be profitable because assistance saves enough effort for lower-ability customers.

Price competition changes both access to assistance and the information conveyed by signals. With two or more providers offering the same already developed technology, a pricing equilibrium has zero membership fees and usage priced at compute cost. Signals fully reveal ability, and providers earn zero operating profits. If providers must finance development before competing on price, the sequential entry game can instead support a sole provider charging an optimal monopoly tariff. A second entrant would not recover its development cost under the specified competitive pricing equilibrium. Hence, potential competition can coexist with pooling.

I also let two established providers improve their technologies before choosing prices. Each can reduce the editing effort required from either or both ability groups, within a bounded set of improvements. Initial development costs are sunk, so this comparison holds entry fixed. Under sufficient productivity and hard-evidence restrictions, and with no limit on assisted production, the game admits equilibria with one profitable improved technology and a rival retaining its baseline. In one such equilibrium, improving both groups' productivity lowers welfare relative to retaining both baseline technologies, although signals fully reveal ability in both outcomes. Making imitation easier for lower-ability senders increases the artifact required from higher-ability senders, and their purchases of AI input. 

AI changes the cost differences that make signals credible, the production required to distinguish ability, and the choice between artifacts and hard evidence. These changes have different welfare implications. At a fixed verification capacity, AI improves welfare relative to manual signaling when resource savings exceed development costs and any loss from inaccurate assessment. Pooling can satisfy this condition, and full separation can fail it, as cheaper imitation can increase the required artifact enough to outweigh production savings. When both outcomes separate, competition can require more signaling resources than monopoly provision of the same technology. Monopoly usage markups can also favor hard evidence over assisted production that uses fewer resources, so the privately cheapest separating outcome need not minimize resource use. Welfare depends on how providers' offers change the information conveyed by signals and the resources spent producing them.

Verification also changes the providers' incentives to improve and price AI assistance. I let a monopolist choose design and pricing, and compare verification chosen before its offer with simultaneous choices. Sequentially, the receiver commits to capacity and anticipates the provider's response. Simultaneously, capacity and the provider's entire offer must be mutual best responses. In a tractable case, the provider reduces artifact quality as capacity increases, encouraging participation and offsetting part of the information gain. Anticipating this response can lead the receiver to choose less verification than in the simultaneous game. A higher externally imposed minimum capacity can also reduce equilibrium information under both timings: the provider switches from assisting higher-ability senders in separation to assisting lower-ability senders in pooling.

The receiver values accuracy and bears verification costs, but omits changes in signaling resources and newly chosen improvement expenditure. Its investment can be socially excessive or insufficient. An accuracy gain can induce investment that replaces a cheap shared outcome with more costly separating production and improvement. Conversely, an expansion that leaves pooling intact can save resources without improving accuracy, giving the receiver no reason to purchase it. Once hard evidence alone deters imitation, separation requires no incremental AI input or costly improvement.

\paragraph{Related literature.}

The closest paper is \citet{CetemenOzdenoren2026}, who introduce an AI token industry into a career-signaling model with skill formation and output trade. Both papers show that cheaper AI production can increase the performance required to distinguish ability. Their analysis follows how token trade affects work, skills, and surplus. Here, ability is fixed, and providers choose technologies and membership and usage prices for privately adopted assistance. Adoption changes the relative cost of producing better signals and can sustain pooling despite unequal costs. The additional hard signal connects these choices to verification and welfare.

\citet{CamboniEtAl2025} study schools choosing attendance fees and monitoring policies, showing that competition can increase signal informativeness and wasteful effort. \citet{Lu2025} studies a seller choosing the price schedule of a signaling good. In the present model, customers buy a private production method and retain the option of producing manually. Membership and usage charges determine both adoption and the cost of departing from a shared signal, linking provider revenue to which ability group purchases assistance.

The signaling analysis builds on \citet{Spence1973} and \citet{Riley1979}, belief refinements \citep{ChoKreps1987,BanksSobel1987,ChoSobel1990}, and multidimensional signaling \citep{Ramey1996}. Hard evidence connects the model to direct assessment and certification.\footnote{\citet{AlonsoCamara2024} study experimentation, reporting, and auditing within organizations. They show that preventing report manipulation can conflict with incentives for informative experimentation. Both papers examine how verification affects incentives behind information provision. Here, verification limits the hard evidence available for signaling, rather than auditing reports, and its informational effect depends on providers' technology and pricing choices and senders' responses.} \citet{KurlatScheuer2021} study how buyer expertise changes reliance on public signals. Certification intermediaries choose what information to disclose or acquire \citep{Lizzeri1999,AsseyerWeksler2024,CelikStrausz2026}. Here, senders produce both signals, and the AI provider changes production costs without learning individual abilities or issuing a quality report.

Recent work examines AI's effects on communication and screening \citep{Gans2024,CowgillEtAl2026}. \citet{GaldinSilbert2025} and \citet{CuiDiasYe2025} document declining informativeness of written applications. The latter also find that employers shift toward work histories. \citet{AbbasNejadEtAl2026} find disproportionate improvements in lower-quality applicants' cover letters, and \citet{Jungbauer2026} shows how better applications can coexist with worse talent allocation. These studies motivate the interaction between assisted writing and alternative evidence.\footnote{\citet{NoyZhang2023} and \citet{BrynjolfssonLiRaymond2025} document heterogeneous productivity gains from AI assistance. Assistance can also improve hiring outcomes: \citet{WilesMunyikwaHorton2025} find increased hiring from nongenerative writing assistance without evidence of lower employer satisfaction.} The present model studies how providers' prices and technology choices shape these informational effects.

\Cref{sec:model,sec:benchmark,sec:two-signal} introduce the signaling environment. \Cref{sec:market,sec:competition,sec:design-competition} study monopoly provision, competition with development, and improvements by established providers. \Cref{sec:welfare,sec:capacity} examine welfare and verification. Proofs appear in the appendix. The supplement contains additional calculations, examples, and robustness results.

\section{The model}
\label{sec:model}

A sender uses costly signals to obtain a favorable assessment of its privately known ability. The receiver observes the signals but does not observe how they were produced. AI can assist production of an observable artifact, such as a written application. A second signal provides harder-to-substitute evidence, such as a supervised work trial, which the sender produces without assistance. An AI provider charges separately for membership and usage. These prices affect who adopts assistance and how costly it is to produce or imitate a signal.


\paragraph{Players, information, and timing.}

There is a unit mass of senders, each evaluated separately by a receiver. A sender privately knows its ability, represented by its type $\theta\in\Th=\{\thL,\thH\}$, where $0<\thL<\thH$. I refer to these as the low and high types. The high-type share is $\prior\in(0,1)$, which is also the receiver's prior probability that a sender has high ability. Define the type gap $\DeltaT\coloneqq\thH-\thL$ and the prior mean $\thbar\coloneqq\prior\thH+(1-\prior)\thL$.

The baseline market has one AI provider. The provider first decides whether to develop a technology and, if so, chooses its features. Then, it posts a tariff specifying membership and usage prices. After observing the offer, senders choose whether to subscribe, how much AI input to buy, and which signals to produce. The receiver observes each sender's signals and chooses an assessment.

Technology features, tariffs, and the capacity of the second signal are public. The receiver does not observe individual subscriptions, AI input, or production methods. The provider knows the type distribution but does not observe individual types.\footnote{\Cref{sec:competition} studies providers offering the same technology and entry before pricing. \Cref{sec:design-competition} holds the number of providers fixed at two and lets them choose improvements before pricing. \Cref{sec:capacity} studies a monopolist choosing improvements and prices, with capacity chosen either beforehand by a committed receiver or simultaneously with the provider's entire offer.}


\paragraph{Production and tariffs.}

Every sender must supply a basic artifact, which can be produced without a paid subscription. Let $q\geq0$ denote observable quality above this baseline: $q=0$ means supplying only the basic artifact. Without paid AI, producing additional quality $q$ costs type $i\in\{\ell,h\}$ the amount $\kappa_i q$, where $0<\kappa_h<\kappa_\ell$. Basic tools, including free assistance, can be incorporated in the basic artifact. The production choices below concern quality above that baseline.

The hard signal is $\hardsig\in[0,\hardQ]$, where $\hardQ\geq0$ limits how much evidence each sender can supply. For example, capacity may be the maximum length of a supervised work trial. I measure the signal in units of the low type's effort cost: producing $\hardsig$ costs the low type $\hardsig$ and the high type $\rhoHard\hardsig$, where $\rhoHard>0$.

A paid AI technology can assist at most $L>0$ units of additional quality.\footnote{Finite coverage summarizes how much of a task the tool can assist. Providers document model-specific token limits and differentiate output limits across subscription plans. See \href{https://platform.claude.com/docs/en/models/overview}{Anthropic's model documentation} and \href{https://claude.com/pricing}{pricing page} (accessed October 13, 2026). These provide examples of limits on assistance. Here, $L$ is measured in units of assisted quality. A periodic subscription allowance would instead constrain usage across tasks.} A subscriber producing quality $q$ chooses AI input $z\in[0,\min\{q,L\}]$. Each unit of input replaces one unit of manual production but requires effort $\epsilon_i$ from type $i$, such as editing or checking the output. The remaining $q-z$ units are produced manually. The provider chooses the technology's physical features $\mathbf d=(\epsilon_\ell,\epsilon_h,L)$ from a feasible design set $\mathcal D$ and incurs compute cost $c>0$ per unit of input. Paying for membership without using input provides no production benefit.

\begin{assumption}
\label{ass:comparative-advantage}
The feasible design set $\mathcal D$ is compact. Every design in $\mathcal D$ has finite positive coverage and satisfies
\begin{equation*}
\begin{aligned}
0<\epsilon_h<\epsilon_\ell<\kappa_h,\\
0<\epsilon_\ell-\epsilon_h<\kappa_\ell-\kappa_h,\\
0<\rhoHard<\min\left\{\frac{\kappa_h}{\kappa_\ell},\frac{\epsilon_h}{\epsilon_\ell}\right\}.
\end{aligned}
\end{equation*}
Manual costs, compute cost, and the hard-signal cost ratio are fixed across designs.
\end{assumption}

AI reduces production effort for both types and narrows the effort difference between them. The low type saves more effort per unit of input, so a given tariff can make adoption worthwhile at a lower artifact quality. The final inequality gives the high type a stronger marginal cost advantage in hard evidence than in either manual or assisted artifact production.

The provider independently chooses a membership fee $F\geq0$ and a usage price $p\geq0$. A subscriber using input $z$ pays $F+pz$.\footnote{AI offerings combine access and usage charges. Claude Enterprise, for example, charges a recurring seat fee plus usage billed at API rates. Individual paid plans include usage allowances and permit separately charged additional usage. See \href{https://claude.com/pricing}{Anthropic's pricing page} and its \href{https://support.claude.com/en/articles/12429409-manage-usage-credits-for-paid-claude-plans}{usage-credit documentation} (accessed October 13, 2026). The linear tariff abstracts from included allowances and model-specific token prices.} The usage price may be below compute cost, $p<c$.\footnote{\Cref{supp:cash-rebates} considers unconditional negative membership fees. A rebate collectible without usage is dominated in monopoly and cannot survive the stated pure competitive pricing incentives.}

After paying the membership fee, a sender can still choose between AI and manual production. Therefore, its cheapest private unit cost within coverage is  $m_i\coloneqq\min\{p+\epsilon_i,\kappa_i\}$.  Optimizing over subscription and input gives type $i$'s private artifact cost\footnote{I write $(y)_+\coloneqq\max\{y,0\}$ for any real $y$.}
\begin{equation}
\label{eq:cost-envelope}
C_i(q\mid\mathbf d,F,p)
=\min\left\{\kappa_i q,\ F+m_i\min\{q,L\}+\kappa_i(q-L)_+\right\}.
\end{equation}
The first option is manual production without a subscription. The second includes the membership fee, the cheaper production method within coverage, and manual production beyond coverage. The sender subscribes when the two options have equal private cost and, conditional on subscribing, uses the most AI input among cost-minimizing choices. I refer to these assumptions as the \emph{subscription rule} and \emph{input rule}, respectively.\footnote{Both rules apply in monopoly and competition.}

I suppress the technology and tariff arguments when they are fixed. Quality $q$ has the same physical meaning across offers, and it is not rescaled when prices change. \Cref{app:implementation} derives the cost function and explains the treatment of the basic artifact. \Cref{sec:competition} specifies production when several offers are available.


\paragraph{Payoffs.}

A sender supplies the signal pair $\sigvec=(q,\hardsig)$. Adding artifact and hard-signal costs gives the two types' private signaling expenditure above the basic artifact,
\begin{align}
\costL(\sigvec)&=C_\ell(q)+\hardsig,\label{eq:low-cost-normalized}\\
\costH(\sigvec)&=C_h(q)+\rhoHard\hardsig.\label{eq:high-cost-normalized}
\end{align}
Let $\action\in[\thL,\thH]$ denote the receiver's assessment. Both types prefer a higher assessment and pay their own signaling costs. Net of the common cost of supplying the basic artifact, type $i$'s payoff is
\begin{equation*}
u_i(\sigvec,\action)=\action-c_i(\sigvec).
\end{equation*}

The receiver prefers assessments close to the sender's true ability. With accuracy weight $\receiverweight>0$, its payoff is
\begin{equation*}
u_R(\action,\theta)=-\frac{\receiverweight}{2}(\action-\theta)^2.
\end{equation*}
After observing $\sigvec$, the receiver assigns posterior probability $\belief(\sigvec)\in[0,1]$ to the high type.

\begin{lemma}
\label{lem:receiver-br}
For every posterior $\belief\in[0,1]$, the receiver's unique best response is
\begin{equation*}
\action^*(\belief)=\belief\thH+(1-\belief)\thL.
\end{equation*}
\end{lemma}

The receiver assesses a sender at its posterior mean ability. Both types gain $\DeltaT$ in gross payoff from being assessed as a high- rather than a low-ability sender. In the application, $\action$ can represent an applicant's evaluation score.\footnote{If the assessment instead represented a monetary payment by the receiver, that payment would also enter the receiver's payoff.}

Developing technology $\mathbf d$ costs the provider $K(\mathbf d)\geq0$, continuous on $\mathcal D$. This expenditure is sunk when the provider sets its tariff. For a developed technology and given hard-signal capacity $\hardQ$, expected net profit is
\begin{equation}
\label{eq:provider-profit}
\profit(\mathbf d,F,p\mid\hardQ)
=\E\!\left[F\ind_{\{\mathrm{subscribed}\}}+(p-c)z\right]-K(\mathbf d),
\end{equation}
where the indicator records subscription and the expectation averages over sender types and equilibrium choices. Capacity $\hardQ$ affects subscription and input demand through the signaling outcome, so the same technology and tariff can generate different profits at different capacities. Each subscriber generates membership revenue $F$ and a usage margin $p-c$ per unit of input. Choosing not to develop yields zero profit.

Membership and usage payments affect senders' private costs but transfer income to the provider. Production effort, compute, and development expenditure consume resources. This distinction matters for the welfare comparisons in \Cref{sec:welfare}.


\paragraph{Equilibrium.}

I use the Perfect Bayesian Equilibrium (PBE) refined by D1 \citep{ChoKreps1987,BanksSobel1987}. Senders optimize their signals and production choices, and the receiver responds optimally to beliefs satisfying Bayes' rule at signals observed on-path. D1 restricts beliefs after unexpected, off-path signals. I restrict attention to equilibria in which each type’s distribution over observable signal pairs has finite support. This allows senders to randomize among finitely many signal pairs, while excluding distributions with infinite or continuous support.

For each public offer or tariff menu and capacity, a \emph{signaling equilibrium} specifies sender strategies, receiver responses, and beliefs satisfying PBE. D1 restricts the admissible off-path beliefs. An \emph{observable signaling outcome} records each type's signal choices and mixing probabilities and the receiver's on-path assessments. Equilibria that differ only in off-path beliefs have the same observable outcome.

I use an additional selection rule to choose among observable outcomes of finite-support signaling equilibria refined by D1. The selected outcome is \emph{sender-Pareto-maximal}: no other such outcome makes both types weakly better off and at least one strictly better off. Among outcomes giving both types the same payoffs, selection favors the greatest population mass at signals shared by both types, then the least artifact quality at the shared signal. Remaining cost ties between separating signals favor the least artifact quality. For each fixed offer or tariff menu and capacity, this equilibrium selection rule selects a unique observable outcome.\footnote{\Cref{app:refinement-details} gives the formal D1 test and discusses the selection rule.}

\section{Signaling with AI assistance}

\subsection{Benchmark: a single AI-assisted signal}
\label{sec:benchmark}

Set $\hardQ=0$ and hold the technology $\mathbf d$ and tariff $(F,p)$ fixed. The sender signals only through artifact quality $q$. This benchmark explains how voluntary AI adoption changes the cost of imitation and can support pooling.

In a separating equilibrium, the low type chooses $q=0$ and receives assessment $\thL$. Imitating the high type would raise its assessment by $\DeltaT$, so the high signal must impose a low-type cost of at least $\DeltaT$. Both cost functions are strictly increasing. The least-cost separating, or Riley, quality $q_R$ makes this constraint bind, and its high-type cost is $\rileyone$. That is, 
\begin{equation}
\label{eq:one-signal-riley-cost}
C_\ell(q_R)=\DeltaT \; \text{ and }\; \rileyone\coloneqq C_h(q_R).
\end{equation}
This allocation supports a separating PBE. Whether it satisfies D1 depends on how adoption changes the high type's cost advantage.

Consider a positive membership fee and a sufficiently low usage price. Let $x$ denote the low type's adoption quality, and suppose that
\begin{equation}
\label{eq:one-signal-adoption-regime}
F>0,\quad p+\epsilon_\ell<\kappa_h,\; \text{ and } \;
x\coloneqq\frac{F}{\kappa_\ell-p-\epsilon_\ell}\in(0,L).
\end{equation}
At $q=x$, the low type's production saving just covers the membership fee, so it subscribes under the subscription rule. The high type saves less from AI, and continues to produce manually. Although they use different methods, both types incur the same private cost as under manual production at this quality. That is, $C_\ell(x)=\kappa_\ell x$ and $C_h(x)=\kappa_h x$.

Let $W_1(q)\coloneqq C_\ell(q)-C_h(q)$ denote the high type's cost advantage. Below $x$, both types produce manually, and this advantage increases with quality. Just above $x$, the low type uses AI while the high type does not. In this case, the low type's marginal cost is $p+\epsilon_\ell<\kappa_h$, so further quality reduces the high type's advantage. The advantage increases again once the high type also adopts assistance or, likewise, when AI coverage is exhausted.

When both types submit quality $x$, D1 attributes a deviation to the high type if it increases this cost advantage and full recognition makes the deviation profitable. Hence, the high type must move beyond the range in which low-type adoption has reduced its advantage. Let $q_E>x$ be the first quality at which the advantage recovers to $W_1(x)$. Define the high type's additional expenditure needed to reach that quality by
\begin{equation}
\begin{aligned}
\label{eq:recovery-cost}
E &\coloneqq  C_h(q_E)-C_h(x),\\ 
W_1(q_E)& =  W_1(x).
\end{aligned}
\end{equation}
At a pooling outcome, the receiver assigns to the sender ability the prior assessment $\thbar$. Full recognition as high ability would raise the high type's gross payoff by
\[
\dpool\coloneqq\thH-\thbar=(1-\prior)\DeltaT.
\]

\begin{figure}[t]
\centering
\includegraphics[width=0.92\linewidth]{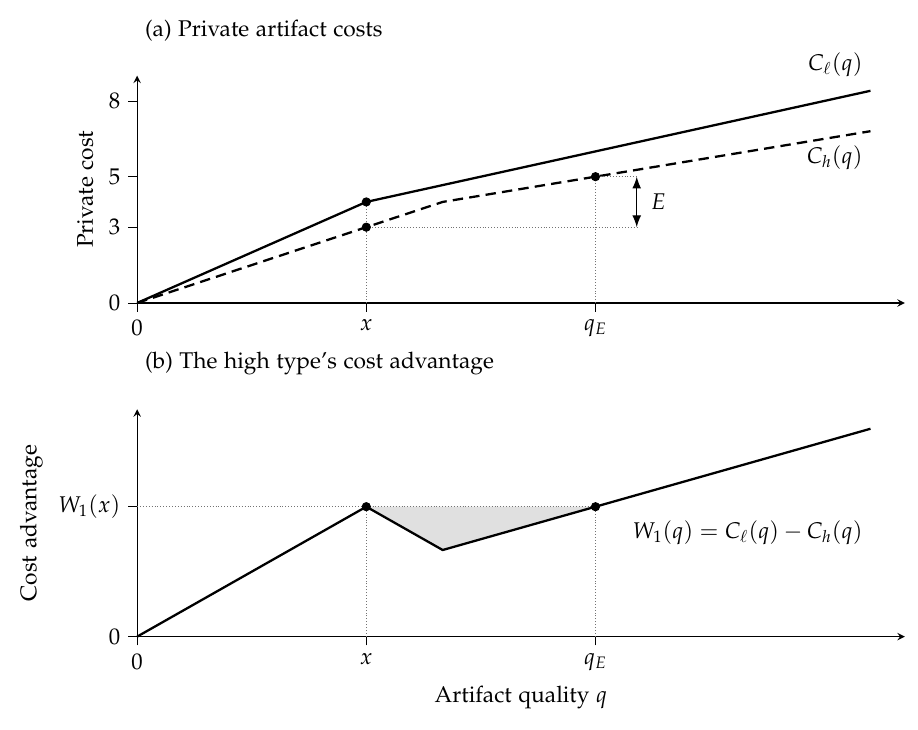}
\caption{\textbf{Adoption and the cost of recognition.} Panel (a) shows private artifact costs; panel (b) shows the high type's cost advantage. Following low-type adoption at $x$, the advantage falls before recovering at $q_E$. Reaching recovery costs the high type an additional $E$.}
\label{fig:ai-signaling-costs}
\end{figure}

\Cref{fig:ai-signaling-costs} illustrates the fall and recovery of the high type's cost advantage following low-type adoption.

\begin{proposition}
\label{prop:one-signal-equilibrium}
Under \cref{eq:one-signal-adoption-regime}, a pure pooling D1 equilibrium exists if and only if
\begin{equation}
\label{eq:one-signal-pooling-condition}
\kappa_\ell x\leq\prior\DeltaT \; \text{ and } \; E\geq\dpool.
\end{equation}
Its common quality is $q=x$. The Riley allocation satisfies D1 if and only if $q_R\leq x$ or $q_R\geq q_E$.
\end{proposition}

The first pooling inequality ensures that the low type is willing to bear the artifact cost in exchange for the prior assessment, rather than choose zero and be identified as low. The second inequality prevents a profitable revealing deviation by the high type. Obtaining a greater cost advantage than at the pool requires quality above $q_E$ and additional expenditure greater than $E$. When $E\geq\dpool$, full recognition cannot compensate for that expenditure.

The equilibrium refinement rule selects the pool whenever these conditions hold. Pooling occurs at positive quality, with the two types using different production methods and bearing different costs.\footnote{\Cref{app:equilibrium-proofs} provides the proof as the zero-capacity case of the two-signal characterization.}

If $x<q_R<q_E$, the Riley allocation fails D1. A deviation to $x$ costs the high type less and gives it a greater cost advantage, so D1 identifies it as high at that cheaper signal. Therefore, eliminating pure pooling need not restore separation. Partial pooling can remain feasible: fewer low types at the shared artifact raise its posterior assessment, allowing their participation to be sustained at a cost that rules out pure pooling. Some tariffs have no pure D1 outcome. 

I call the equality $p+\epsilon_\ell=\kappa_h$ the \emph{price boundary}. At this price, assisted production for the low type and manual production for the high type have equal marginal private costs. \Cref{sec:two-signal} characterizes partial pooling and the selected outcome for every tariff, including offers at the price boundary.

\subsection{The harder-to-substitute signal}
\label{sec:two-signal}

The second signal gives the receiver another way to distinguish types. Under \Cref{ass:comparative-advantage}, hard evidence imposes a given cost on a mimicking low type at a lower private cost to the high type than artifact production does. Its capacity affects both the signals used in separation and the survival of pooling and partial pooling. Throughout this section, the technology $\mathbf d$ and tariff $(F,p)$ are fixed and exogenous.


\paragraph{Least-cost separation.}

In a pure separating PBE, the low type is identified as low and chooses $(0,0)$. Any positive-cost signal receiving the low assessment could be replaced by $(0,0)$, which guarantees an assessment of at least $\thL$. Imitating the high type would raise the low type's assessment by $\DeltaT$, so the high signal must impose a low-type cost of at least $\DeltaT$.

To separate at least private cost, the high type meets this requirement through hard evidence as far as capacity allows. Let $r\coloneqq(\DeltaT-\hardQ)_+$ denote the deterrence requirement remaining after exhausting useful hard capacity.

\begin{lemma}
\label{cor:riley}
The unique least-cost separating signal pair for the high type satisfies
\begin{equation}
\label{eq:riley-signal}
\hardsig_R=\min\{\hardQ,\DeltaT\} \; \text{ and } \; C_\ell(q_R)=r.
\end{equation}
Its private cost is
\begin{equation}
\label{eq:riley-cost}
\rileycost\coloneqq C_h(q_R)+\rhoHard\hardsig_R<\DeltaT.
\end{equation}
Together with the low type at $(0,0)$, these choices support a separating PBE.
\end{lemma}

If artifact production supplies part of the deterrence while hard capacity remains unused, then the high type can replace some artifact quality with hard evidence. This substitution leaves the low type's imitation cost unchanged and reduces the high type's private expenditure. Least-cost separation uses hard evidence up to capacity or until it alone deters imitation, with the artifact supplying any leftover. The imitation constraint binds, and the high type strictly prefers its separating signal to $(0,0)$.

Greater hard capacity reduces the Riley artifact quality and private cost until $\hardQ=\DeltaT$. At and above that capacity, hard evidence alone separates the types. That is, $q_R=0$, $\hardsig_R=\DeltaT$, and $\rileycost=\rhoHard\DeltaT$. At $\hardQ=0$, the allocation reduces to the single-signal benchmark in \cref{eq:one-signal-riley-cost}. Tariff payments enter the private-cost objective, so least-cost separation need not minimize resource use. \Cref{sec:welfare} examines this distinction.


\paragraph{Separation, pooling, and partial pooling.}

This part shows how only the Riley allocation can be a pure separating D1 outcome, but it need not satisfy D1. A shared signal can instead survive when obtaining a greater cost advantage requires the high type to spend more than it gains from recognition.

The benchmark considered offers for which the low type’s marginal private cost of producing artifact quality with AI, $p+\epsilon_\ell$, is strictly below the high type’s marginal cost of manual production, $\kappa_h$. I now also include offers for which these costs are equal, i.e.,
\begin{equation}
\label{eq:shared-signal-regime}
F>0,\quad p+\epsilon_\ell\leq\kappa_h,\; \text{ and } \;
x\coloneqq\frac{F}{\kappa_\ell-p-\epsilon_\ell}\in(0,L).
\end{equation}
For $p+\epsilon_\ell<\kappa_h$, retain the quality $q_E$ and expenditure $E$ defined in \cref{eq:recovery-cost}. At the equality $p+\epsilon_\ell=\kappa_h$, the cost advantage $W_1$ is flat on $[x,q_E]$, where $q_E=\min\{L,F/(\epsilon_\ell-\epsilon_h)\}$. Define $E=C_h(q_E)-C_h(x)$ in this case as well: it is the high type's expenditure across the flat interval. In both cases, quality must exceed $q_E$ to yield a cost advantage greater than $W_1(x)$.

A signal shared by both types in a D1 equilibrium must exhaust hard capacity. Otherwise, a small increase in hard evidence would raise the high type's cost advantage, and D1 would identify it as high at that profitable deviation. In the artifact channel, only the peak at $x$, or the flat interval at the price boundary, can block small revealing deviations. Therefore, the least-quality candidate shared signal is 
\begin{equation}
\label{eq:canonical-pool}
\sigvec^P=(x,\hardQ).
\end{equation}
At quality $x$, the low type subscribes and the high type produces manually. Let $v$ denote the low type's total private cost at this signal. The two types' costs are
\begin{align*}
v\coloneqq\costL(\sigvec^P)&=\kappa_\ell x+\hardQ,\\
\costH(\sigvec^P) & =\kappa_h x+\rhoHard\hardQ.
\end{align*}

When $\prior\DeltaT<v<\DeltaT$, consider a candidate partial pool in which all high types and a fraction $\alpha\in(0,1)$ of low types use $(x,\hardQ)$. The remaining low types choose the origin $(0,0)$, and are identified as low. Low-type indifference and Bayes' rule determine the shared posterior $\belief$ and participation probability,
\begin{equation}
\begin{aligned}
\label{eq:semi-posterior}
\belief&=\frac{v}{\DeltaT},\\
\alpha&=\frac{\prior(1-\belief)}{(1-\prior)\belief}.
\end{aligned}
\end{equation}

The shared assessment $\thL+\belief\DeltaT$ compensates participating low types for a cost exceeding their gain from pure pooling.

\begin{theorem}
\label{thm:pure-d1}
Under \Cref{ass:comparative-advantage}, the following statements hold.
\begin{enumerate}[label=\textup{(\roman*)}]
\item For offers that do not satisfy \cref{eq:shared-signal-regime}, Riley separation is the unique finite-support D1 observable outcome. For offers satisfying that condition, when $p+\epsilon_\ell<\kappa_h$, it satisfies D1 if and only if $q_R\leq x$ or $q_R\geq q_E$. At $p+\epsilon_\ell=\kappa_h$, it always satisfies D1. Every more costly pure separating outcome fails D1;

\item For offers satisfying \cref{eq:shared-signal-regime}, a pure pooling D1 equilibrium exists if and only if
\begin{equation}
\label{eq:pool-existence}
v\leq\prior\DeltaT\; \text{ and } \; E\geq\dpool=(1-\prior)\DeltaT.
\end{equation}
Its least-quality shared signal is $\sigvec^P$, with posterior $\prior$.

\item For offers satisfying \cref{eq:shared-signal-regime}, partial pooling at $(x,\hardQ)$ with probabilities in \cref{eq:semi-posterior} satisfies D1 if and only if
\begin{equation}
\label{eq:semi-condition}
\prior\DeltaT<v<\DeltaT \; \text{ and } \; E\geq\DeltaT-v.
\end{equation}
If also $p+\epsilon_\ell<\kappa_h$ and $E>\DeltaT-v$, this is the unique finite-support D1 observable outcome.
\end{enumerate}
\end{theorem}

The pure-pooling conditions combine participation and deviation incentives. The low type must be willing to bear cost $v$ in exchange for the prior assessment. The high type must spend more than $E$ to obtain a greater cost advantage than at the pool. If its recognition gain $\dpool$ does not exceed $E$, that expenditure cannot be worthwhile.

At a partial pool, the shared assessment is $\thL+v$, so the high type's remaining gain from full recognition is $\DeltaT-v$. A higher shared assessment reduces this gain and can make revelation unprofitable even when pure pooling cannot survive. Reducing hard evidence cannot provide a cheaper revealing deviation because it sacrifices the high type's strongest marginal comparative advantage.\footnote{\Cref{app:equilibrium-proofs} provides the complete deviation checks, including offers at the price boundary and equality in the participation and revealing-deviation conditions.}

Partial pooling can survive after pure pooling becomes infeasible. At some price schedules, it is the only finite-support D1 outcome: low-type mixing supplies a signaling equilibrium when no pure signaling equilibrium survives D1. At the price boundary, the flat cost-advantage interval can support shared outcomes alongside Riley separation.


\paragraph{Selection and interpretation.}

Whenever pure pooling is feasible under D1, the pool at its least-quality shared signal weakly Pareto dominates all other finite-support D1 outcomes for senders. The selection rule outlined in \Cref{sec:model} selects that pool.

If no pure pooling D1 equilibrium exists, the rule selects partial pooling whenever it satisfies D1. When partial pooling and Riley
separation coexist, they give both types the same payoffs, so sender-Pareto maximality does not distinguish them. The additional tie-breaking criterion favors the greatest population mass at signals shared by both types and, therefore, selects partial pooling. If neither pure pooling nor partial pooling satisfies D1, the rule selects Riley separation. The selected signals and on-path receiver posteriors are summarized in \Cref{tab:signaling-outcomes}.

\begin{table}[htbp]
\centering
\renewcommand{\arraystretch}{1.15}
\begin{tabular}{@{}llll@{}}
\toprule
Outcome & Low type & High type & Posterior \\
\midrule
Separation
& $(0,0)$
& $(q_R,\hardsig_R)$
& \begin{tabular}[c]{@{}l@{}}
    $0$ at the origin;\\
    $1$ at the high signal
  \end{tabular}
\\[6pt]
Pooling
& $(x,\hardQ)$
& $(x,\hardQ)$
& $\prior$
\\[6pt]
Partial pooling
& \begin{tabular}[c]{@{}l@{}}
    $(x,\hardQ)$ with probability $\alpha$;\\
    $(0,0)$ otherwise
  \end{tabular}
& $(x,\hardQ)$
& \begin{tabular}[c]{@{}l@{}}
    $\belief$ at the shared signal;\\
    $0$ at the origin
  \end{tabular}
\\
\bottomrule
\end{tabular}
\caption{\textbf{Selected signaling outcomes.} Signal choices and on-path posterior probabilities of high ability under the selection rule, for a fixed technology, tariff, and hard-signal capacity.}
\label{tab:signaling-outcomes}
\end{table}

This rule supplies a signaling outcome after every admissible tariff. Some offers have a unique finite-support D1 observable outcome. The provider anticipates these responses and influences incentives through its technology and tariff. It has no contractual control over sender behavior or receiver beliefs. \Cref{supp:alternative-selection} discusses alternative selection assumptions.

Hard capacity changes both the cost of sharing a signal and the expenditure required for separation. Eliminating pure pooling alone need not restore full information, since partial pooling may remain. Capacity $\hardQ\geq\DeltaT$ guarantees separation throughout the technology class. \Cref{sec:market} allows the provider to choose its tariff and technology. \Cref{sec:capacity} studies verification when the provider chooses both design and prices.


\section{The market for AI-assisted signaling}

\subsection{A monopolistic provider}
\label{sec:market}

The provider chooses both the technology and its two-part tariff. Pricing changes adoption and the cost of imitation, and can affect the information conveyed by signals. I first solve the pricing problem for a developed technology $\mathbf d=(\epsilon_\ell,\epsilon_h,L)$ at a given hard capacity $\hardQ$, then consider feature choice and launch. Operating profit excludes the development cost $K(\mathbf d)$, which is sunk when prices are set.

Let $w_i$ denote the resource saving from replacing one unit of type $i$'s manual production with AI. That is, for $i\in\{\ell,h\}$,
\begin{equation}
\label{eq:production-savings}
w_i\coloneqq\kappa_i-\epsilon_i-c.
\end{equation}
Because $w_\ell>w_h$, AI can save resources for the low type even when it does not save resources for the high type. These savings bound profitable provision. The provider must also choose prices that support the intended signaling outcome, subject to the selection rule outlined in \Cref{sec:model}.


\paragraph{Pricing a separating product.}

At separation, the low type supplies only the basic artifact and uses no paid AI input, so any revenue must come from high types. The artifact they supply depends on the cost of low-type imitation. Clearly, the provider cannot choose how much input to sell independently of signaling incentives.

Retain the remaining deterrence requirement $r=(\DeltaT-\hardQ)_+$ from \Cref{sec:two-signal}, and let $\delta\coloneqq\epsilon_\ell-\epsilon_h>0$ denote the difference in editing effort.

\begin{proposition}
\label{prop:monopoly-separation}
For a fixed technology and capacity, the maximum operating profit among tariffs selecting separation is
\begin{equation}
\label{eq:profit-separation-general}
R(\mathbf d,\hardQ)
=\prior(w_h)_+\min\left\{L,\frac{r}{\kappa_h+\delta}\right\}.
\end{equation}
If $w_h>0$ and $r>0$, let $z_*\coloneqq\min\{L,r/(\kappa_h+\delta)\}$. Every tariff satisfying
\begin{equation}
\label{eq:separating-tariff}
\kappa_h-\epsilon_\ell<p<\kappa_h-\epsilon_h \;\text{ and }\;
F=(\kappa_h-p-\epsilon_h)z_*
\end{equation}
attains this value. The high type subscribes, uses input $z_*$, and supplies
\begin{equation}
\begin{aligned}
\label{eq:monopoly-separating-output}
q_R&=z_*+\frac{r-(\kappa_h+\delta)z_*}{\kappa_\ell},\\
\hardsig_R&=\min\{\hardQ,\DeltaT\}.
\end{aligned}
\end{equation}
The low type does not subscribe.
\end{proposition}

The high type can avoid subscribing by producing the same observable artifact manually. Its adoption decision compares the costs of the two production methods at that artifact. At a profitable optimum, the fee extracts its production saving after usage payments. The high type is indifferent and subscribes under the subscription rule. Its total payment is $(\kappa_h-\epsilon_h)z_*$. Subtracting compute cost leaves operating profit $w_hz_*$ per high type.

Although the low type does not subscribe in the selected outcome, it would use AI to imitate the high type's artifact: its production saving is greater than the high type's. With the high type's adoption constraint binding, the mimicking low type's private cost of the assisted part is $(\kappa_h+\delta)z_*$. The binding imitation constraint limits input to $r/(\kappa_h+\delta)$, while coverage limits it to $L$. When $w_h>0$, profit rises with input, so optimal pricing induces the greatest quantity consistent with these limits. Any remaining deterrence is supplied through manual artifact production.

The usage-price interval in \cref{eq:separating-tariff} ensures that Riley separation is selected. Several fee--price pairs implement the same optimum. Their relation at the optimum follows from the high type's adoption constraint, even though membership and usage prices are independent choices.\footnote{\Cref{app:market-proofs} establishes the profit bound over all separating offers.}


\paragraph{Pricing for pooling and partial pooling.}

At the least-quality shared signal $(x,\hardQ)$ characterized in \Cref{sec:two-signal}, participating low types subscribe and high types produce manually. The low type is indifferent between the two production methods at quality $x$, so
\begin{equation}
\label{eq:shared-payments}
F=(\kappa_\ell-p-\epsilon_\ell)x \implies F+px=(\kappa_\ell-\epsilon_\ell)x.
\end{equation}
The total payment equals the low type's avoided manual effort less its editing effort. Subtracting compute cost leaves operating profit $w_\ell x$ per subscriber. For a given shared quality, both the total payment and this profit are unchanged by the division between membership and usage charges.\footnote{Adoption at indifference matters differently for the two operating values. A small fee reduction makes high-type adoption strict at a separating offer and approaches $R$. At a selected shared outcome, reducing the fee also moves the canonical quality to the new low-type adoption threshold, where assisted and manual production are again tied. Selecting manual production at these ties instead gives zero shared sales and $H=0$ under the signaling selection rule. \Cref{app:production-indifference} characterizes the minimizing production choices and their profit and welfare consequences.}

However, that division matters for revealing deviations. Changing $p$ and adjusting $F$ to preserve $x$ changes the cost of larger artifacts and the high type's recovery expenditure $E$. Prices must satisfy the pooling or partial-pooling deviation condition in \cref{eq:pool-existence,eq:semi-condition}. A lower usage price combined with a higher membership fee can change whether a shared outcome survives, even though the bill at the shared artifact stays fixed.

Let $\mathcal X(\mathbf d,\hardQ)$ denote the set of shared qualities $x\in(0,L)$ for which some tariff satisfying \cref{eq:shared-signal-regime} selects pooling or partial pooling. To support such a quality, the provider chooses $p\in[0,\kappa_h-\epsilon_\ell]$, sets the fee according to \cref{eq:shared-payments}, and satisfies the appropriate participation and deviation conditions in \Cref{thm:pure-d1}. These conditions include $\kappa_\ell x+\hardQ<\DeltaT$. Offers at the price boundary, $p=\kappa_h-\epsilon_\ell$, are allowed.\footnote{ \Cref{app:market-proofs} derives the feasible quality set and implementing prices.}

For $x\in\mathcal X(\mathbf d,\hardQ)$, the mass of subscribing low types is
\begin{equation}
\label{eq:low-type-demand}
D_\ell(x,\hardQ)
=\min\left\{1-\prior,\frac{\prior(\DeltaT-\kappa_\ell x-\hardQ)}{\kappa_\ell x+\hardQ}\right\}.
\end{equation}
In pure pooling, all low types subscribe, giving the first term. In partial pooling, the subscribing mass is $(1-\prior)\alpha$, with $\alpha$ determined by \cref{eq:semi-posterior}, giving the second term.

\begin{proposition}
\label{prop:monopoly-shared-value}
The best operating value from pooling or partial pooling, including the option of earning zero, is
\begin{equation}
\label{eq:profit-pooling-general}
H(\mathbf d,\hardQ)
=(w_\ell)_+\max_{x\in\mathcal X(\mathbf d,\hardQ)}xD_\ell(x,\hardQ).
\end{equation}
Set $H=0$ if $\hardQ\geq\DeltaT$ or $\mathcal X$ is empty. Whenever $H>0$, a maximizing quality and a feasible fee--price pair attain this value under the specified equilibrium selection and private-production rules.
\end{proposition}

When $w_\ell>0$, a larger shared artifact raises profit per subscriber. In pure pooling, every low type continues to subscribe, so profit increases with quality. In partial pooling, higher quality reduces the subscribing mass as fewer low types are willing to bear the cost of joining the shared signal. The provider balances profit per subscriber against participation. Either pooling or partial pooling can maximize the shared operating value, subject to coverage and revealing-deviation constraints.\footnote{\Cref{supp:shared-quality-calculations} gives the complete constrained calculation.}

Optimal pricing can include a usage subsidy. For example, an optimal tariff for a fixed technology can set $p=0<c$ and $F=0.34965$, supporting pooling at $x=0.035$ and earning operating profit $0.01323$.\footnote{Take $(\kappa_\ell,\kappa_h)=(10,8)$, $(\epsilon_\ell,\epsilon_h)=(0.01,0.002)$, $L=0.5$, $c=8.1$, $\prior=0.8$, $\DeltaT=1$, $\hardQ=0.45$, and $\rhoHard=0.02$. Then $w_h=-0.102$, $w_\ell=1.89$, and $E=0.33465>\dpool=0.2$. \Cref{supp:usage-subsidy} verifies global optimality for this technology.} Membership revenue finances the compute bill.\footnote{Zero monetary usage charges occur in AI markets. Google's \href{https://ai.google.dev/gemini-api/docs/pricing}{Gemini API pricing page}, for example, lists free input and output tokens for selected models within its free tier (accessed October 13, 2026). The model allows compute expenditure despite a zero usage price and shows how positive membership revenue can finance it.} The low type's avoided manual effort exceeds editing and compute costs, allowing positive operating profit despite usage priced below resource cost. An unconditional cash rebate has different incentives because it can be collected without using AI.\footnote{\Cref{supp:cash-rebates} examines that comparison.}


\paragraph{Feature choice, launch, and the supplied outcome.}

For a developed technology, the provider's best operating return is the larger of the separating and shared values, i.e., 
\begin{equation}
\label{eq:monopoly-operating-value}
G(\mathbf d,\hardQ)\coloneqq\max\{R(\mathbf d,\hardQ),H(\mathbf d,\hardQ)\}.
\end{equation}
If $R>H$, an optimal offer selects separation. If $H>R$, it selects pooling or partial pooling at a maximizing shared quality. When the values coincide, optimal offers can support different outcomes.

Feature choice compares each design's operating return with its development cost $K(\mathbf d)$. This cost may be zero. Once paid, it is sunk. A provider maximizes operating profit without subtracting development cost again.

\begin{proposition}
\label{thm:product-bifurcation}
Under \Cref{ass:comparative-advantage}, the monopolist's maximum profit is
\begin{equation}
\label{eq:monopoly-design-choice}
\profit^*(\hardQ)
=\max\left\{0,\max_{\mathbf d\in\mathcal D}\big[G(\mathbf d,\hardQ)-K(\mathbf d)\big]\right\}.
\end{equation}
A strictly positive value induces the development of a maximizing design and a tariff attaining its operating value. A strictly negative best net development return induces no launch. At a zero best net return, launch and no launch are both optimal.
\end{proposition}

Editing improvements affect the two operating values differently. The comparison below holds manual costs, compute cost, coverage, hard capacity, the hard-signal cost ratio, and the type distribution fixed. Each change must remain feasible under \Cref{ass:comparative-advantage}.

\begin{corollary}
\label{prop:monopoly-editing-incentives}
A reduction in $\epsilon_\ell$, holding $\epsilon_h$ fixed, weakly raises both $R$ and $H$. A reduction in $\epsilon_h$, holding $\epsilon_\ell$ fixed, weakly raises $R$ and weakly lowers $H$. If $R>0$, reducing $\epsilon_\ell$ strictly raises expected signaling resource cost at the optimal separating allocation in \cref{eq:monopoly-separating-output}.
\end{corollary}

Improving low-type productivity increases the production savings available from shared customers and makes revealing deviations more expensive. At separation, it also lowers the editing cost of imitation, requiring a larger artifact and increasing resource use. Improving high-type productivity increases the savings available from separating customers, but makes revelation from a shared signal cheaper. These operating incentives do not rank net design returns without specifying development costs.\footnote{\Cref{app:market-proofs} gives the existence argument and tariff constructions. Attainment uses adoption at indifference and includes offers at the price boundary and equality in the participation and revealing-deviation conditions. \Cref{app:monopoly-editing-incentives} proves the comparisons. \Cref{supp:monopoly-improvements} solves a bounded quadratic-cost monopoly example.}

Hard capacity changes both operating values, and the provider can respond by adjusting prices. At $\hardQ\geq\DeltaT$, hard evidence alone deters imitation, so $R=H=0$ and no technology with positive development cost is launched. Below that level, the outcome depends on which separating and shared offers remain feasible and profitable. \Cref{sec:competition} next examines pricing and development with competing providers. \Cref{sec:capacity} allows a provider to choose design and prices as capacity changes.

\begin{figure}[t]
\centering
\includegraphics[width=0.92\linewidth]{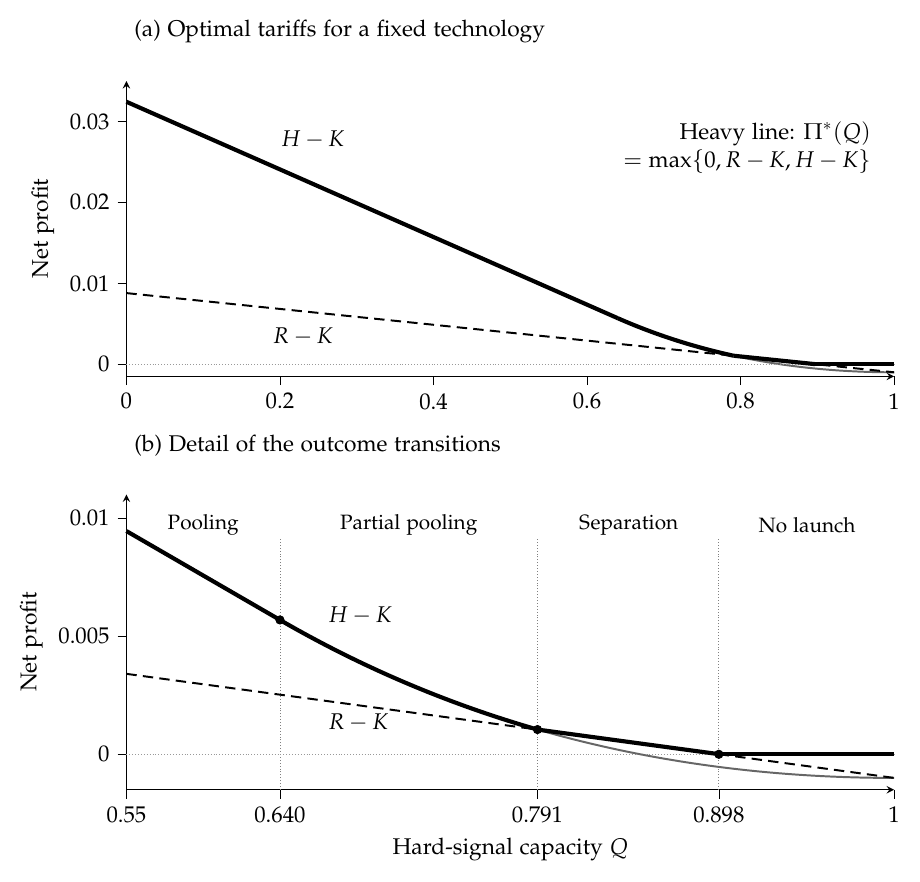}
\caption{\textbf{Optimal pricing and prospective launch.} The thin curves show $R-K$ (dashed) and $H-K$ (gray). The heavy curve includes the no-launch option. Panel (b) enlarges the outcome transitions. Both prices are reoptimized at each capacity. The launch decision concerns prospective development.}
\label{fig:provider-profit-envelope}
\end{figure}

\Cref{fig:provider-profit-envelope} illustrates optimal pricing and the launch decision for one fixed technology. In this illustration, pooling gives way to partial pooling before separation becomes more profitable. At still greater capacity, operating profit no longer covers the cost of fresh development.\footnote{\Cref{fig:provider-profit-envelope}'s parameter values are given in \Cref{app:figure-parameters}.}

\subsection{Competition among AI providers}
\label{sec:competition}

Competition affects both prices and incentives to develop a technology. I first study pricing among providers whose identical technologies are already available, then consider their decisions to develop those technologies before setting tariffs. Physical features are fixed in both comparisons. \Cref{sec:design-competition} subsequently lets two established providers choose improvements before pricing.


\paragraph{Pricing with identical technologies.}

Suppose a finite set $\mathcal J$ of at least two providers has developed the same technology $\mathbf d=(\epsilon_\ell,\epsilon_h,L)$, with common compute cost $c$. Each provider $j\in\mathcal J$ independently chooses $F_j\geq0$ and $p_j\geq0$. Development costs have already been incurred and are sunk. Offers are public, while individual subscriptions and AI input remain private.

Subscriptions are nonexclusive. Identical tools assist the same parts of the artifact, so subscribing to several providers does not expand coverage beyond $L$. Coverage thus represents a technological limit, such as the range of tasks a tool can perform. A sender with several subscriptions can concentrate input in the plan with the lowest usage price and drop the others without increasing production cost. With nonnegative membership fees, one plan suffices to attain minimum cost.

Let $\mathcal M=\{F_j,p_j\}_{j\in\mathcal J}$ denote the public menu. Type $i$'s minimized artifact cost is
\begin{equation}
\label{eq:identical-menu-cost}
C_i(q\mid\mathcal M)
=\min_{j\in\mathcal J}C_i(q\mid\mathbf d,F_j,p_j),
\end{equation}
where each single-offer cost in \cref{eq:cost-envelope} already includes the option of manual production. The subscription rule in \Cref{sec:model} applies, and ties between identical providers can be split symmetrically. The equilibrium selection rule applies to the resulting cost functions.\footnote{ \Cref{app:competition-proofs} establishes finite-support D1 signaling equilibria after every such menu, including those induced by unilateral tariff deviations.}

\begin{proposition}
\label{prop:provider-competition}
With at least two developed identical technologies, the tariffs
\begin{equation}
\label{eq:competitive-price}
F_j=0\; \text{ and }\; p_j=c,
\end{equation}
for $j\in\mathcal J$, form a pure pricing Nash equilibrium. Each provider earns zero operating profit. The unique finite-support D1 observable outcome is Riley separation, and its signal choices minimize resource expenditure among separating allocations for this technology menu.
\end{proposition}

The pricing argument rests on the availability of an identical rival tool. A sender can switch providers without changing artifact quality, editing effort, or the observed signal. To attract that sender, a deviating provider must charge a total bill no greater than the rival's compute-cost bill. Its revenue cannot exceed its own compute expense. This bound holds even if the deviation changes signaling behavior. Indeed, at any resulting signal, the sender can still substitute the rival's tool.

At the tariffs in \cref{eq:competitive-price}, private artifact cost becomes
\begin{equation}
\label{eq:competitive-artifact-cost}
C_i(q\mid\mathcal M)
=\min\{\kappa_i,c+\epsilon_i\}\min\{q,L\}
+\kappa_i(q-L)_+.
\end{equation}
Private production costs coincide with resource costs, and the high type's cost advantage increases with quality. A small increase in quality from a shared signal would induce D1 to attribute the deviation to the high type, which would gain from full recognition. Shared signals cannot survive. Under \Cref{ass:comparative-advantage}, separation uses hard evidence first, with $\hardsig_R=\min\{\hardQ,\DeltaT\}$ and $C_\ell(q_R\mid\mathcal M)=(\DeltaT-\hardQ)_+$.

The resource-efficiency result holds for the available technology menu and its imitation constraints. Cheaper production also makes imitation cheaper, which can increase the artifact quality required for separation. Welfare comparisons with no paid AI or with optimal monopoly pricing must account for this response. \Cref{sec:welfare} develops those comparisons. At the competitive tariffs, providers earn no operating return to cover development costs. The next part examines whether they undertake development in the first place.


\paragraph{Development before pricing.}

Development costs can prevent the competitive pricing outcome from arising. Suppose potential providers first decide whether to develop a copy of the fixed technology at cost $K>0$, then observe entry and set tariffs. If maximum monopoly operating profit exceeds $K$, an equilibrium has one entrant charging an optimal monopoly tariff. Additional entry would lead to zero operating profit under the competitive pricing equilibrium in \Cref{prop:provider-competition}, leaving the entrant unable to recover its development cost. Potential competition can therefore coexist with pooling or partial pooling.

\Cref{app:provider-entry} gives the formal construction, the cases $G=K$ and $K=0$, and a pooling example. Other pricing equilibria may support different entry outcomes. The construction allows repricing after entry; simultaneous entry and binding tariff offers change the entry incentives, as shown in \Cref{supp:simultaneous-entry}. These comparisons hold physical features fixed. Differentiated technologies need not support compute-cost pricing, and the next section allows two established providers to improve their tools before choosing prices.


\subsection{Design competition between providers}
\label{sec:design-competition}

I next let two established providers choose technology improvements before setting prices. Initial development costs are sunk, and either provider can retain its baseline technology at zero additional cost. Holding entry and exit fixed isolates improvement incentives from the entry problem studied in \Cref{sec:competition}.

Both providers initially possess editing requirements $(\epsilon_h^0,\epsilon_\ell^0)$. Write $\delta_0=\epsilon_\ell^0-\epsilon_h^0$. Provider $j$ independently improves the two requirements through
\begin{equation}
\begin{aligned}
\label{eq:ed-design}
(u_j,v_j)&\in[0,\bar u]\times[0,\bar v],\\
\epsilon_{hj}&=\epsilon_h^0-u_j,\\
\epsilon_{\ell j}&=\epsilon_\ell^0-v_j.
\end{aligned}
\end{equation}
Its assisted cost gap is $\delta_j=\delta_0+u_j-v_j$. This permits improvements to either requirement, both, or neither. Incremental expenditure is $K(u,v)=k_hu^2/2+k_\ell v^2/2$, with $k_h,k_\ell>0$, and is incurred regardless of subsequent sales.

Providers simultaneously choose improvements, observe both designs, and then simultaneously choose $F_j\geq0$ and $p_j\geq0$. Usage below compute cost remains feasible. Designs and tariffs are public. Senders subsequently choose private subscriptions, production methods, and signals. The payoffs and equilibrium selection rule in \Cref{sec:model} remain unchanged.

I hold coverage fixed at $L=\infty$ to study editing improvements and pricing when coverage limits do not bind.\footnote{Sufficiently large finite coverage gives the same relevant signaling and pricing game. Under nonnegative tariffs, artifact production costs either type at least $(\epsilon_h^0-\bar u)q$. A quality exceeding $\DeltaT/(\epsilon_h^0-\bar u)$ costs more than the largest possible assessment gain and cannot be used in equilibrium or provide a profitable deviation. Taking $L>\DeltaT/(\epsilon_h^0-\bar u)$ makes coverage slack throughout the relevant choices.} Overlapping coverage and nonexclusive subscriptions remain. Each tool can assist the entire artifact, so a sender can concentrate production in its cheapest subscribed tool and discard unnecessary positive-fee plans. Binding coverage could instead require some quality to be produced manually. Its artifact cost is
\begin{equation}
\label{eq:ed-menu-cost}
C_i(q)=\min\left\{\kappa_iq,\ F_1+(p_1+\epsilon_{i1})q,\ F_2+(p_2+\epsilon_{i2})q\right\}.
\end{equation}
At equal costs, the sender uses the tool requiring less editing for its type, and identical ties are split symmetrically. This additional purchase tie-rule permits the attainment of the price below. Adoption relative to manual production follows the existing rule.

The feasible improvements satisfy
\begin{equation}
\label{eq:ed-feasible}
\begin{aligned}
0<\epsilon_h^0<\epsilon_\ell^0<\kappa_h,\\
0<\bar u<\epsilon_h^0,\\
0<\bar v<\delta_0,\\
\delta_0+\bar u<\kappa_\ell-\kappa_h.
\end{aligned}
\end{equation}
These inequalities preserve the editing-cost restrictions in \Cref{ass:comparative-advantage}. Two further sufficient conditions make the differentiated pricing game tractable, i.e.,
\begin{equation}
\label{eq:ed-separating-restrictions}
\begin{aligned}
c+\epsilon_h^0<\epsilon_\ell^0-\bar v,\\
0<\rhoHard<\frac{\epsilon_h^0-\bar u}{\kappa_\ell}.
\end{aligned}
\end{equation}
The productivity condition compares the baseline high type's compute-plus-editing cost with the smallest feasible low-type editing requirement. In the pricing equilibrium below, a rival with zero membership fee bounds the high type's private unit cost by $c+\epsilon_h^0$, while nonnegative usage prices keep a deviating plan's low-type marginal cost at least $\epsilon_\ell^0-\bar v$. Their strict ordering rules out a decreasing cost-advantage interval after unilateral tariff deviations. This condition is stronger than requiring AI to save resources relative to manual production. The hard-evidence condition makes hard evidence privately preferable at the margin across heterogeneous menus. Neither restriction is imposed on the monopoly analysis.

\paragraph{Prices and improvements.}

For $\hardQ<\DeltaT$, let $r=\DeltaT-\hardQ$ and $m_i^0=c+\epsilon_i^0$. After designs are chosen, order providers so that $u_1\geq u_2$, and set $\delta_*=\min\{\delta_1,\delta_2\}$. When $u_1>u_2$, I call provider~1 the \emph{leader}, referring to its lower high-type editing requirement.

\begin{proposition}
\label{prop:ed-pricing}
Suppose that $\hardQ<\DeltaT$, and \cref{eq:ed-feasible,eq:ed-separating-restrictions} hold. Every feasible design pair has a pure pricing equilibrium
\begin{equation*}
\begin{aligned}
F_1=F_2=0,\\
p_1=c+u_1-u_2,\\
p_2=c.
\end{aligned}
\end{equation*}
Its unique finite-support D1 observable outcome separates, with the low type at the origin and
\begin{equation}
\label{eq:ed-signals}
q_R=\frac{r}{m_h^0-u_2+\delta_*}
\; \text{ and } \; \hardsig_R=\hardQ.
\end{equation}
If $u_1>u_2$, provider 1 supplies high types, and operating profits are
\begin{equation*}
\label{eq:ed-profits}
\profit_1^{\mathrm{op}}=\frac{\prior r(u_1-u_2)}{m_h^0-u_2+\delta_*}
\; \text{ and } \; \profit_2^{\mathrm{op}}=0.
\end{equation*}
If $u_1=u_2$, both operating profits are zero. The separating allocation minimizes real signaling expenditure among separating allocations for this tariff menu and its private imitation constraints.
\end{proposition}

The leader's usage markup offsets its high-type editing advantage. High types face equal private costs from the two tools and use the leader because the purchase rule favors the tool with lower high-type editing effort at equal private cost. The rival constrains the price despite making no sales. Low types use the origin $(0,0)$, but their cheapest available method of imitation determines the artifact needed for separation. Their cost depends on the smaller assisted gap, which need not belong to the leader.

The proof checks positive fees and usage subsidies and constructs the displayed pricing equilibrium after every design history. With a rival charging no membership fee, \cref{eq:ed-separating-restrictions} rules out shared signals at the proposed menu and after unilateral tariff deviations. Other tariff menus can pool or partially pool.\footnote{\Cref{app:ed-proofs} provides their signaling equilibria. Pricing uniqueness is not established.}

Design choices anticipate these pricing decisions. Pure asymmetric improvement equilibria exist even though the providers have identical primitives. For example, when low-type improvement is sufficiently costly, one provider chooses $u^*=\prior r/(k_hm_\ell^0)$ and $v=0$, and its rival retains the baseline.\footnote{\Cref{prop:ed-pure-design} gives sufficient conditions and checks the entire design choice. \Cref{supp:ed-mixed-design} gives a symmetric mixed design equilibrium under those conditions.} Positive-cost identical designs earn zero operating profit, while identical baseline designs admit a profitable small improvement. Hence, no pure symmetric design equilibrium exists.

\paragraph{Information and resource expenditure.}

Design incentives can also support improvements to both editing requirements. Compare the design equilibrium with the pricing equilibrium in which both AI tools are held at their baseline and charge $F=0$, $p=c$. This benchmark retains paid AI and holds $\hardQ$ fixed. Initial establishment costs are common and sunk, so they cancel from the welfare comparison in \Cref{prop:welfare-identity}. I suppress these common costs here, in \Cref{tab:ed-resources}, and in the corresponding welfare calculation in \Cref{app:ed-proofs}. Thus, $K_{\mathrm{tot}}=K(u_1,v_1)+K(u_2,v_2)$ records only incremental improvement expenditure in this comparison.

\begin{proposition}
\label{prop:ed-harmful-improvement}
There are primitives satisfying \cref{eq:ed-feasible,eq:ed-separating-restrictions} for which the sequential game has an equilibrium with pure provider strategies in which one provider improves both editing requirements and earns positive net profit, while its rival retains the baseline. Welfare is lower than at the fixed-design baseline for every $\receiverweight>0$, although both outcomes fully reveal type.
\end{proposition}

In the example constructed in \Cref{app:ed-welfare-example}, higher productivity increases the artifact required for separation and total resource expenditure. \Cref{tab:ed-resources} summarizes the example's comparison.

\begin{table}[!htbp]
\centering
\small
\begin{tabular}{lrr}
\toprule
& Fixed baseline & Chosen designs \\
\midrule
Separating artifact $q_R$ & .458333 & .534520 \\
Signaling resources $C^{\mathrm{real}}$ & .263867 & .272767 \\
Improvement expenditure $K_{\mathrm{tot}}$ & .000000 & .015743 \\
Leader's net profit & .000000 & .018022 \\
\bottomrule
\end{tabular}
\caption{\textbf{Design improvements and resources.} Expected costs average over sender types. Both outcomes separate and use $\hardsig_R=.45$. Establishment and capacity costs are common.}
\label{tab:ed-resources}
\end{table}

Low-type improvement makes imitation cheaper. The high type must produce a larger artifact to separate, increasing the leader's usage revenue even though low types purchase no input on path. Here, the increased artifact and improvement expenditure outweigh the production saving, lowering welfare by approximately $.024643$. Conditional resource efficiency in \Cref{prop:ed-pricing} holds the technology menu fixed. It does not make the privately chosen improvements socially desirable. The example establishes a loss relative to retaining the baseline tools. Total signaling and improvement resources nevertheless remain below those required for manual separation in this example.

Verification can save resources even when ability is already fully revealed. With designs installed, more hard capacity replaces artifact production and lowers resource expenditure under \cref{eq:ed-separating-restrictions}. \Cref{supp:ed-resources-verification} gives the resource calculations and a game in which capacity precedes improvements and the receiver chooses no additional verification, although an expansion improves welfare. These comparisons concern separating outcomes with provider numbers fixed. Competitive pooling and a general welfare ranking remain uncharacterized.

\section{Welfare}
\label{sec:welfare}

AI changes both the resources needed to produce signals and the information those signals convey. Provider profit captures neither effect in full. I evaluate welfare by adding sender utility, provider profits, and receiver utility, then subtracting the resource cost of providing hard-signal capacity. The common basic-artifact cost is suppressed, as in \Cref{sec:model}. Signal quality has no direct productive benefit beyond its effect on assessment in this welfare criterion.


\paragraph{Accounting for information and resources.}

Let $C^{\mathrm{real}}$ denote the expected resource cost of manual production, editing, compute, and hard-signal effort actually incurred. Let $K_{\mathrm{tot}}$ be total development and improvement expenditure on all available technologies, including unused ones, and let $\capcost(\hardQ)\geq0$ be the resource cost of making hard capacity available. Define the \emph{expected squared assessment error} as
\begin{equation}
\label{eq:assessment-error}
\mathrm{MSE}\coloneqq\E\left[\left(\theta-\E\left[\theta\mid\sigvec\right]\right)^2\right].
\end{equation}
The receiver's expected loss is $\receiverweight\mathrm{MSE}/2$. Full separation has zero error, whereas pooling and partial pooling leave uncertainty about types at the shared signal.

\begin{lemma}
\label{prop:welfare-identity}
For every Bayes-consistent signaling outcome with posterior-mean assessments, including mixed outcomes and multiple providers, welfare satisfies
\begin{equation}
\label{eq:welfare-master}
\welfare
=\thbar-C^{\mathrm{real}}-K_{\mathrm{tot}}-\capcost(\hardQ)
-\frac{\receiverweight}{2}\mathrm{MSE}.
\end{equation}
\end{lemma}

The posterior-mean assessment has an unconditional mean of $\thbar$. More informative signals improve accuracy, but leave aggregate gross assessment unchanged. Membership and usage payments transfer surplus from senders to providers and cancel when their payoffs are added. Manual production, editing, compute, and hard-signal effort remain resource costs. Tariffs affect welfare through adoption and signaling, while the payments themselves are transfers.

I use subscripts $R$, $P$, and $M$ for Riley separation, pooling, and partial pooling, respectively, and $0$ for the economy without paid incremental AI. Under an optimal monopoly separating offer with positive operating profit, as in \Cref{prop:monopoly-separation}, the low type uses the origin $(0,0)$. The high type uses AI input $z_*$ and produces the remaining quality $q_R-z_*$ manually. Expected resource cost and assessment error are
\begin{equation}
\begin{aligned}
\label{eq:separating-resource-cost}
C_R^{\mathrm{real}}
&=\prior\left[(\epsilon_h+c)z_*+\kappa_h(q_R-z_*)+\rhoHard\hardsig_R\right],\\
 \mathrm{MSE}_R&=0.
\end{aligned}
\end{equation}
This expression uses the optimal production quantities in \cref{eq:monopoly-separating-output}. At another tariff, resource accounting must use the actual input and effort choices induced by that offer.

For pooling or partial pooling at $(x,\hardQ)$, retain the subscribing low-type mass $D_\ell=D_\ell(x,\hardQ)$ from \cref{eq:low-type-demand}, and the shared posterior $\belief$. Participating low types use AI input $x$, whereas high types produce the artifact manually. Resource cost and assessment error are
\begin{align}
\label{eq:shared-resource-cost}
C_{P,M}^{\mathrm{real}}
&=D_\ell\left[(\epsilon_\ell+c)x+\hardQ\right]
+\prior\left[\kappa_hx+\rhoHard\hardQ\right],\\
\label{eq:shared-assessment-error}
\mathrm{MSE}_{P,M}&=\prior(1-\belief)\DeltaT^2.
\end{align}
In pure pooling, $D_\ell=1-\prior$ and $\belief=\prior$. In partial pooling, $D_\ell=(1-\prior)\alpha$ and $\belief=(\kappa_\ell x+\hardQ)/\DeltaT$. Low types at the origin incur no incremental signaling cost and are identified. The shared group has mass $\prior/\belief$ and conditional type variance $\belief(1-\belief)\DeltaT^2$, yielding \cref{eq:shared-assessment-error}.

For comparison, consider the economy without paid incremental AI. Write $r=(\DeltaT-\hardQ)_+$. Manual Riley separation has artifact quality $r/\kappa_\ell$ and hard signal $\hardsig_R=\min\{\hardQ,\DeltaT\}$, giving resource cost
\begin{equation}
\label{eq:baseline-resource-cost}
C_0^{\mathrm{real}}
=\prior\left[\frac{\kappa_h}{\kappa_\ell}r+\rhoHard\hardsig_R\right].
\end{equation}
This outcome reveals type perfectly and incurs no incremental development cost. The comparisons below hold capacity and its cost fixed. 


\paragraph{Separating products and Riley efficiency.}

Fix a technology with $w_h>0$ and $\hardQ<\DeltaT$, and compare its optimal monopoly separating offer with the economy without paid incremental AI. Define 
\[
s\coloneqq\kappa_\ell-\kappa_h-\delta>0,
\]
where $\delta=\epsilon_\ell-\epsilon_h$ was defined in \Cref{sec:market}. The quantity $s$ measures how much assistance narrows the effort difference between types. The optimal artifact in \cref{eq:monopoly-separating-output} can be written as
\begin{equation*}
q_R=\frac{r}{\kappa_\ell}+\frac{s}{\kappa_\ell}z_*.
\end{equation*}
The first term is the artifact quality required for manual separation. Assistance makes imitation cheaper, so separation requires the additional quality in the second term.

Let $U_S^R$ and $U_S^0$ be aggregate sender payoffs with the separating product and without paid incremental AI, respectively. Their difference, and the welfare effect of a fresh separating launch, are
\begin{align}
\label{eq:separating-sender-surplus-change}
U_S^R-U_S^0
&=-\prior\frac{\kappa_hs}{\kappa_\ell}z_*,\\
\label{eq:social-value-separation-product}
\welfare_R-\welfare_0
&=\prior\left(w_h-\frac{\kappa_hs}{\kappa_\ell}\right)z_*-K(\mathbf d).
\end{align}
The provider's net return is $\prior w_hz_*-K(\mathbf d)$. Its tariff leaves the high type indifferent between assisted and manual production at the required artifact. The increase in quality raises the high type's private expenditure relative to manual separation, and reduces aggregate sender payoff. In the welfare comparison, the production saving $w_hz_*$ must be weighed against this additional signaling and the development cost. Both outcomes reveal type perfectly, so the comparison is independent of $\receiverweight$.

Even at a tariff that maximizes monopoly profit for a given technology, a profitable separating launch can lower welfare relative to manual signaling at the same capacity.\footnote{\Cref{supp:welfare-examples} gives an example in which both outcomes fully reveal ability but assisted signaling uses more resources.}

A second efficiency question arises within a given tariff. Riley minimizes the high type's private cost of deterring imitation. With a usage markup, private costs and resource costs can rank separating allocations differently.

\begin{proposition}
\label{prop:riley-resource-inefficiency}
Under \Cref{ass:comparative-advantage}, there exist technologies and capacities for which a tariff maximizing monopoly profit for the given technology selects Riley separation, yet another pure separating PBE under the same tariff uses strictly fewer resources. The alternative has a greater high-type private signaling cost and fails D1.
\end{proposition}

When AI input is cheap to supply but expensive to purchase, replacing some hard evidence with a larger assisted artifact can reduce resource use while raising the high type's private expenditure. The low type's imitation constraint can remain binding, so the alternative still separates.\footnote{\Cref{app:welfare-proofs} constructs such an equilibrium and gives the local cost comparison. The comparison concerns resource use among separating PBEs under the same tariff. The selected D1 outcome remains Riley separation.}

With free membership and compute-cost usage pricing, private and resource costs coincide, and the conditional efficiency result in \Cref{prop:provider-competition} applies. Under a monopoly tariff, D1 selection of the privately least-cost separating outcome alone does not establish resource efficiency.


\paragraph{Pooling, partial pooling, and competition.}

Pooling and partial pooling leave the receiver uncertain about the types using the shared signal. Any resource saving must be weighed against assessment error. By \cref{eq:welfare-master}, the welfare effect of a fresh shared product relative to no paid incremental AI is
\begin{equation}
\label{eq:social-value-pooling-product}
\welfare_{P,M}-\welfare_0
=C_0^{\mathrm{real}}-C_{P,M}^{\mathrm{real}}-K(\mathbf d)
-\frac{\receiverweight}{2}\prior(1-\belief)\DeltaT^2.
\end{equation}
A less informative market improves welfare if its resource saving, net of development cost, exceeds the assessment loss. Shared signaling can nevertheless be profitable when welfare falls. The provider's operating value in \cref{eq:profit-pooling-general} excludes receiver losses and need not rank products as welfare does.

Partial pooling has less assessment error than pure pooling: its shared posterior exceeds the prior, and some low types are identified. Its resource cost depends on both the shared artifact and the subscribing mass, so changes in these quantities must be weighed alongside the information gain. At a fixed technology and capacity, the weight $\receiverweight$ changes the welfare value of accuracy without changing the receiver's posterior-mean best response or the provider's pricing problem.

Competition introduces a further distinction between cheaper access and resource use. Compare the same technology under an optimal monopoly separating offer and the pricing equilibrium among developed identical technologies in \Cref{prop:provider-competition}. Let $C_C^{\mathrm{real}}$ denote expected signaling resource cost in the competitive market.

\begin{corollary}
\label{prop:competition-resource-cost}
Fix the same physical technology, compute cost, and capacity $\hardQ<\DeltaT$, with $w_h>0$. If the monopoly optimum selects separation, then
\begin{equation}
\label{eq:competition-resource-order}
C_C^{\mathrm{real}}>C_R^{\mathrm{real}}.
\end{equation}
Both outcomes reveal type perfectly. This comparison concerns signaling resources and excludes differences in incurred development and capacity costs.
\end{corollary}

Free membership and compute-cost usage pricing lower the low type's imitation cost. The high type must then supply a larger artifact to be recognized. With the physical costs of production unchanged, this increase in quality raises total resource expenditure. Competitive separation minimizes resource use under its own imitation constraints, which differ from those under monopoly pricing. The receiver is indifferent between the two fully informative outcomes at the same capacity.\footnote{\Cref{app:welfare-proofs} proves the resource ordering, including cases in which AI coverage binds.}

This comparison holds physical features fixed and requires separation under monopoly. Comparisons involving different designs or shared monopoly outcomes must account for their resource use and assessment error. Fresh-entry comparisons must also count every incurred development cost. The one-provider equilibria in \Cref{prop:provider-entry} can retain the monopoly signaling outcome despite potential competitors.

\Cref{prop:ed-harmful-improvement} shows that equilibrium improvements can lower welfare relative to retaining baseline tools. Both outcomes fully reveal type, and initial establishment costs cancel. The comparison counts signaling resources and incremental improvement expenditure. The chosen menu induces resource-efficient separation under its private imitation constraints, so the example identifies a welfare loss from endogenous design. \Cref{sec:capacity} next allows hard capacity to change and providers to reprice and redesign.

\begin{figure}[t]
\centering
\includegraphics[width=0.92\linewidth]{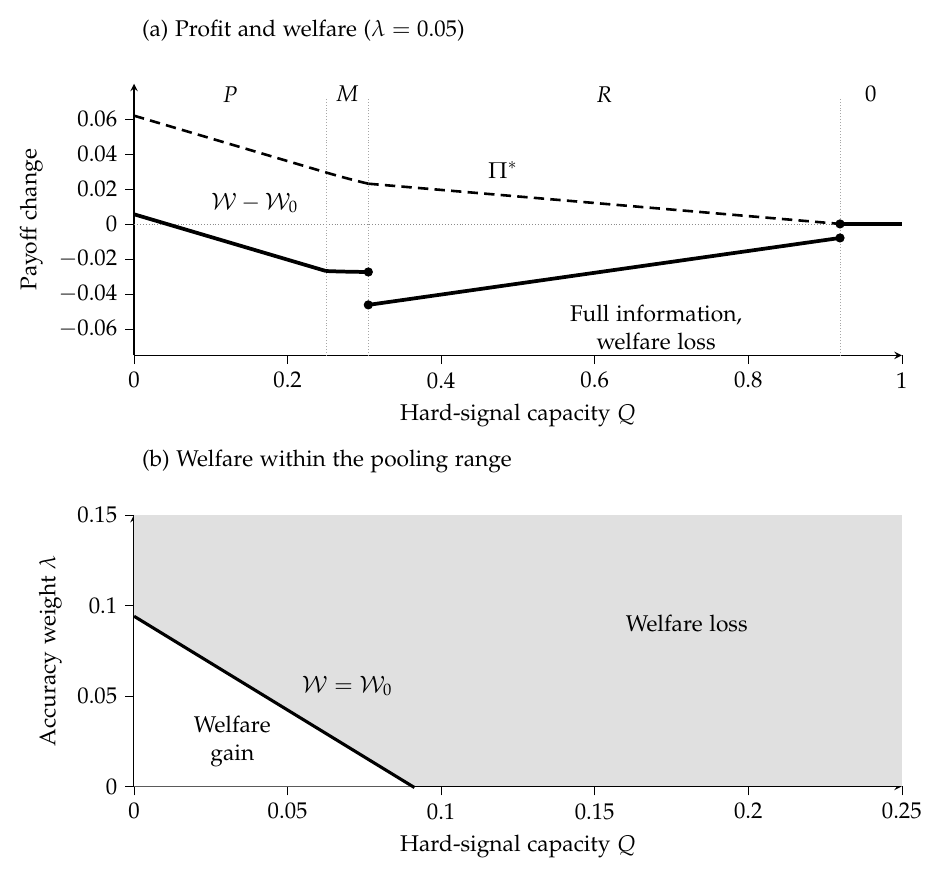}
\caption{\textbf{Profit and welfare under optimal pricing.} Panel (a) compares maximum net profit (dashed) with the welfare change relative to manual separation at the same capacity (solid). Dots mark distinct welfare values at provider profit ties. Panel (b) shows welfare within pooling as the accuracy weight varies. Both prices are reoptimized. Parameter values are given in \Cref{app:figure-parameters}.}
\label{fig:welfare-regions}
\end{figure}

\Cref{fig:welfare-regions} illustrates the distinction between profitable provision and welfare relative to no paid incremental AI. Separation can remain profitable while lowering welfare, and assessment losses can outweigh resource savings under pooling.

\section{Endogenous verification}
\label{sec:capacity}

Verification gives senders another way to demonstrate ability, but it also changes the market for assisted artifacts. An established provider can respond by improving its technology, changing prices, or serving a different group of senders. I compare two timings that capture different opportunities to commit. In the sequential game, the receiver sets verification capacity before the provider chooses improvements and prices. This describes assessment infrastructure chosen in advance of the provider's response. In the simultaneous game, capacity and the provider's entire offer are chosen at the same stage and must be mutual best responses.

The comparison asks how provider adaptation changes the return to verification. Sequentially, the receiver anticipates that an expansion will induce changes in technology and prices. Simultaneously, it evaluates a unilateral capacity change holding the provider's offer fixed, although subsequent signaling adjusts. In the shared-provision case studied below, the provider reduces artifact quality as capacity increases, encouraging low-type participation and offsetting part of the information gain. Anticipating this response can discourage verification relative to the simultaneous benchmark.

Studying both timings also clarifies which conclusions depend on commitment. Information can deteriorate under a higher externally imposed capacity requirement in either game. The comparison of voluntary investment instead depends on how the receiver evaluates provider responses. A simultaneous equilibrium supplies a test of unilateral incentives, but need not exist in pure strategies or describe convergence under repeated adjustments.

\paragraph{Improvements and capacity choices.}

The provider begins with editing requirements $(\epsilon_h^0,\epsilon_\ell^0)$ and fixed coverage $L$. Holding ample coverage fixed isolates how verification changes editing improvements and prices. As in \Cref{sec:design-competition}, it can improve either requirement, choosing $(u,v)\in[0,\bar u]\times[0,\bar v]$, where $\epsilon_h=\epsilon_h^0-u$ and $\epsilon_\ell=\epsilon_\ell^0-v$, with cost
\[
K(u,v)=\frac{k_hu^2+k_\ell v^2}{2}.
\]
Here, $k_h,k_\ell>0$. Retaining the baseline costs nothing, and further improvement has increasing marginal cost. Initial development expenditure is sunk. Each capacity comparison begins from the same installed baseline and counts newly chosen improvements. 

Every feasible design satisfies \Cref{ass:comparative-advantage}. I assume that baseline assistance saves resources for the low type, so
\[
w_\ell^0\coloneqq\kappa_\ell-\epsilon_\ell^0-c>0.
\]
Coverage is sufficiently large, and the assisted cost gap is sufficiently small, that the profit-maximizing shared quality satisfies the adoption and D1 deviation constraints at every feasible design. These restrictions keep input limits and the feasibility of signaling deviations from determining that quality. A high type can use enough input to consider separating from a shared signal, but doing so remains sufficiently costly.\footnote{Write $g=\kappa_\ell-\kappa_h$, $\delta_0=\epsilon_\ell^0-\epsilon_h^0$, $\underline\delta=\delta_0-\bar v$, and $\overline\delta=\delta_0+\bar u$. Sufficient conditions are $0<\underline\delta\leq\overline\delta<g$, $L>g\DeltaT/(\kappa_\ell\underline\delta)$, and $\overline\delta\leq\prior\kappa_hg/(\kappa_\ell+\prior\kappa_h)$. Coverage remains finite. These conditions apply to this extension. The stronger competitive restrictions in \cref{eq:ed-separating-restrictions} are not imposed.}

The receiver chooses $\hardQ\geq\hardQ_0$, where $0\leq\hardQ_0<\DeltaT$ is the initial or externally required minimum capacity. Making additional capacity available costs $\capcost(\hardQ)-\capcost(\hardQ_0)$, with $\capcost$ continuous and strictly increasing. Senders retain control of their signals, and a minimum capacity requirement does not compel them to produce a minimum amount of hard evidence.

Write $\mathbf o=(u,v,F,p)$ for the provider's offer, and $\mathrm{MSE}(\hardQ,\mathbf o)$ for the assessment error in the selected signaling outcome. The receiver's capacity payoff is
\begin{equation*}
-\frac{\receiverweight}{2}\mathrm{MSE}(\hardQ,\mathbf o)
-\bigl[\capcost(\hardQ)-\capcost(\hardQ_0)\bigr].
\end{equation*}
Sequentially, the receiver anticipates a profit-maximizing offer $\mathbf o^*(\hardQ)$ after each capacity choice. Simultaneously, the receiver holds the offer fixed when considering a capacity deviation, and the provider holds capacity fixed when considering a design or tariff deviation. All choices are public before senders produce signals. Both timings retain the signaling and adoption structure in \Cref{sec:model}. The simultaneous game tests unilateral stability of prospective choices.

\paragraph{How the provider adapts.}

Under the restrictions above, the optimal shared artifact is
\begin{equation}
\label{eq:capacity-shared-output}
x_H(\hardQ)=
\begin{cases}
\displaystyle\frac{\prior\DeltaT-\hardQ}{\kappa_\ell},
&0\leq\hardQ\leq\prior^2\DeltaT,\\[5pt]
\displaystyle\frac{\sqrt{\hardQ\DeltaT}-\hardQ}{\kappa_\ell},
&\prior^2\DeltaT<\hardQ<\DeltaT.
\end{cases}
\end{equation}
The first interval pools both types. Low-type participation binds, and increasing hard capacity reduces the artifact while leaving the posterior at $\prior$. The second interval partially pools them, with shared posterior $\belief=\sqrt{\hardQ/\DeltaT}$. Some low types choose the origin, and assessment error falls as capacity increases. These are conditional shared optima, which the provider must still compare with separation.

Let $A(\hardQ)=x_H(\hardQ)D_\ell(x_H(\hardQ),\hardQ)$ be aggregate low-type AI input, using the demand in \cref{eq:low-type-demand}, and set $A(\hardQ)=0$ for $\hardQ\geq\DeltaT$. Shared provision yields optimal improvements $u_H=0$ and $v_H=\min\{\bar v,A(\hardQ)/k_\ell\}$. Only subscribing low types use AI, so high-type improvement adds expenditure without increasing shared profit. Improving low-type production saves $A(\hardQ)$ per unit of improvement. The provider balances this saving against marginal improvement cost.

Write $\widehat R$ and $\widehat H$ for maximal net profit from separation and shared signaling, respectively. The fixed-design values in \Cref{sec:market} give
\begin{align*}
\widehat R(\hardQ)
&=\max_{(u,v)\in[0,\bar u]\times[0,\bar v]}
\left\{\frac{\prior(\DeltaT-\hardQ)_+
(\kappa_h-\epsilon_h^0+u-c)_+}
{\kappa_h+\delta_0+u-v}-K(u,v)\right\},\\
\widehat H(\hardQ)
&=(w_\ell^0+v_H)A(\hardQ)-\frac{k_\ell v_H^2}{2}.
\end{align*}
The provider chooses the larger value. Under separation, high types buy assistance, and easier low-type imitation can increase the artifact they need to distinguish themselves. Improvements to either requirement can be profitable. Provider ties must be specified separately from the sender selection rule.

Within a shared optimum, capacity weakly improves information. The comparison between separation and sharing can reverse that conclusion. Additional capacity reduces the artifact market under separation, and can make high-type improvements less attractive. Then, the provider may earn more from assisting low types at a shared signal.

\begin{proposition}
\label{prop:verification-information-loss}
There are primitives satisfying the restrictions above and two minimum capacities, $\hardQ_0<\widetilde{\hardQ}_0$, for which each timing has a pure equilibrium at the respective minimum. The lower-capacity equilibrium separates. The higher-capacity equilibrium pools, and has strictly greater assessment error. The provider optimally chooses both improvements and prices.
\end{proposition}

The construction has minimum capacities $0$ and $0.9$, with $\DeltaT=1$ and $\prior=0.95$. At zero, the provider improves high-type production and sells assistance in a separating market. At the higher requirement, the return from this market falls below the return from assisting low types. The provider switches to pooling, leaving assessment error $0.0475$. In each game, the receiver prefers its minimum capacity. Both outcomes also pass the simultaneous best-response test.\footnote{\Cref{app:welfare-proofs} gives the parameters, optimal offers, and global deviation checks.}

This comparison changes an external capacity requirement. A receiver that can retain its original capacity will not voluntarily pay for an expansion that it correctly anticipates will worsen accuracy. More available verification can reduce equilibrium information without being the receiver's preferred investment.

\paragraph{Commitment and voluntary verification.}

A comparison of capacity choices is particularly transparent when serving high types cannot save resources at any feasible design, i.e.,
\begin{equation*}
\kappa_h-\epsilon_h^0+\bar u-c\leq0.
\end{equation*}
Then, $\widehat R=0$ and shared provision is optimal for every $\hardQ<\DeltaT$. This additional restriction applies only to the timing comparison in this subsection. It retains endogenous low-type improvement and both tariff instruments.

Suppose that $\prior>1/2$ and $\hardQ_0\geq\prior^2\DeltaT$, so the relevant shared optima partially pool. Holding a shared offer fixed, its posterior is $\belief=(\kappa_\ell x+\hardQ)/\DeltaT$. Using \cref{eq:shared-assessment-error}, the marginal accuracy benefit of capacity is $\receiverweight\prior\DeltaT/2$. When the provider adjusts optimally after capacity changes, the benefit is instead
\begin{equation}
\label{eq:verification-marginal-information}
\frac{\receiverweight\prior\DeltaT}{4}
\sqrt{\frac{\DeltaT}{\hardQ}}.
\end{equation}
The provider reduces shared artifact quality as capacity increases in this region. This makes low-type participation easier and offsets part of the information gain that would arise with the offer fixed.

If $\capcost$ is continuously differentiable and strictly convex, a pure simultaneous equilibrium with $\hardQ_0<\hardQ^{\mathrm{sim}}<\DeltaT$ must satisfy $\capcost'(\hardQ^{\mathrm{sim}})=\receiverweight\prior\DeltaT/2$. If such an equilibrium exists, the sequential receiver chooses a strictly lower capacity. It anticipates the smaller information return in \cref{eq:verification-marginal-information}. An interior sequential choice equates that return to $\capcost'(\hardQ^{\mathrm{seq}})$, while the receiver may instead retain $\hardQ_0$. Existence of the simultaneous equilibrium also requires that downward deviations do not induce cheaper, fully informative separation. Different optimal implementing tariffs can affect this requirement even when they induce the same current outcome.

\begin{figure}[t!]
\centering
\includegraphics[width=0.92\linewidth]{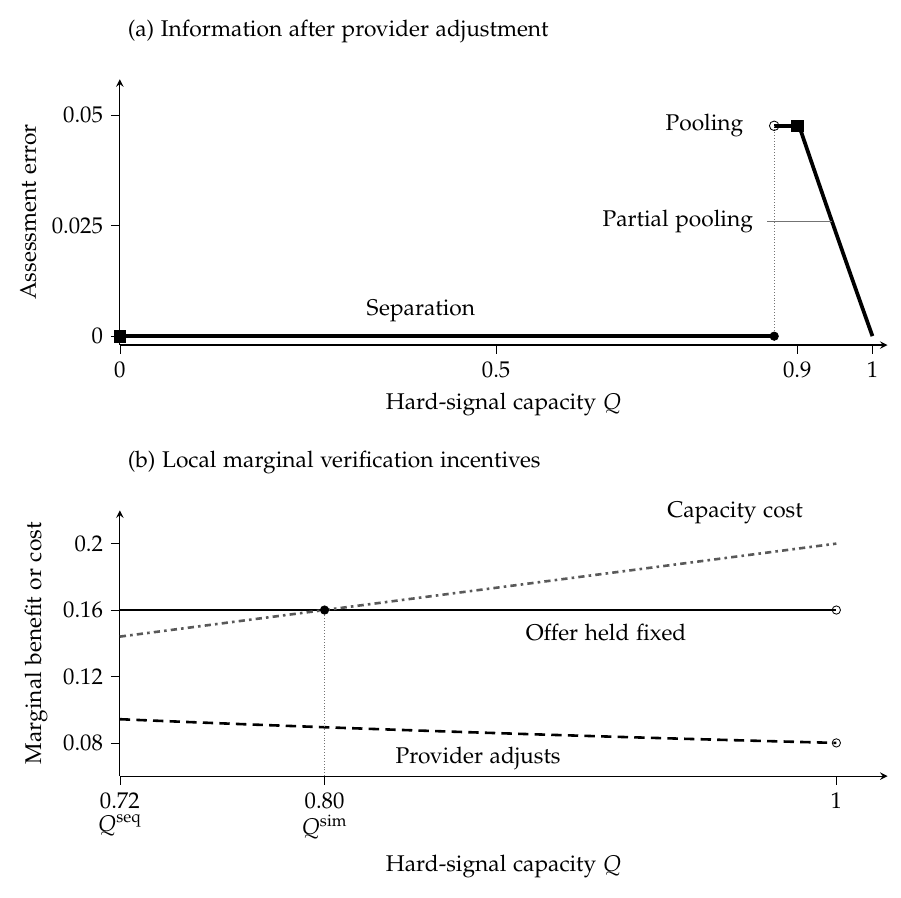}
\caption{\textbf{Verification, provider adaptation, and timing.} Panel (a) plots assessment error after optimal provider responses. Squares mark equilibria under both timings. Panel (b) compares marginal accuracy gains with the offer fixed or reoptimized against marginal capacity cost. Returns are evaluated at each capacity's optimal offer. Open markers denote an unselected outcome in (a) and left-hand limits in (b).}
\label{fig:verification-design-timing}
\end{figure}

The comparison is attained in the example illustrated in panel (b) of \Cref{fig:verification-design-timing}. The sequential receiver retains its minimum capacity, while simultaneous choices support a pure equilibrium with expansion and lower assessment error.\footnote{\Cref{app:verification-timing} gives the parameters, equilibrium quantities, and global best-response checks.}

A pure simultaneous equilibrium is not guaranteed. \Cref{app:verification-no-pure} gives a specification with no such equilibrium, while \Cref{app:verification-timing} gives the global existence test for shared provision.

\Cref{fig:verification-design-timing} illustrates the two capacity comparisons. Panel (a) traces the provider's response to externally varied capacity, with the two equilibria under different minimum requirements marked. Panel (b) compares local verification returns at each capacity's optimal offer. Anticipating provider adjustment lowers the return to expansion in this example.

\paragraph{Information gains and resource costs.}

The receiver's objective excludes production and improvement resources. Fix a profit-maximizing provider offer $\mathbf o^*(\hardQ)$ at each capacity, including a specified choice at profit ties. Evaluate signaling outcomes under the equilibrium selection rule in \Cref{sec:model}, suppress the offer argument, and write the receiver's investment gain as
\begin{equation}
\label{eq:receiver-restoration-value}
\begin{aligned}
V_R(\hardQ_1\mid\hardQ_0)
={}&\underbrace{\frac{\receiverweight}{2}
\bigl[\mathrm{MSE}(\hardQ_0)-\mathrm{MSE}(\hardQ_1)\bigr]}_{\text{accuracy gain}} 
-\underbrace{\bigl[\capcost(\hardQ_1)-\capcost(\hardQ_0)\bigr]}_{\text{additional capacity cost}}.
\end{aligned}
\end{equation}
Let $K_{\mathrm{tot}}(\hardQ)=K(u^*(\hardQ),v^*(\hardQ))$ record incremental improvement expenditure, with common sunk development costs suppressed. By \cref{eq:welfare-master}, the welfare gain is
\begin{equation}
\label{eq:social-restoration-value}
\begin{aligned}
\welfare(\hardQ_1)-\welfare(\hardQ_0)
={}&\underbrace{V_R(\hardQ_1\mid\hardQ_0)}_{\text{receiver gain}}\\
&+\underbrace{C^{\mathrm{real}}(\hardQ_0)-C^{\mathrm{real}}(\hardQ_1)}_{\text{signaling resource saving}}
+\underbrace{K_{\mathrm{tot}}(\hardQ_0)-K_{\mathrm{tot}}(\hardQ_1)}_{\text{improvement expenditure saving}}.
\end{aligned}
\end{equation}
These comparisons count the provider's optimal response at each capacity. They do not refund improvements already made. The two resource-saving terms can be negative, as verification may increase signaling resources or improvement expenditure.

\begin{proposition}
\label{prop:receiver-restoration-distortion}
There are primitives satisfying the restrictions above for which the sequential game has a pure equilibrium in which the receiver optimally expands capacity although welfare is maximized at the initial capacity. There are also such primitives for which the sequential game has a pure equilibrium in which the receiver optimally retains the initial capacity although welfare is maximized at a higher capacity. In both constructions, the provider chooses globally optimal improvements and prices.
\end{proposition}

Excessive investment can arise when an accuracy gain induces the receiver to replace a cheap shared outcome with more costly separating production and improvement. Insufficient investment can arise when the provider reduces artifact production as capacity increases while pooling persists. The receiver gains no accuracy from that expansion, but the resource saving can justify it socially.\footnote{\Cref{app:welfare-proofs} verifies both examples over all capacities and specifies the provider's choice at profit ties.}

At $\hardQ\geq\DeltaT$, the high type separates using hard evidence alone, with $q=0$ and $\hardsig=\DeltaT$. No incremental AI input is purchased, and the provider chooses no costly improvement. Below that capacity, information, signaling resources, and design expenditure must be considered together. Capacity that improves accuracy need not improve welfare, and capacity that saves resources need not improve accuracy.

\section{Conclusion}
\label{sec:conclusion}

AI assistance changes who can imitate whom and how much signaling is needed to establish ability. Providers' prices and technology choices can sustain shared signals or increase the artifact required for separation. Relative to manual signaling at the same verification capacity, assistance improves welfare when signaling resource savings exceed development costs and any loss from inaccurate assessment. Induced pooling can satisfy this condition, and full separation can fail it. Profitable AI provision can lower welfare even when ability remains fully revealed, because easier imitation increases signaling enough to outweigh production savings.

Prices and improvement incentives also shape the cost of signaling with existing tools. Competition among providers offering identical technologies can support full information through free membership and compute-cost usage pricing. When monopoly provision also separates types, this competitive allocation requires more signaling resources. Monopoly usage markups can instead favor hard evidence that is privately cheaper but uses more resources than assisted production. Providers can also profitably improve their tools while lowering welfare relative to retaining baseline technologies. Improving lower-ability senders' productivity makes imitation easier, increasing the artifact and AI purchases required from higher-ability senders.

The prices that make installed tools informative need not finance their development. When development precedes pricing, an additional entrant anticipates price competition and may fail to recover its development cost. The game can sustain a sole provider whose monopoly offer pools. Development incentives connect market structure to the information conveyed by signals.

Verification offers another way to demonstrate ability, but providers adjust their technologies and prices to its availability. A higher externally required capacity can shift provision from assisting higher-ability senders in separation to assisting lower-ability senders in pooling, reducing equilibrium information under either timing. Anticipating provider adjustment can also reduce the receiver's incentive to expand capacity. The receiver values accuracy and bears verification costs, but omits changes in signaling resources and improvement expenditure. As a result, its investment can be socially excessive or insufficient: restoring information can cost more than it is worth, while resource-saving verification can be worthwhile without improving accuracy. 


\section*{Disclosure}

I used ChatGPT for general-purpose research assistance and language editing, and \href{https://www.refine.ink/}{Refine.ink} for technical feedback. I reviewed and revised all content and take full responsibility for the accuracy, integrity, and originality of the manuscript.


\clearpage
\phantomsection
\addcontentsline{toc}{section}{References}
\bibliographystyle{aer}
\bibliography{ai_signaling_references}



\clearpage
\appendix
\numberwithin{proposition}{section}
\numberwithin{corollary}{section}
\numberwithin{assumption}{section}
\numberwithin{lemma}{section}
\phantomsection
\hypersetup{bookmarksdepth=0}
\addcontentsline{toc}{section}{Appendix}
\hypersetup{bookmarksdepth=}

\crefalias{section}{appendix}
\crefalias{subsection}{appendix}
\crefalias{subsubsection}{appendix}

\numberwithin{equation}{section}

\etocdepthtag.toc{appendix}
\pdfbookmark[0]{Appendix}{appendix-overview}

\begin{center}
{\LARGE\bfseries Appendix\par}
\medskip
{\large The Market for AI-Assisted Signaling \par}
\end{center}

\bigskip
\begingroup
\singlespacing
\small
\hypersetup{linkcolor=black}
\etocsettagdepth{supplement}{none}
\renewcommand{\contentsname}{Contents of the appendix}
\etocsettagdepth{main}{none}
\etocsettagdepth{appendix}{subsection}
\tableofcontents
\endgroup
\clearpage


\section{Production choices and signaling preliminaries}
\label{app:implementation}

This appendix derives the private cost schedules in \Cref{sec:model}, records the production choices needed for profit and welfare accounting, and verifies the reduction to observable signals. It also states the D1 belief restriction and equilibrium selection rule. Throughout, the technology $\mathbf d=(\epsilon_\ell,\epsilon_h,L)$ and tariff $(F,p)$ are public, with $F,p\geq0$. The receiver observes $(q,\hardsig)$, while subscription and AI input remain private.


\subsection{Subscription and input choice}

Fix sender type $i\in\{\ell,h\}$, and additional artifact quality $q\geq0$. Without a subscription, production costs $\kappa_iq$. With a subscription and AI input $z\in[0,\min\{q,L\}]$, private production cost is
\begin{equation*}
F+\kappa_i(q-z)+(\epsilon_i+p)z
=F+\kappa_iq-(\kappa_i-p-\epsilon_i)z.
\end{equation*}
Conditional on subscribing, the sender uses the largest feasible input when $p+\epsilon_i<\kappa_i$, and zero input when $p+\epsilon_i>\kappa_i$. At equality, every feasible input minimizes cost, and the input rule in \Cref{sec:model} selects the largest. Minimized cost conditional on subscription is
\begin{equation*}
F+\kappa_iq-(\kappa_i-p-\epsilon_i)_+\min\{q,L\}.
\end{equation*}
Comparing this cost with manual production gives \cref{eq:cost-envelope}. Equivalently,
\begin{equation*}
C_i(q)=\kappa_iq-
\left[(\kappa_i-p-\epsilon_i)_+\min\{q,L\}-F\right]_+.
\end{equation*}
The subtracted term is the saving relative to manual production after paying the membership fee.

Let $I_i(q)\in\{0,1\}$ indicate subscription, and let $z_i(q)$ denote the resulting AI input. Subscription at indifference and maximal input at a cost tie give
\begin{equation}
\label{eq:private-production-rule}
\begin{aligned}
I_i(q)&=\ind_{\{F\leq(\kappa_i-p-\epsilon_i)_+\min\{q,L\}\}},\\
z_i(q)&=I_i(q)\min\{q,L\}\ind_{\{p+\epsilon_i\leq\kappa_i\}}.
\end{aligned}
\end{equation}
In particular, a free subscription is accepted even when input is too expensive to use. Input is then zero, generating neither usage revenue nor compute expenditure.

If $p+\epsilon_i<\kappa_i$ and $0\leq F\leq(\kappa_i-p-\epsilon_i)L$, define the adoption threshold
\begin{equation*}
x_i\coloneqq\frac{F}{\kappa_i-p-\epsilon_i}\in[0,L].
\end{equation*}
The cost schedule becomes
\begin{equation*}
C_i(q)=
\begin{cases}
\kappa_iq,&0\leq q\leq x_i,\\
F+(p+\epsilon_i)q,&x_i\leq q\leq L,\\
F+(p+\epsilon_i)L+\kappa_i(q-L),&q\geq L.
\end{cases}
\end{equation*}
The expressions agree at $q=x_i$ and $q=L$. At $q=x_i$, manual and assisted production cost the same, and the sender subscribes. If $F=(\kappa_i-p-\epsilon_i)L$, the two methods remain tied at every $q\geq L$, and subscription still occurs.

If $p+\epsilon_i<\kappa_i$ but the fee exceeds the maximum saving $(\kappa_i-p-\epsilon_i)L$, the sender never subscribes. When $p+\epsilon_i\geq\kappa_i$, a positive fee also prevents adoption. At $F=0$, subscription occurs. Input is zero when $p+\epsilon_i>\kappa_i$, and maximal when $p+\epsilon_i=\kappa_i$, as specified in \cref{eq:private-production-rule}.

Under \Cref{ass:comparative-advantage}, $\kappa_\ell-\epsilon_\ell>\kappa_h-\epsilon_h$. At a given quality and usage price, the low type's saving from AI input before paying the membership fee is weakly greater. It can begin using AI at a lower quality than the high type. \Cref{sec:benchmark} studies how this difference changes the cost advantage conveyed by an artifact.


\subsection{Private costs and resource expenditure}

Each $C_i$ starts at zero and is continuous, strictly increasing, and unbounded. These properties follow from \cref{eq:cost-envelope}: both candidate schedules have positive slopes and eventually increase at rate $\kappa_i$. Therefore, the inverse cost used to construct Riley separation in \cref{eq:riley-signal} is well defined.

The high type also has strictly lower artifact cost at every positive quality. For any common subscription and input choice, the low type's additional effort is
\begin{equation*}
(\kappa_\ell-\kappa_h)(q-z)+(\epsilon_\ell-\epsilon_h)z
\geq(\epsilon_\ell-\epsilon_h)q.
\end{equation*}
The high type can use the low type's minimizing production method and pay the same membership and usage charges. Its effort cost is at least $(\epsilon_\ell-\epsilon_h)q$ lower, and optimizing its own method can only reduce its cost further. Hence, $C_\ell(q)-C_h(q)\geq(\epsilon_\ell-\epsilon_h)q>0$ for $q>0$. This cost advantage can nevertheless decline over an interval when only the low type adopts.

Resource accounting uses the input selected at the private-cost minimum in \cref{eq:private-production-rule}. Define artifact resource cost as
\begin{equation*}
C_i^{\mathrm{real}}(q)
=\kappa_i\bigl(q-z_i(q)\bigr)+(\epsilon_i+c)z_i(q).
\end{equation*}
Adding $\hardsig$ for the low type and $\rhoHard\hardsig$ for the high type gives real signaling cost $c_i^{\mathrm{real}}(\sigvec)$. For the same subscription and input choice, the difference between private and resource cost is
\begin{equation*}
c_i(\sigvec)-c_i^{\mathrm{real}}(\sigvec)
=FI_i(q)+(p-c)z_i(q).
\end{equation*}
The right side is provider revenue from this sender less compute expenditure. The identity holds at every signal and, therefore, also in expectation, including mixed signaling. \Cref{sec:welfare} adds development and capacity costs to this accounting.

At $F=0$ and $p=c$, private and resource costs coincide. At other tariffs, membership and usage payments can make them differ. At an adoption threshold, manual and assisted production have the same private cost but may use different resources. Therefore, private cost is continuous at adoption, while resource cost can jump when the selected method changes. Profit and welfare calculations must retain $I_i(q)$ and $z_i(q)$ alongside $C_i(q)$.

\subsection{D1 and equilibrium selection}
\label{app:refinement-details}


\subsubsection{The D1 belief restriction}

Fix an equilibrium payoff $U_i$ for each type and an unused signal $\sigvec$. Define the assessment threshold
\begin{equation}
\label{eq:d1-threshold}
\gamma_i(\sigvec)=U_i+c_i(\sigvec).
\end{equation}
Type $i$ is indifferent between deviating and retaining its equilibrium payoff when the assessment equals this threshold, and strictly prefers deviating when the assessment exceeds it. Production choices are already optimized in $c_i$.

Let $B_i^W(\sigvec)$ and $B_i^S(\sigvec)$ denote the sets of feasible assessments that make type $i$ weakly and strictly prefer deviating, respectively
\begin{align*}
B_i^W(\sigvec)
&\coloneqq\{\action\in[\thL,\thH] \mid \action-c_i(\sigvec)\geq U_i\},\\
B_i^S(\sigvec)
&\coloneqq\{\action\in[\thL,\thH] \mid \action-c_i(\sigvec)>U_i\}.
\end{align*}
D1 requires the receiver to assign zero probability to type $i$ if the other type $j$ satisfies
\[
B_i^W(\sigvec)\subseteq B_j^S(\sigvec),
\]
\[
B_j^S(\sigvec)\neq\varnothing.
\]
Every assessment that makes the deviation weakly profitable for type $i$ must make it strictly profitable for type $j$, and type $j$ must strictly gain under some assessment. If neither type can strictly gain, D1 imposes no restriction.\footnote{This is the weak-versus-strict incentive comparison in \citet{ChoKreps1987,BanksSobel1987,ChoSobel1990}.}

In terms of the threshold in \cref{eq:d1-threshold}, these sets become
\[
B_i^W(\sigvec)=[\gamma_i(\sigvec),\thH]\cap[\thL,\thH],
\]
\[
B_i^S(\sigvec)=(\gamma_i(\sigvec),\thH]\cap[\thL,\thH].
\]
Under our nonnegative-fee restriction, signaling costs are nonnegative and $U_i\geq\thL$, so each threshold is at least $\thL$. The type with the lower threshold rules out the other type whenever it can strictly benefit from deviating. Equal thresholds do not distinguish the types. Every assessment in $[\thL,\thH]$ is the receiver's best response to some posterior, and random responses affect sender incentives only through their mean. Comparing assessment thresholds gives the relevant D1 test directly. \Cref{eq:d1-cost-advantage-test,eq:d1-revealing-deviation} apply this test to signaling costs.

I use D1 because it strengthens the Intuitive Criterion and permits this direct comparison of deviation incentives. In generic finite signaling games, strategically stable outcomes survive D1 \citep{ChoSobel1990}. That result does not establish strategic stability in the continuous game studied here. The existence and classification arguments in \Cref{app:equilibrium-proofs,app:competition-proofs,app:ed-proofs} apply to the stated PBE concept with finite observable support, refined by D1.\footnote{I do not impose the Never-a-Weak-Best-Response (NWBR) test or claim that pooling survives it. Applying NWBR would require a separate argument for the relevant continuous-game equilibrium notion. \citet{Manelli1997} identifies conditions under which NWBR outcomes of finite approximating games converge to an equilibrium outcome of the infinite game. A discrete interpretation of real choices alone does not establish those conditions.}

\subsubsection{Equilibrium selection}

The equilibrium selection rule in \Cref{sec:model} is an additional assumption, applied to finite-support signaling equilibria that survive D1. It excludes an observable outcome if another such equilibrium gives both sender types weakly higher payoffs and at least one strictly higher payoff.

Among equilibria giving both types the same payoffs, selection favors the greatest mass at a signal shared by both types, then the least artifact quality at that shared signal. Remaining cost ties between separating signals favor the least artifact quality. These selection criteria are additional assumptions, not implications of D1. Providers choose tariffs anticipating the ensuing sender strategies and receiver beliefs.

The selected outcomes are Riley separation, pooling, and partial pooling, as characterized in \Cref{sec:two-signal}.

For some offers, D1 determines a unique observable outcome among equilibria with finite support. Offers outside \cref{eq:shared-signal-regime} yield Riley separation. Within that regime, a strict price inequality $p+\epsilon_\ell<\kappa_h$, together with $\prior\DeltaT<v<\DeltaT$ and $E>\DeltaT-v$, yields unique partial pooling by \Cref{thm:pure-d1}. A strict-price offer satisfying
\begin{equation}
\label{eq:unique-pooling-region}
v\leq\prior\DeltaT \; \text{ and } \; E>\DeltaT-v
\end{equation}
also yields unique pure pooling. To see this, $C_\ell(q_E)+\hardQ=v+E>\DeltaT$, so $x<q_R<q_E$ and Riley separation fails D1. Coexistence of a shared signal and a high-only signal would require $U_\ell-\thL=\DeltaT-v-E<0$, contradicting the payoff that the low type can guarantee at the origin. The support restrictions in \Cref{lem:finite-support-signaling} leave only pooling at $(x,\hardQ)$. In these regions, the observable outcome is independent of selection among D1 survivors.

\Cref{supp:alternative-selection} discusses alternative equilibrium selection rules and their implications for pricing and welfare.


\subsection{Units and the reduced signaling game}

Costs and payoffs are measured net of the common cost of supplying the required basic artifact. Its production is held fixed across signals, technologies, and tariffs, so this normalization preserves production choices, equilibrium comparisons, and welfare differences. Free assistance in producing the basic artifact can be included in this baseline. Free assistance for additional quality would have to enter the outside production method explicitly.

Hard evidence can be expressed in low-type cost units without changing the information it conveys. Suppose physical evidence is $y\in[0,\bar y]$, the low type's effort cost $C_\ell^y(y)$ is continuous and strictly increasing with $C_\ell^y(0)=0$, and the high type's effort cost is $\rhoHard C_\ell^y(y)$. With these costs fixed across AI offers, the change of units
\begin{equation}
\begin{aligned}
\label{eq:hard-signal-units}
\hardsig&=C_\ell^y(y),\\
\hardQ&=C_\ell^y(\bar y)
\end{aligned}
\end{equation}
maps feasible physical evidence one-to-one onto $[0,\hardQ]$. It preserves information and gives effort costs $\hardsig$ and $\rhoHard\hardsig$. Artifact quality $q$ remains in the same physical units across tariffs, while prices enter its private cost directly.

\begin{proof}[Proof of \Cref{lem:receiver-br}]
At posterior $\belief$, expected squared error can be written as
\begin{equation*}
\E\bigl[(\action-\theta)^2\mid\sigvec\bigr]
=\bigl(\action-\E[\theta\mid\sigvec]\bigr)^2
+\Var(\theta\mid\sigvec).
\end{equation*}
The variance term is independent of the assessment. Since $\receiverweight>0$ and the posterior mean lies in $[\thL,\thH]$, the unique best response is $\action=\E[\theta\mid\sigvec]=\belief\thH+(1-\belief)\thL$.
\end{proof}

Subtracting $\thL$ from both assessments and types would express this response as $\belief\DeltaT$ and leave assessment errors unchanged. Sender payoffs would fall by the same constant $\thL$. The main text retains the original assessment units and prior mean $\thbar$.

\begin{lemma}
\label{lem:private-production-reduction}
For a fixed public offer and capacity, under the subscription and input rules in \Cref{sec:model}, the production game and the reduced signaling game with costs $c_i$ have corresponding PBEs with the same observable signals, beliefs, and receiver actions. Sender payoffs and the D1 deviation comparisons are preserved.
\end{lemma}

\begin{proof}[Proof of \Cref{lem:private-production-reduction}]
At any observable signal, a private change in subscription or input leaves the receiver's response unchanged. A sender minimizes production cost conditional on that signal. The choices in \cref{eq:private-production-rule} attain $C_i(q)$ and select a minimizing method at every tie. Any strictly more expensive method is dominated at the same observation.

Sender best responses in observable signals consequently coincide with those in the reduced game. An equilibrium of the reduced game lifts to the production game by assigning the prescribed subscription and input to every chosen signal. Conversely, minimizing private production choices in a production equilibrium yields its reduced equilibrium. Observable strategies, including their finite supports, are unchanged, so Bayes consistency and receiver optimality are preserved.

For D1, let $U_i$ denote equilibrium payoff. At an unused signal $\sigvec$, minimizing private production cost gives deviation payoff $\action-c_i(\sigvec)$ in both games. Therefore, the assessment threshold is $U_i+c_i(\sigvec)$ in both games. The assessment sets yielding weak and strict gains coincide, and D1 excludes the same types. Identical sender payoffs also preserve sender-Pareto rankings and the observable tie-breaking rules in \Cref{sec:model}. The hard-evidence change of units in \cref{eq:hard-signal-units} simply re-indexes the same observations and deviations.
\end{proof}

This correspondence holds for a fixed offer. Retaining the minimizing production choices also recovers usage demand, provider revenue, and resource expenditure. Providers can compare offers in the common quality units used in \Cref{sec:market,sec:competition}.


\section{Proofs of the signaling results}
\label{app:equilibrium-proofs}

Throughout this appendix, the technology $\mathbf d$, tariff $(F,p)$, and hard capacity $\hardQ$ are fixed. Costs are those in \cref{eq:cost-envelope,eq:low-cost-normalized,eq:high-cost-normalized}. \Cref{app:implementation} justifies minimizing over private production choices before analyzing signals. The proofs establish least-cost separation, the D1 conditions in \Cref{sec:benchmark,sec:two-signal}, and the sender-payoff comparisons underlying the equilibrium selection rule.

\subsection{Least-cost separation}

\begin{proof}[Proof of \Cref{cor:riley}]
A low type identified as low receives $\thL$. Any positive-cost signal with that assessment yields less than choosing the origin, which guarantees an assessment of at least $\thL$. Therefore, the low type chooses $(0,0)$ in every pure separating PBE. Its imitation constraint at the high signal is $\costL(\sigvec_h)\geq\DeltaT$.

Hard evidence is cheaper for the high type per unit of cost imposed on the low type than any artifact-production segment. To verify this comparison, write $a_i\coloneqq p+\epsilon_i$. The low type's per-unit saving from AI exceeds the high type's by $(\kappa_\ell-\kappa_h)-(\epsilon_\ell-\epsilon_h)>0$, so the low type can begin using AI no later than the high type. Wherever artifact costs are differentiable, their marginal high-to-low cost ratio is one of $\kappa_h/\kappa_\ell$, $\kappa_h/a_\ell$, and $a_h/a_\ell$. The middle ratio applies when only the low type uses AI. In that case, $a_\ell\leq\kappa_\ell$, so the ratio is at least $\kappa_h/\kappa_\ell$. The last ratio applies when both types use AI and satisfies $a_h/a_\ell\geq\epsilon_h/\epsilon_\ell$ because $p\geq0$. Thus, every displayed ratio exceeds $\rhoHard$ under \Cref{ass:comparative-advantage}. Cost continuity extends this comparison at qualities where a type begins using AI or exhausts coverage.

For any deterrence requirement $t>0$, minimizing $\costH(\sigvec)$ subject to $\costL(\sigvec)\geq t$ makes the constraint bind. Otherwise, a positive component of the signal could be reduced, lowering both types' costs while preserving deterrence. If $q>0$ and $\hardsig<\min\{\hardQ,t\}$, replacing a small amount of low-type artifact cost with the same amount of hard-signal cost lowers high-type cost. The unique minimizer and its cost are
\begin{equation}
\begin{aligned}
\label{eq:deterrence-private-frontier}
\hardsig(t)&=\min\{\hardQ,t\},\\
q(t)&=C_\ell^{-1}\bigl((t-\hardQ)_+\bigr),\\
\detval(t)&=C_h(q(t))+\rhoHard\hardsig(t).
\end{aligned}
\end{equation}
Set $\detval(0)=0$. The inverse exists by arguments in \Cref{app:implementation}. Taking $t=\DeltaT$ gives \cref{eq:riley-signal,eq:riley-cost}.

The high type's minimizing cost is strictly below $\DeltaT$. Any positive artifact contribution costs it less than the low type, since $C_h(q)<C_\ell(q)$ for $q>0$. Each unit of hard evidence also costs it less than one. Assign posterior one at the Riley signal and zero everywhere else. The low type is indifferent between the origin and imitation. The high type receives $\thH-\rileycost>\thL$, whereas every other signal is assessed at $\thL$ and incurs a nonnegative cost. These beliefs support the stated separating PBE.
\end{proof}

For a fixed offer, the same marginal comparison establishes the capacity statement in \Cref{sec:two-signal}. When $\hardQ<\DeltaT$, the condition $C_\ell(q_R)=\DeltaT-\hardQ$ makes $q_R$ decrease with capacity. On each production segment, the derivative of private separating cost is $\rhoHard-C_h'(q_R)/C_\ell'(q_R)<0$. This cost is continuous when separating quality crosses an adoption threshold or $L$ and constant once $\hardQ\geq\DeltaT$.


\subsection{Cost advantage and recovery expenditure}

Define the high type's total cost advantage by
\begin{equation}
\label{eq:total-cost-advantage}
W(q,\hardsig)\coloneqq\costL(q,\hardsig)-\costH(q,\hardsig)
=W_1(q)+(1-\rhoHard)\hardsig,
\end{equation}
where $W_1(q)=C_\ell(q)-C_h(q)$ is its artifact cost advantage. Retain $a_i=p+\epsilon_i$, and let $g\coloneqq\kappa_\ell-\kappa_h$, $\delta\coloneqq\epsilon_\ell-\epsilon_h$, $s\coloneqq g-\delta>0$, and $b_i\coloneqq\kappa_i-a_i$.

In the strict shared-signal regime, $a_\ell<\kappa_h$, both $b_i$ are positive, $b_\ell>b_h$, and $x=F/b_\ell\in(0,L)$. Both types produce manually up to $x$, where $W_1(x)=gx$. The low type adopts, and $W_1$ decreases until the high type adopts or coverage is exhausted. Its subsequent increase returns the advantage to $W_1(x)$ at the recovery quality $q_E$.

\begin{lemma}
\label{lem:recovery-expenditure}
In the strict shared-signal regime, the recovery expenditure in \cref{eq:recovery-cost} is
\begin{equation}
\label{eq:recovery-expenditure-piecewise}
E=
\begin{cases}
\displaystyle \frac{a_\ell s}{\delta}x,
& 0<x\leq\displaystyle\frac{\delta L}{g},\\[6pt]
\displaystyle b_\ell x+\left(a_h-\frac{\kappa_h\delta}{g}\right)L,
& \displaystyle\frac{\delta L}{g}\leq x\leq\frac{b_hL}{b_\ell},\\[6pt]
\displaystyle \frac{\kappa_hb_\ell}{g}(L-x),
& \displaystyle\frac{b_hL}{b_\ell}\leq x<L.
\end{cases}
\end{equation}
At $a_\ell=\kappa_h$, the advantage is flat on $[x,q_E]$, where $q_E=\min\{L,F/\delta\}$ and $E=\kappa_h(q_E-x)$. In either case,
\begin{equation}
\label{eq:recovery-low-cost}
C_\ell(q_E)=\kappa_\ell x+E.
\end{equation}
Every signal with $W(q,\hardsig)>W(x,\hardQ)$ has high-type cost greater than $C_h(q_E)+\rhoHard\hardQ$. This bound is the infimum of such costs.
\end{lemma}

\begin{proof}[Proof of \Cref{lem:recovery-expenditure}]
If recovery occurs within coverage after both types have adopted, their artifact cost difference is $\delta q$. Equating this difference to $W_1(x)=gx$ gives $q_E=gx/\delta$. Recovery within coverage requires $x\leq\delta L/g$. Substituting $q_E$ into $E=C_h(q_E)-\kappa_hx$ gives the first expression in \cref{eq:recovery-expenditure-piecewise}.

If both types adopt but recovery occurs beyond coverage, their cost difference is $\delta L+g(q-L)$. Equating it to $gx$ gives $q_E=L+x-\delta L/g$. Substitution into the high type's cost gives the second expression. This case requires $\delta L/g\leq x\leq b_hL/b_\ell$.

If $x\geq b_hL/b_\ell$, the high type never obtains a strict saving from subscription. Recovery occurs at $q_E=x+b_\ell(L-x)/g$, and the high type's additional manual cost gives the third expression. The thresholds satisfy $\delta L/g<b_hL/b_\ell<L$ when $a_\ell<\kappa_h$, and the expressions agree at $x=\delta L/g$ and $x=b_hL/b_\ell$. At $F=b_hL$, high-type subscription at indifference leaves these private costs unchanged.

At $a_\ell=\kappa_h$, the segment on which only the low type uses AI has equal marginal costs for the two types. The flat advantage ends when the high type adopts, at $F/b_h=F/\delta$, or when coverage is exhausted, whichever occurs first. The high type produces manually throughout this interval, giving $E=\kappa_h(q_E-x)$. In all cases, equality of the cost advantages at $x$ and $q_E$ implies \cref{eq:recovery-low-cost}.

It remains to establish the cost bound for signals with a greater advantage. Let $w\coloneqq W(x,\hardQ)$. Since $\hardsig\leq\hardQ$, obtaining an advantage greater than $w$ requires $q>q_E$. To compare the high type's costs across different combinations of artifact quality and hard evidence, define
\begin{equation*}
J(q)\coloneqq C_h(q)-\frac{\rhoHard}{1-\rhoHard}W_1(q).
\end{equation*}
On each production segment,
\[
J'(q)=\frac{C_h'(q)-\rhoHard C_\ell'(q)}{1-\rhoHard}>0
\]
by the marginal comparison in the proof of \Cref{cor:riley}. Continuity makes $J$ strictly increasing at qualities where either type's marginal artifact cost changes. If $W(q,\hardsig)>w$, then
\begin{align*}
\costH(q,\hardsig)
&>J(q)+\frac{\rhoHard}{1-\rhoHard}w\\
&>J(q_E)+\frac{\rhoHard}{1-\rhoHard}w
=C_h(q_E)+\rhoHard\hardQ.
\end{align*}
Taking $q$ down to $q_E$ from above with $\hardsig=\hardQ$ gives signals with a greater advantage whose costs approach the bound. This establishes the infimum and shows that reducing hard evidence cannot yield a cheaper deviation with a greater cost advantage.
\end{proof}


\subsection{D1 and finite-support outcomes}

For equilibrium payoffs $U_\ell,U_h$, define $D\coloneqq U_h-U_\ell$. The threshold comparison in \cref{eq:d1-threshold} becomes
\begin{equation}
\label{eq:d1-cost-advantage-test}
\gamma_\ell(\sigvec)-\gamma_h(\sigvec)=W(\sigvec)-D.
\end{equation}
Every sender can guarantee $\thL$ at the origin, so $U_i\geq\thL$. A deviation assigned posterior zero cannot improve either payoff. Therefore, an otherwise incentive-compatible outcome fails D1 precisely when an unused signal satisfies
\begin{equation}
\label{eq:d1-revealing-deviation}
W(\sigvec)>D \; \text{ and } \; \costH(\sigvec)<\thH-U_h.
\end{equation}
D1 forces the posterior to one, and the high type gains from deviating. If no such signal exists, assign posterior one wherever D1 requires it and posterior zero at all other unused signals. Neither type can gain at these beliefs. At a forced high posterior, the high type cannot strictly gain, and the low type's threshold is higher still.

Finite support allows an unused signal to be chosen arbitrarily close to any on-path signal. This observation yields the following support restrictions, including those needed to characterize mixed outcomes.

\begin{lemma}
\label{lem:finite-support-signaling}
In every finite-support D1 equilibrium, a low-only signal is the origin. There is at most one high-only signal. If it is used, it is the unique minimizer in \cref{eq:deterrence-private-frontier} at $t=\DeltaT-(U_\ell-\thL)$. Every shared signal exhausts hard capacity. Under a strict shared-signal price inequality, its artifact quality is $x$. At the price boundary, its quality lies in $[x,q_E]$. Offers outside \cref{eq:shared-signal-regime} have no shared signal.
\end{lemma}

\begin{proof}[Proof of \Cref{lem:finite-support-signaling}]
A low-only signal receives assessment $\thL$, so any positive-cost choice is dominated by the origin. A high-only signal receives assessment $\thH$ and must impose low-type cost of at least $t=\DeltaT-(U_\ell-\thL)>0$. To see why $t$ is positive, note that assessments are at most $\thH$ and costs are nonnegative. Equality $U_\ell=\thH$ would require a low type to receive the high assessment at the zero-cost origin, contrary to Bayes' rule.

Suppose a high-only signal costs the high type more than $\detval(t)$. At the minimizing signal for $t$, the low type's deviation threshold is $\thH$, while the high type's threshold is strictly lower. This minimizer cannot already be a shared signal: the low type's threshold would exceed its shared assessment. Nor can it be another high-only signal with lower cost, since the high type would prefer that signal on path. Therefore, it is unused. D1 forces full recognition there, and the high type gains. Uniqueness of the minimizer gives the high-only restriction.

At a shared signal, both types receive the same assessment and attain their equilibrium payoffs, so $W(\sigvec)=D$. Its posterior is below one. If hard capacity is unused, a sufficiently small increase in hard evidence raises $W$ and costs the high type less than its gain from full recognition. An unused nearby signal satisfies \cref{eq:d1-revealing-deviation}. Shared signals must use $\hardsig=\hardQ$.

The same argument rules out any shared quality from which a small quality change raises $W_1$. A shared quality must be a local maximum or lie on a flat interval of $W_1$. Below low-type adoption, the advantage increases at rate $g$. While only the low type uses AI, its slope is $a_\ell-\kappa_h$. After both types adopt or coverage is exhausted, it increases again. A strict local maximum occurs only at $x$ under $a_\ell<\kappa_h$. At $a_\ell=\kappa_h$, the possible shared qualities lie on the plateau $[x,q_E]$. Outside \cref{eq:shared-signal-regime}, the advantage is strictly increasing, so no shared signal survives. A shared origin is also eliminated because a small positive quality raises the advantage.
\end{proof}

\begin{proof}[Proof of \Cref{thm:pure-d1}] I consider separately the cases of separation, partial pooling, and pure pooling.

\textit{Separation.} At Riley separation, $U_\ell=\thL$, $U_h=\thH-\rileycost$, and $D=\DeltaT-\rileycost=W(q_R,\hardsig_R)$. Any more costly pure separating PBE is eliminated by a deviation to the Riley signal: the low type's threshold there is $\thH$, and the high type's threshold is lower.

First suppose $\hardQ<\DeltaT$. A signal with $q>q_R$ and high-type cost below $\rileycost$ satisfies
\begin{equation*}
W(q,\hardsig)
<C_\ell(q)-\frac{C_h(q)}{\rhoHard}
+\frac{1-\rhoHard}{\rhoHard}\rileycost
<W(q_R,\hardQ).
\end{equation*}
The second inequality follows because $C_\ell(q)-C_h(q)/\rhoHard$ is strictly decreasing on every production segment. Such a signal cannot satisfy the D1 deviation test.

Among qualities $q\leq q_R$, Riley survives if and only if $W_1(q_R)$ is at least as great as every earlier value of $W_1$. In the strict shared-signal regime, this holds exactly when $q_R\leq x$ or $q_R\geq q_E$. If $x<q_R<q_E$, the signal $(x,\hardQ)$ has a greater advantage and a lower high-type cost, so Riley fails D1. At the price boundary, $W_1$ is nondecreasing and Riley always survives. Outside the shared-signal regime, it is strictly increasing and Riley also survives.

If $\hardQ\geq\DeltaT$, Riley uses $q_R=0$ and $\hardsig_R=\DeltaT$. The same decreasing-function argument rules out a positive-quality deviation with lower high-type cost and greater advantage. A zero-quality deviation with lower high-type cost has $\hardsig<\DeltaT$ and a smaller advantage. Therefore, Riley satisfies D1. Outside the shared-signal regime, \Cref{lem:finite-support-signaling} also establishes uniqueness of the finite-support observable outcome: all low types use the origin, and the sole high-only signal is Riley.

\textit{Pure pooling.} At the canonical pool $(x,\hardQ)$, let $v=\kappa_\ell x+\hardQ$, $h_P\coloneqq\kappa_hx+\rhoHard\hardQ$, and $w\coloneqq v-h_P$. Payoffs are $U_\ell=\thbar-v$ and $U_h=\thbar-h_P$, so $D=w$. Pooling is a PBE if and only if $v\leq\prior\DeltaT$, which keeps both types' payoffs weakly above $\thL$. By \Cref{lem:recovery-expenditure}, the infimum additional high-type cost of a greater-advantage deviation is $E$. The high type's gain from full recognition is $\dpool$. Hence, D1 permits the pool if and only if $E\geq\dpool$. At equality, every greater-advantage deviation costs strictly more than $E$, so the weak inequality is sufficient.

Under a strict price inequality, \Cref{lem:finite-support-signaling} leaves no other pooling quality. At the price boundary, a pool at $(y,\hardQ)$, with $y\in[x,q_E]$, incurs costs $v+\kappa_h(y-x)$ and $h_P+\kappa_h(y-x)$. Its remaining recovery expenditure is $E-\kappa_h(y-x)$. It survives exactly when
\[
v+\kappa_h(y-x)\leq\prior\DeltaT,
\]
\[
E-\kappa_h(y-x)\geq\dpool.
\]
Existence of any such pool implies existence of the pool at $x$.

\textit{Partial pooling.} If all high types and a fraction $\alpha$ of low types use $(x,\hardQ)$, Bayes' rule gives $\belief=\prior/[\prior+(1-\prior)\alpha]$. Low-type indifference with the origin requires $\belief\DeltaT=v$, yielding \cref{eq:semi-posterior}. The restrictions $\prior\DeltaT<v<\DeltaT$ ensure $\alpha\in(0,1)$. Payoffs are $U_\ell=\thL$ and $U_h=\thL+w$. Since $w>0$, the high type prefers sharing to the origin. Its gain from full recognition is $\DeltaT-v$, so the same deviation bound gives \cref{eq:semi-condition}, including equality.

To establish uniqueness in the strict partial-pooling region, \Cref{lem:finite-support-signaling} leaves only the origin $(0,0)$, $(x,\hardQ)$, and possibly a high-only signal. If a high-only signal coexists with sharing at posterior $\belief$, its deterrence requirement is $t=(1-\belief)\DeltaT+v>\hardQ$. Hence, its minimizing signal also exhausts hard capacity. Both signals must have cost advantage equal to the equilibrium payoff difference. The high-only quality exceeds $x$ and has first-channel advantage $W_1(x)$, so its signal is $(q_E,\hardQ)$. Indifference requires
\begin{equation}
\label{eq:coexistence-posterior-payoff}
\belief_E=1-\frac{E}{\DeltaT} \; \text{ and } \;
U_\ell-\thL=\DeltaT-v-E.
\end{equation}
When $v>\prior\DeltaT$ and $E>\DeltaT-v$, this low-type payoff is below $\thL$, which is impossible. Pure pooling is also infeasible. Riley fails D1 because \cref{eq:recovery-low-cost} places $q_R$ strictly between $x$ and $q_E$. Only the stated partial pool remains.
\end{proof}


\subsection{Mixed signaling outcomes and equilibrium selection}

Before applying the equilibrium selection rule, I characterize the additional finite-support mixed outcomes permitted by D1.

Under a strict shared-signal price inequality, the support restrictions also characterize the remaining mixtures. Suppose the shared signal and the high-only recovery signal coexist. Let $\beta$ be the high type's probability of sharing and $\alpha$ the low type's probability of sharing. With $\belief_E$ as in \cref{eq:coexistence-posterior-payoff}, Bayes' rule requires
\begin{equation}
\label{eq:coexistence-probabilities}
\alpha=\frac{\prior(1-\belief_E)}{(1-\prior)\belief_E}\,\beta.
\end{equation}
If $\DeltaT-v-E>0$, the low type's payoff exceeds $\thL$, so the origin is unused and $\alpha=1$. Such coexistence requires $\dpool\leq E<\DeltaT-v$; at $E=\dpool$, it reduces to pure pooling. If $E=\DeltaT-v$, both types' mixing probabilities can vary subject to \cref{eq:coexistence-probabilities} and $0\leq\alpha,\beta\leq1$. The limit $\beta=0$ is Riley separation. For $v>\prior\DeltaT$, shared mass is greatest at $\beta=1$, which gives the partial pool in \cref{eq:semi-posterior}.

Low-type indifference between the shared and recovery signals follows from \cref{eq:recovery-low-cost}, and high-type indifference follows from the definition of $E$. If the origin is used, \cref{eq:coexistence-posterior-payoff} equates its low-type payoff to the shared payoff. Greater-advantage deviations cost the high type strictly more than the recovery signal, so full recognition cannot improve its payoff. The belief construction following \cref{eq:d1-revealing-deviation} supplies the remaining off-path beliefs. Conversely, the support restrictions and indifference determine these same signals, payoffs, and probabilities. The described mixtures exhaust the additional finite-support outcomes under a strict price inequality.

At the price boundary, several shared qualities can coexist on the plateau. The following construction characterizes these outcomes before applying the equilibrium selection rule. Let $u\coloneqq U_\ell-\thL\geq0$, and retain $v$, $h_P$, and $w=v-h_P$ from the preceding proof. Every shared quality $y$ has posterior
\begin{equation*}
\belief_y=\frac{u+v+\kappa_h(y-x)}{\DeltaT}.
\end{equation*}
D1 permits sharing only if
\begin{equation}
\label{eq:plateau-shared-payoff-bound}
u\geq\DeltaT-v-E.
\end{equation}
Indeed, the high type's payoff is $\thL+u+w$, and the infimum cost of a greater-advantage deviation is $h_P+E$. Comparing this cost with $\thH-U_h$ gives the inequality.

Choose a nonempty finite set of shared qualities $y_j\in[x,q_E]$, with posteriors $\belief_{y_j}\in(0,1)$ and population masses $m_j>0$. Let $m_H\geq0$ and $m_0\geq0$ be the high-only and low-only population masses. Bayes' rule and the type totals require
\[
\sum_j\belief_{y_j}m_j+m_H=\prior,
\]
\[
\sum_j(1-\belief_{y_j})m_j+m_0=1-\prior.
\]
A positive $m_0$ requires $u=0$, and uses the origin. A positive $m_H$ uses $(q_u,\hardQ)$, where
\begin{equation*}
q_u=x+\frac{\DeltaT-u-v}{\kappa_h}.
\end{equation*}
The posterior restrictions and \cref{eq:plateau-shared-payoff-bound} place $q_u$ in $(x,q_E]$ and every shared quality $y_j$ below $q_u$.

All low types receive $\thL+u$, and all high types receive $\thL+u+w$. The high-only signal minimizes high-type cost at deterrence requirement $\DeltaT-u$. Shared signals satisfy the required indifferences, and the deviation bound supplies D1-consistent beliefs as above. Conversely, the support restrictions, indifference, and Bayes' rule imply exactly these conditions. At the price boundary, pure Riley separation is the remaining outcome with no shared signal.

The payoff comparisons now determine which surviving outcome the equilibrium selection rule selects.

\begin{lemma}
\label{lem:signaling-continuation-selection}
Whenever the canonical pure pool is feasible, it weakly Pareto dominates all finite-support D1 outcomes for senders. If pure pooling is unavailable and the canonical partial pool is feasible, it is sender-Pareto-maximal. Otherwise, the selected outcome is Riley separation. In particular, $\hardQ\geq\DeltaT$ guarantees separation for every admissible offer.
\end{lemma}

\begin{proof}[Proof of \Cref{lem:signaling-continuation-selection}]
For any outcome with sharing, let $u=U_\ell-\thL$. Every shared signal has payoff difference $w=W(x,\hardQ)$, including all plateau qualities. Sender payoffs are $\thL+u$ and $\thL+u+w$.

If the origin $(0,0)$ is used, then $u=0$. If it is unused, every on-path posterior is at least $(u+v)/\DeltaT$. Shared qualities are no smaller than $x$, and any high-only signal has posterior one. Since the average on-path posterior equals the prior, $u\leq\prior\DeltaT-v$. A feasible canonical pure pool attains $u_P=\prior\DeltaT-v\geq0$ and thus weakly dominates every outcome with sharing.

Compare this pool with Riley separation. Pooling feasibility implies $v\leq\prior\DeltaT<\DeltaT$, so $\hardQ<\DeltaT$ and $q_R>x$. Under a strict price inequality, any surviving Riley signal has $q_R\geq q_E$. Its additional high-type cost above $h_P$ is at least $E\geq\dpool$, giving $\thH-\rileycost\leq\thbar-h_P$. At the price boundary, Riley either lies on the plateau and gives the high type payoff $\thL+w$, or lies beyond it and satisfies the same expenditure comparison. Its low-type payoff is $\thL$. Both types weakly prefer the canonical pool. The ties select full shared mass and its least quality, $x$.

Now suppose pure pooling is unavailable. Under a strict price inequality, the coexistence formulas show that a shared outcome with $u>0$ would require the pure-pooling inequalities. At the price boundary, such an outcome must satisfy both $u\leq\prior\DeltaT-v$ and \cref{eq:plateau-shared-payoff-bound}. These imply $v\leq\prior\DeltaT$ and $E\geq\dpool$, again making pure pooling feasible. All remaining shared outcomes have $u=0$.

Existence of such a shared outcome requires $v<\DeltaT$ and $E\geq\DeltaT-v$. If $v\leq\prior\DeltaT$, then $E\geq\DeltaT-v\geq\dpool$ would make pure pooling feasible. Sharing is consequently possible precisely in the partial-pooling region stated in \cref{eq:semi-condition}.

A feasible partial pool gives payoffs $\thL$ and $\thL+w$. Under a strict price inequality, Riley coexists only at $E=\DeltaT-v$ and gives the same payoffs. At the price boundary, partial-pooling feasibility places $q_R$ on the plateau, so Riley again gives those payoffs. All additional shared mixtures have the same payoff pair.

It remains to compare shared mass. If its total is $M$, shared posteriors are at least $v/\DeltaT$, and the high-type mass using shared signals is at most $\prior$. Thus, $M\leq\prior\DeltaT/v$. Equality is attained by placing all high types and the low-type fraction in \cref{eq:semi-posterior} at $x$. This establishes the shared-mass and quality selection.

If neither canonical shared outcome is feasible, the classification leaves only Riley separation. It survives D1, as under a strict price inequality, failure would require $x<q_R<q_E$, implying $v<\DeltaT$ and $E>\DeltaT-v$. These inequalities would make either pure pooling or partial pooling feasible. All other offers already satisfy the Riley test. Finally, $\hardQ\geq\DeltaT$ makes every possible shared signal cost the low type at least $\DeltaT$, exceeding its gain from any assessment below $\thH$. Only separation remains.
\end{proof}

The classification concerns finite-support equilibria and does not characterize atomless equilibria. D1 permits the payoff-equivalent mixed equilibria described above. The equilibrium selection rule selects among them.

\begin{proof}[Proof of \Cref{prop:one-signal-equilibrium}]
Set $\hardQ=0$ in \Cref{cor:riley,thm:pure-d1}. Then $C_\ell(q_R)=\DeltaT$, $v=\kappa_\ell x$, and the canonical shared signal has quality $x$. The pure-pooling conditions reduce to \cref{eq:one-signal-pooling-condition}, and the strict-price Riley test is $q_R\leq x$ or $q_R\geq q_E$. This proves the benchmark proposition. \Cref{lem:signaling-continuation-selection} supplies its stated pooling selection.
\end{proof}


\section{Proofs of monopoly pricing and design}
\label{app:market-proofs}

This appendix proves the pricing and development results in \Cref{sec:market}. The technology $\mathbf d=(\epsilon_\ell,\epsilon_h,L)$ is fixed until the feature-choice proof. Operating profit excludes development cost, which is sunk at the pricing stage. \Cref{app:equilibrium-proofs} determines the selected signaling equilibrium at each tariff.

\subsection{The separating profit bound}

\begin{proof}[Proof of \Cref{prop:monopoly-separation}]
At a selected separating outcome, the low type chooses the origin $(0,0)$, and generates no revenue. Hard evidence supplies $\min\{\hardQ,\DeltaT\}$ units of deterrence, leaving the artifact to impose low-type imitation cost $r=(\DeltaT-\hardQ)_+$.

Suppose that the high type subscribes and uses positive AI input $z$. Cost minimization gives $z=\min\{q_R,L\}$, including at an input-cost tie. Write $e\coloneqq q_R-z\geq0$ for the remaining manual production. Subscription requires
\begin{equation*}
F\leq(\kappa_h-p-\epsilon_h)z.
\end{equation*}
The low type saves more from assistance at the same artifact and also subscribes when imitating the high type. Its binding artifact imitation cost is
\begin{equation*}
r=F+(p+\epsilon_\ell)z+\kappa_\ell e.
\end{equation*}
These relations also apply at $F=0$. If the high type uses no AI input, it pays no positive fee and generates zero operating profit.

Let $P_h\coloneqq F+(p-c)z$ be operating profit per subscribing high type. The adoption constraint and imitation cost imply
\begin{equation}
\label{eq:separating-revenue-bounds}
\begin{aligned}
r&\leq(\kappa_h+\delta)z+\kappa_\ell e,\\
P_h&\leq w_hz.
\end{aligned}
\end{equation}
If $w_h\leq0$, then no separating offer earns positive operating profit. If $w_h>0$, let $t_c\coloneqq\epsilon_\ell+c$. Then, $\kappa_h+\delta-t_c=w_h>0$, and so
\begin{align*}
t_cz+\kappa_\ell e
&\geq\frac{t_c}{\kappa_h+\delta}
\bigl[(\kappa_h+\delta)z+\kappa_\ell e\bigr],\\
&\geq\frac{t_c}{\kappa_h+\delta}r.
\end{align*}
Rewriting profit as $P_h=r-t_cz-\kappa_\ell e$ gives $P_h\leq w_hr/(\kappa_h+\delta)$. The second bound in \cref{eq:separating-revenue-bounds}, together with $z\leq L$, also gives $P_h\leq w_hL$. Multiplying by the high-type mass $\prior$ establishes \Cref{eq:profit-separation-general} as an upper bound over all separating tariffs.

To attain the bound when $w_h>0$ and $r>0$, choose any usage price in \cref{eq:separating-tariff}, let $z_*=\min\{L,r/(\kappa_h+\delta)\}$, and set $F=(\kappa_h-p-\epsilon_h)z_*>0$. Define
\[
e_*=\frac{r-(\kappa_h+\delta)z_*}{\kappa_\ell},
\]
\[
q_R=z_*+e_*.
\]
Either $e_*=0$ or $z_*=L$, so $\min\{q_R,L\}=z_*$. These are the production quantities in \cref{eq:monopoly-separating-output}.

At this quality, the high type's saving from assistance equals the fee, so it subscribes under the subscription rule. The low type has a larger saving and subscribes when imitating. Substitution gives $C_\ell(q_R)=r$. Since the price satisfies $p+\epsilon_\ell>\kappa_h$, then \Cref{thm:pure-d1} makes Riley separation the unique finite-support D1 observable outcome. The operating profit is $\prior w_hz_*$. On path, the low type remains at the origin $(0,0)$, and does not subscribe.

If $r=0$ or $w_h\leq0$, then the maximum operating value is zero. It is attained by a positive membership fee and a usage price making AI more costly than manual production for both types. Nobody subscribes, and the selected signaling outcome separates. This establishes both the upper bound and its attainment.
\end{proof}


\subsection{Prices supporting a shared signal}

A shared quality $x$ fixes the low type's total bill, but its division between membership and usage payments affects revealing deviations. Maximizing recovery expenditure determines whether any admissible tariff can support that quality.

Write $g\coloneqq\kappa_\ell-\kappa_h$, $\delta\coloneqq\epsilon_\ell-\epsilon_h$, and $s\coloneqq g-\delta>0$, as in \Cref{app:equilibrium-proofs}. For a usage price $p\in[0,\kappa_h-\epsilon_\ell]$, set $F=(\kappa_\ell-p-\epsilon_\ell)x$. Let $E(x,p)$ denote the recovery expenditure in \Cref{lem:recovery-expenditure}, including offers at the price boundary.

Two quality thresholds determine the price maximizing this expenditure, that is,
\begin{equation}
\label{eq:entry-maximization-thresholds}
\begin{aligned}
x_1&\coloneqq\frac{\delta L}{g},\\
x_2&\coloneqq\frac{(\kappa_h-\epsilon_h)L}{\kappa_\ell-\epsilon_\ell}.
\end{aligned}
\end{equation}

Under \Cref{ass:comparative-advantage}, $0<x_1<x_2<L$.

\begin{lemma}
\label{lem:maximum-recovery-expenditure}
For $0<x<L$, the maximum recovery expenditure over admissible shared-signal prices is
\begin{equation}
\label{eq:maximum-recovery-expenditure}
\overline E(x)\coloneqq\max_{0\leq p\leq\kappa_h-\epsilon_\ell}E(x,p)
=\begin{cases}
\displaystyle\frac{\kappa_hs}{\delta}x,
& 0<x\leq x_1,\\[6pt]
\displaystyle\frac{\kappa_hs}{g}L,
& x_1\leq x\leq x_2,\\[6pt]
\displaystyle\frac{\kappa_h(\kappa_\ell-\epsilon_\ell)}{g}(L-x),
& x_2\leq x<L.
\end{cases}
\end{equation}
An attaining price is
\begin{equation}
\label{eq:recovery-maximizing-price}
p_E(x)=\begin{cases}
\kappa_h-\epsilon_\ell,
& 0<x\leq x_1,\\[6pt]
\displaystyle\frac{(\kappa_h+\delta)L-\kappa_\ell x}{L-x}-\epsilon_\ell,
& x_1\leq x\leq x_2,\\[6pt]
0,
& x_2\leq x<L.
\end{cases}
\end{equation}
Set $F_E(x)=(\kappa_\ell-p_E(x)-\epsilon_\ell)x$. The expenditure function extends continuously to $\overline E(0)=\overline E(L)=0$.
\end{lemma}

\begin{proof}[Proof of \Cref{lem:maximum-recovery-expenditure}]
For $x\leq x_1$, recovery occurs within coverage. \Cref{eq:recovery-expenditure-piecewise} gives $E(x,p)=(p+\epsilon_\ell)sx/\delta$, which increases with $p$. The maximum is attained at the price boundary, $p=\kappa_h-\epsilon_\ell$.

For $x>x_1$, let $a_i=p+\epsilon_i$ and $b_i=\kappa_i-a_i$. If $F=b_\ell x\leq b_hL$, then the high type adopts by the coverage limit, and
\begin{equation*}
E(x,p)=b_\ell x+\left(a_h-\frac{\kappa_h\delta}{g}\right)L.
\end{equation*}
Its derivative with respect to $p$ is $L-x>0$. If $F\geq b_hL$, the high type produces manually, and recovery expenditure is $E(x,p)=\kappa_hb_\ell(L-x)/g$, which decreases with $p$. The expressions agree at $F=b_hL$. Solving this equality for $p$ gives the middle price in \cref{eq:recovery-maximizing-price}.

This price falls from $\kappa_h-\epsilon_\ell$ at $x_1$ to zero at $x_2$. Between these thresholds, it maximizes recovery expenditure. The equality $b_\ell x=b_hL$ implies $b_\ell(L-x)=(b_\ell-b_h)L=sL$, yielding the constant middle value in \cref{eq:maximum-recovery-expenditure}. For $x>x_2$, the equality would require a negative usage price, so the maximum is attained at $p=0$.

The formulas agree at both thresholds and have the stated endpoint limits. All attaining prices are nonnegative and satisfy the shared-signal price restriction. Their membership fees are positive because $\kappa_\ell-p_E(x)-\epsilon_\ell\geq g>0$.
\end{proof}

For $0<x\leq x_1$, the maximizing price creates the flat cost-advantage interval described in \Cref{app:equilibrium-proofs}. For $x_1\leq x\leq x_2$, the maximizing tariff makes the high type indifferent about adoption at the coverage limit. These ties affect off-path production costs even though the high type produces the shared artifact manually.


\subsection{The shared profit bound and quality choice}

\begin{proof}[Proof of \Cref{prop:monopoly-shared-value}]
\textit{Revenue and feasibility.} At a selected shared signal, participating low types subscribe and use input $x$. High types produce manually, and low types at the origin do not subscribe. By \cref{eq:shared-payments}, revenue per subscriber is $(\kappa_\ell-\epsilon_\ell)x$ and compute cost is $cx$. Thus, operating profit is $w_\ell xD_\ell(x,\hardQ)$.

For $\hardQ<\DeltaT$, \Cref{thm:pure-d1,lem:signaling-continuation-selection} give the feasible quality set
\begin{equation}
\label{eq:shared-feasible-quality-set}
\mathcal X(\mathbf d,\hardQ)
=\left\{x\in(0,L) \mid \kappa_\ell x+\hardQ<\DeltaT,\;\text{ and }\;
\overline E(x)\geq\min\{\dpool,\DeltaT-\kappa_\ell x-\hardQ\}\right\}.
\end{equation}
When $v=\kappa_\ell x+\hardQ\leq\prior\DeltaT$, the high type's gain from full recognition is $\dpool$, and pure pooling is selected. When $\prior\DeltaT<v<\DeltaT$, that gain is $\DeltaT-v$, and partial pooling is selected. \Cref{lem:maximum-recovery-expenditure} establishes necessity of the recovery inequality. Its attaining tariff $(F_E(x),p_E(x))$ establishes sufficiency. Subscription at indifference and the equilibrium selection rule also apply to offers at the price boundary and to equality in the participation and revealing-deviation conditions, so every feasible quality implements the operating profit above.

If $w_\ell\leq0$, serving these subscribers cannot earn positive operating profit, and an inactive offer earns zero. If $w_\ell>0$, maximizing over feasible qualities gives \cref{eq:profit-pooling-general}. An empty feasible set also gives zero shared value. At $\hardQ\geq\DeltaT$, sharing is impossible.

\textit{Attainment.} To establish attainment despite the strict quality restrictions, define $B_Q\coloneqq\min\{L,(\DeltaT-\hardQ)/\kappa_\ell\}$ and the compact auxiliary set
\begin{equation}
\label{eq:closed-shared-quality-set}
\widehat{\mathcal X}(\mathbf d,\hardQ)
=\{0\}\cup\left\{x\in[0,B_Q] \mid
\overline E(x)\geq\min\{\dpool,\DeltaT-\kappa_\ell x-\hardQ\}\right\}.
\end{equation}
The added point zero represents the inactive option. Let $A_\ell(x,\hardQ)\coloneqq xD_\ell(x,\hardQ)$ denote aggregate low-type AI input. Extend it to the auxiliary domain using the expression in \cref{eq:low-type-demand} for $x>0$, and set $A_\ell(0,\hardQ)=0$. This function is continuous, including when $\hardQ=0$.

At $\kappa_\ell x+\hardQ=\DeltaT$, low-type demand is zero. At $x=L$, recovery expenditure is zero, so the auxiliary feasibility inequality can hold only at the equality $\kappa_\ell x+\hardQ=\DeltaT$. Every point added to the genuine feasible set has aggregate input zero. If $\mathcal X$ is nonempty, the continuous function $A_\ell$ has a positive maximum on the compact auxiliary set, and that maximum lies in $\mathcal X$. The attaining tariff implements it. This proves attainment whenever $H>0$.

\textit{Unconstrained qualities.} For partial pooling, aggregate input and its quality derivative are
\begin{align*}
A_\ell(x,\hardQ)
&=\frac{\prior x(\DeltaT-\kappa_\ell x-\hardQ)}{\kappa_\ell x+\hardQ},\\
\frac{\partial A_\ell}{\partial x}
&=\prior\left[\frac{\DeltaT\hardQ}{(\kappa_\ell x+\hardQ)^2}-1\right].
\end{align*}
The objective is concave, and strictly concave when $\hardQ>0$. Its unconstrained stationary quality is, for $0<\hardQ<\DeltaT$,
\begin{equation}
\label{eq:partial-pooling-optimal-quality}
x_M\coloneqq\frac{\sqrt{\hardQ\DeltaT}-\hardQ}{\kappa_\ell}.
\end{equation}
At $\hardQ=0$, the partial-pooling objective decreases with $x$.

Dropping coverage and recovery restrictions gives an upper bound $\overline H$ on shared operating value. Let $x_P=(\prior\DeltaT-\hardQ)/\kappa_\ell$. Where pure pooling is available without those restrictions, its objective increases up to $x_P$. The partial-pooling derivative at $x_P$ changes sign at $\hardQ=\prior^2\DeltaT$. The unconstrained maximizing quality is $x_P$ for $0\leq\hardQ\leq\prior^2\DeltaT$ and $x_M$ for $\prior^2\DeltaT<\hardQ<\DeltaT$. Substitution gives
\begin{equation}
\label{eq:unconstrained-shared-value}
\overline H(\mathbf d,\hardQ)
=\begin{cases}
\displaystyle\frac{(1-\prior)(w_\ell)_+}{\kappa_\ell}
(\prior\DeltaT-\hardQ),
&0\leq\hardQ\leq\prior^2\DeltaT,\\[6pt]
\displaystyle\frac{\prior(w_\ell)_+}{\kappa_\ell}
\left(\sqrt{\DeltaT}-\sqrt{\hardQ}\right)^2,
&\prior^2\DeltaT\leq\hardQ<\DeltaT.
\end{cases}
\end{equation}
For $w_\ell>0$, $H=\overline H$ if and only if the corresponding maximizing quality belongs to \cref{eq:shared-feasible-quality-set}. It must lie below coverage and provide sufficient recovery expenditure. At $\hardQ=\prior^2\DeltaT$, both quality formulas coincide and the shared posterior is $\prior$, so the outcome is pure pooling. Above that capacity, the unconstrained optimum is partial pooling with posterior $\sqrt{\hardQ/\DeltaT}$. These formulas solve the shared-profit problem. Global provider optimality also requires comparison with $R$.
\end{proof}


\subsection{Feature choice and fresh development}

\begin{proof}[Proof of \Cref{thm:product-bifurcation}]
By \Cref{lem:signaling-continuation-selection}, every selected tariff yields separation or a canonical shared outcome. \Cref{prop:monopoly-separation,prop:monopoly-shared-value} give the largest operating profit in each class, including the inactive zero-profit option. The operating optimum for a developed technology is therefore $G=\max\{R,H\}$. The tariff constructions above attain every positive value. An inactive offer attains zero.

To establish an optimal design, it remains to show that the operating value cannot fall at the limit of a sequence of designs. The separating value $R$ is continuous in $\mathbf d$. For fixed $\hardQ<\DeltaT$, the recovery envelope in \cref{eq:maximum-recovery-expenditure} is jointly continuous in design and quality because its expressions agree at the moving thresholds. Compactness of $\mathcal D$ bounds coverage, and the strict technology restrictions keep the denominators bounded away from zero. The graph of the auxiliary feasible set in \cref{eq:closed-shared-quality-set} is consequently closed and bounded. The objective $(w_\ell)_+A_\ell(x,\hardQ)$ is jointly continuous.

Consider any convergent sequence of designs and choose a maximizing auxiliary quality for each. Take a subsequence whose shared values approach the original sequence's limit superior, then a further subsequence whose qualities converge. The closed graph makes the limiting quality feasible for the limiting design. By continuity, its objective equals that limit superior. The shared value at the limiting design is at least as large, establishing upper semicontinuity of $H$.

For $\hardQ\geq\DeltaT$, $H=0$ throughout the design set. Thus, for every fixed capacity, $G=\max\{R,H\}$ is upper semicontinuous. Subtracting continuous development cost $K$ preserves this property. A maximizing sequence on compact $\mathcal D$ has a limit attaining the best net development return.

Development incurs $K(\mathbf d)$ before pricing. Comparing the attained return $G(\mathbf d,\hardQ)-K(\mathbf d)$ with the no-launch payoff zero gives \cref{eq:monopoly-design-choice} and its launch cases. At a positive operating optimum, $R>H$ selects separation, whereas $H>R$ selects pooling or partial pooling at an optimal shared quality. At $R=H>0$, both attaining tariff families are optimal. Different shared maximizers can also remain tied. These provider profit ties are preserved.

For an installed technology, development cost has already been incurred, so the provider maximizes $G$ when repricing. At $\hardQ\geq\DeltaT$, $R=H=0$, making fresh development with positive $K$ unprofitable. If $K=0$, launch and no launch can tie without inducing positive incremental AI usage. Attainment throughout uses subscription at indifference, the included price boundary, and the specified ties in the equilibrium selection rule. The provider anticipates the selected signaling outcome and has no commitment power over sender choices or receiver beliefs.
\end{proof}

\subsection{Editing improvements and monopoly design}
\label{app:monopoly-editing-incentives}

\begin{proof}[Proof of \Cref{prop:monopoly-editing-incentives}]
Write $r=(\DeltaT-\hardQ)_+$, $\delta=\epsilon_\ell-\epsilon_h$, $g=\kappa_\ell-\kappa_h$, and $s=g-\delta$. The separating value is
\begin{equation*}
R=\prior(w_h)_+\min\left\{L,\frac{r}{\kappa_h+\delta}\right\}.
\end{equation*}
Reducing $\epsilon_\ell$ leaves $w_h$ fixed and lowers the denominator. If $w_h>0$ and coverage does not bind, a high-type improvement of size $u$ raises $w_h$ and $\delta$ by $u$, with derivative
\begin{equation*}
\frac{dR}{du}
=\frac{\prior r(\epsilon_\ell+c)}{(\kappa_h+\delta)^2}>0.
\end{equation*}
when $r>0$. When coverage binds, the derivative is $\prior L>0$. The profit expressions are continuous at the coverage threshold. The positive-part operator gives weak monotonicity when $w_h\leq0$. If $r=0$, then $R$ remains zero.

For shared provision, the recovery envelope in \cref{eq:maximum-recovery-expenditure} can be written as
\begin{equation*}
\overline E(x)
=\min\left\{
\frac{\kappa_hs}{\delta}x,
\frac{\kappa_hs}{g}L,
\frac{\kappa_h(\kappa_\ell-\epsilon_\ell)}{g}(L-x)
\right\}.
\end{equation*}
for $0\leq x\leq L$. The first expression is smallest up to $x_1$, the second from $x_1$ to $x_2$, and the third thereafter. A low-type improvement lowers $\delta$, raises $s$ and $\kappa_\ell-\epsilon_\ell$, and thus weakly raises every expression in the minimum. It expands the feasible shared-quality set and raises $w_\ell$. Aggregate input $xD_\ell(x,\hardQ)$ is unchanged at a fixed quality, so $H$ weakly rises. A high-type improvement raises $\delta$ and lowers $s$. The first two expressions weakly fall and the third is unchanged. The feasible shared-quality set contracts while $w_\ell$ remains fixed, so $H$ weakly falls. The zero-value cases follow from the inactive option.

Finally, suppose $R>0$, and put $b=\kappa_h+\delta$. The optimal separating resource cost in \cref{eq:separating-resource-cost} is
\begin{equation*}
C_R^{\mathrm{real}}
=\begin{cases}
\displaystyle\prior\left[(\epsilon_h+c)\frac{r}{b}+\rhoHard\hardQ\right],&r\leq bL,\\[6pt]
\displaystyle\prior\left[(\epsilon_h+c)L+\frac{\kappa_h}{\kappa_\ell}(r-bL)+\rhoHard\hardQ\right],&r\geq bL.
\end{cases}
\end{equation*}
For a low-type improvement of size $u$, the derivative is $\prior(\epsilon_h+c)r/b^2$ when coverage is slack and $\prior\kappa_hL/\kappa_\ell$ when it binds. Both are positive. Continuity at $r=bL$ establishes the strict resource comparison across the threshold.
\end{proof}

\section{Proofs of competition and entry results}
\label{app:competition-proofs}

The pricing and entry proofs require signaling equilibria after every tariff menu, including menus induced by deviations from equilibrium offers. I first establish these signaling equilibria for identical technologies. \Cref{supp:differentiated-technologies,supp:simultaneous-entry} consider differentiated tools and simultaneous entry and tariff posting.


\subsection{Signaling after a tariff menu}

Fix a nonempty finite menu $\mathcal M$ of identical technologies with nonnegative membership fees and usage prices. Subscribing to several providers does not expand coverage. A sender can concentrate input in its subscribed plan with the lowest usage price and drop the other plans without increasing its cost. One plan suffices to minimize production cost, giving \cref{eq:identical-menu-cost}. Additional unused zero-fee subscriptions affect neither production cost nor provider profit.

Write $C_i(q)$ for the resulting artifact cost and $c_i(q,\hardsig)$ for total signaling cost, as in \cref{eq:low-cost-normalized,eq:high-cost-normalized}. Artifact costs are continuous, strictly increasing, and piecewise affine, with $C_i(0)=0$. They are unbounded because production beyond $L$ has manual slope $\kappa_i$. For $q>0$, we have $C_h(q)<C_\ell(q)$: the high type can use the low type's minimizing production method, pay the same tariff, and incur less effort. This advantage need not increase with $q$, since the types can choose different plans.

The existence and selection argument below uses only continuity, strict monotonicity, unboundedness, zero cost at the origin, and $C_h(q)<C_\ell(q)$ for $q>0$, together with $0<\rhoHard<1$ and finite $\hardQ$. Thus, it applies to other tariff menus once these cost properties have been verified.

To test deviations across all possible signals, retain the cost advantage from \cref{eq:total-cost-advantage},
\[
W(\sigvec)=c_\ell(\sigvec)-c_h(\sigvec).
\]
For a high-type expenditure budget $t\geq0$, define the greatest attainable advantage as
\begin{equation}
\label{eq:menu-advantage-frontier}
V(t)\coloneqq\max_{c_h(\sigvec)\leq t}W(\sigvec).
\end{equation}
The maximum exists because the budget bounds $q$ and $\hardsig\in[0,\hardQ]$. The function $V$ is continuous and nondecreasing, with $V(0)=0$. For continuity from smaller budgets, slightly reduce a maximizing positive signal until it fits the lower budget. Strictly increasing costs make this feasible, and its advantage converges to the original value. Compactness gives continuity from larger budgets, including at zero.

\begin{lemma}
\label{lem:menu-signaling-continuation}
Every nonempty finite tariff menu of identical technologies has a finite-support D1 signaling equilibrium satisfying the equilibrium selection rule in \Cref{sec:model}. It can be chosen as pooling, partial pooling, or least-cost separation.
\end{lemma}

\begin{proof}[Proof of \Cref{lem:menu-signaling-continuation}]
\textit{D1 tests.} Let $u_i\coloneqq U_i-\thL$ denote equilibrium payoff above the low assessment. The origin guarantees $u_\ell\geq0$, and the high type can imitate the low type at weakly lower cost, so $u_h\geq u_\ell$. Set $D=u_h-u_\ell$.

The threshold calculation in \cref{eq:d1-cost-advantage-test,eq:d1-revealing-deviation} applies to the menu costs. A profitable deviation that D1 attributes to the high type requires $W(\sigvec)>D$ and $c_h(\sigvec)<\DeltaT-u_h$. At every other unused signal, a low-type belief can deter both types. Together with on-path incentives, the D1 test is therefore
\begin{equation}
\label{eq:menu-global-d1}
V(\DeltaT-u_h)\leq u_h-u_\ell.
\end{equation}
Using a closed budget in $V$ does not change the test. If an advantage above $D$ were attainable at a signal satisfying $c_h(\sigvec)=\DeltaT-u_h>0$, then a slight reduction in the signal would preserve that strict advantage and make the deviation profitable. Also, $u_h<\DeltaT$ in equilibrium: equality would require full recognition at the zero-cost origin, which the low type would also prefer.

Let $\psi$ be the high type's minimum cost subject to $c_\ell(\sigvec)\geq\DeltaT$. A minimum exists, and the constraint binds. Separation at a minimizing signal gives payoffs $(u_\ell,u_h)=(0,\DeltaT-\psi)$ and passes D1 if
\begin{equation}
\label{eq:menu-separation-test}
V(\psi)\leq\DeltaT-\psi.
\end{equation}
A more costly separating signal fails D1. Indeed, deviation to a minimizing signal gives both a greater advantage than the proposed payoff difference and a high-type cost below the deviation budget.

Pooling at a signal $v$ gives $u_i=\prior\DeltaT-c_i(v)$. On-path incentives and D1 hold if and only if
\begin{equation}
\label{eq:menu-pooling-test}
c_\ell(v)\leq\prior\DeltaT\;\text{ and }\;
V\bigl(c_h(v)+(1-\prior)\DeltaT\bigr)\leq W(v).
\end{equation}
If $\prior\DeltaT<c_\ell(v)<\DeltaT$, partial pooling instead has posterior $\belief=c_\ell(v)/\DeltaT$ and low-type participation probability
\[
\alpha=\frac{\prior(1-\belief)}{(1-\prior)\belief}.
\]
All high types use $v$, and the remaining low types use the origin. Bayes' rule and low-type indifference give payoffs $(0,W(v))$. Its D1 test is
\begin{equation}
\label{eq:menu-partial-pooling-test}
V(\DeltaT-W(v))\leq W(v).
\end{equation}
These tests also supply off-path beliefs: assign the high type whenever D1 requires it and the low type otherwise.

\textit{Existence.} There is a unique $D_*\in(0,\DeltaT)$ satisfying
\begin{equation}
\label{eq:menu-payoff-fixed-point}
D_*=V(\DeltaT-D_*).
\end{equation}
Indeed, $V(\DeltaT-D)-D$ is continuous and strictly decreasing, positive at zero, and negative at $\DeltaT$. Choose a maximizing signal $v_*$ at budget $t_*\coloneqq\DeltaT-D_*$. Then, $W(v_*)=D_*$ and $c_\ell(v_*)\leq\DeltaT$.

If $c_\ell(v_*)=\DeltaT$, then $c_h(v_*)=t_*$ and $\psi\leq t_*$. At a minimizing separating signal, $V(\psi)\geq\DeltaT-\psi$. If $\psi<t_*$, monotonicity would imply $V(\psi)\leq V(t_*)=D_*<\DeltaT-\psi$, a contradiction. Hence, $\psi=t_*$, and \cref{eq:menu-separation-test} holds.

If $c_\ell(v_*)\leq\prior\DeltaT$, pooling passes \cref{eq:menu-pooling-test}, since $c_h(v_*)+(1-\prior)\DeltaT\leq t_*$. In the remaining case, partial pooling passes \cref{eq:menu-partial-pooling-test}. Therefore, at least one of the three outcomes exists.

\textit{Selection.} First, every shared signal exhausts hard capacity. Otherwise, a small increase in hard evidence raises $W$ while preserving a profitable high-type deviation to full recognition. Thus, shared signals have $\hardsig=\hardQ$. A low-only signal must be the origin, since a positive-cost choice would yield less than the payoff guaranteed there.

On-path posteriors have unconditional mean $\prior$. If an equilibrium uses several signals, retain at most two whose posteriors bracket $\prior$ and choose their masses to preserve that mean. Bayes' rule determines feasible type-specific probabilities. Each retained signal gives a using type its original payoff, and the other type cannot gain by switching to it. Removed signals can be deterred: shared signals have $W=D$, high-only signals have $c_h=\DeltaT-u_h$, and the only possible low-only signal is the origin, i.e., $(0,0)$. Thus, the support can be reduced to at most two signals without changing sender payoffs or violating D1.

If both retained signals are shared, pool at the one with the lower posterior. Both types used that signal, so replacing its assessment by the prior mean weakly raises both payoffs. If one retained signal is shared and the other high-only, pooling at the shared signal also weakly raises both payoffs. In either case, the shared signal's advantage $W=D$ is unchanged and the high-type deviation budget weakly falls, preserving D1. A low-only signal paired with a shared signal gives partial pooling. Paired with a high-only signal, it gives least-cost separation. Thus, every finite-support D1 equilibrium is weakly sender-Pareto dominated by one of the three stated forms.

For any D1 partial pool, $c_\ell(v)<\DeltaT$ implies $V(\DeltaT-W(v))\geq W(v)$. Its D1 test and \cref{eq:menu-payoff-fixed-point} give $W(v)=D_*$. The separating test likewise gives $\DeltaT-\psi=D_*$. A D1 pool has $u_h\geq D_*$. For otherwise, its deviation budget exceeds $t_*$, while \cref{eq:menu-global-d1} would require $V(\DeltaT-u_h)\leq u_h-u_\ell<D_*$. Its low-type payoff is nonnegative. Every D1 pool weakly dominates all D1 separating and partial-pooling outcomes.

Among D1 pools, the least quality $q$ gives both types the highest payoffs, since every pool uses $\hardsig=\hardQ$ and both artifact costs increase with $q$. A least quality exists, as pooling candidates lie in a compact set and satisfy the closed inequalities in \cref{eq:menu-pooling-test}.

If no D1 pool exists, the support reduction implies that every equilibrium has payoffs $(0,D_*)$. Then, every shared signal has posterior $\belief=c_\ell(v)/\DeltaT>\prior$. A shared posterior at or below the prior would give a D1 pool. Among shared candidates, the least quality has the least posterior, say $\belief_*$. Total shared mass cannot exceed $\prior/\belief_*$, because the high-type mass is $\prior$. Partial pooling with all high types at this signal attains the bound.

A least shared quality exists whenever there is a shared candidate. Fix one at quality $q_0$ and restrict attention to $q\leq q_0$. Limits preserve the closed D1 inequality, while the upper bound $q_0$ keeps low-type cost strictly below $\DeltaT$. A limit cannot have $c_\ell\leq\prior\DeltaT$, as its pooling deviation budget would be no larger than its partial-pooling budget, so the same signal would satisfy the pooling test, contradicting the absence of a D1 pool. Selection favors partial pooling at the least shared quality. If no shared candidate exists, it favors a minimum-cost separating signal with the least $q$. These choices follow the equilibrium selection rule in \Cref{sec:model}.
\end{proof}


\subsection{Pricing with identical developed technologies}

\begin{proof}[Proof of \Cref{prop:provider-competition}]
Hold every rival's offer at $F=0$, $p=c$, and let provider $j$ deviate to any $F_j\geq0$, $p_j\geq0$. A subscriber using $z_j$ units from $j$ can obtain the same input from a rival without changing artifact quality, editing effort, or the observed signal. Hence, cost minimization requires $F_j+p_jz_j\leq cz_j$. The deviator's operating profit from this subscriber satisfies
\begin{equation}
\label{eq:competitive-deviation-bound}
F_j+(p_j-c)z_j\leq0.
\end{equation}
The same inequality holds when input is split across plans. The sender can drop $j$ and move its input to a cost-priced rival. An unused subscription with a positive fee cannot minimize cost, while an unused zero-fee subscription generates no revenue. Averaging over subscribers preserves the profit bound, regardless of the signaling equilibrium induced by the deviation. Every provider earns zero operating profit at the proposed profile, so no unilateral deviation is profitable. \Cref{lem:menu-signaling-continuation} supplies a signaling equilibrium after every deviation.

At this profile, \cref{eq:competitive-artifact-cost} gives initial slopes $m_i=\min\{\kappa_i,c+\epsilon_i\}$ and manual slopes beyond $L$. The high type has a lower marginal cost under both production methods, so $m_h<m_\ell$. Hence, its artifact cost advantage $C_\ell(q)-C_h(q)$ increases throughout. A shared signal fails D1, as a sufficiently small increase in quality raises the advantage and induces full type recognition, which the high type prefers to its shared assessment. Every finite-support D1 outcome must consequently separate, and its high signal must minimize expenditure subject to $c_\ell\geq\DeltaT$.

By \Cref{ass:comparative-advantage}, the high type's marginal cost under either production method exceeds $\rhoHard$ times the low type's. For assisted production,
\[
c+\epsilon_h>\rhoHard(c+\epsilon_\ell).
\]
Taking the minimum across methods preserves this inequality, giving $m_h>\rhoHard m_\ell$. Hard evidence is the cheaper way to deter imitation until capacity is exhausted. The unique minimizing signal is $\hardsig_R=\min\{\hardQ,\DeltaT\}$ and $C_\ell(q_R)=(\DeltaT-\hardQ)_+$.

Write $\psi$ for the high type's cost at this signal. To verify D1, suppose a signal had $W>\DeltaT-\psi$ and $c_h<\psi$. If its low-type cost were already at least $\DeltaT$, it would contradict the definition of $\psi$. Otherwise, increase its quality until low-type cost reaches $\DeltaT$. The advantage increases with quality, so the resulting high-type cost is less than $\DeltaT-W<\psi$, again contradicting minimality. Thus, \cref{eq:menu-separation-test} holds.

At $F=0$, $p=c$, each type's minimized private cost equals its minimized resource cost. Any separating allocation satisfies the low type's imitation constraint
\[
c_\ell(\sigvec_h)\geq\DeltaT+c_\ell(\sigvec_\ell)\geq\DeltaT.
\]
Its high-type resource cost is at least $\psi$, and its low-type resource cost is nonnegative. Riley separation attains this lower bound with the low type at the origin. The efficiency statement concerns separating allocations for this fixed menu and its imitation constraints. Comparisons involving development costs, different features, or an economy without paid AI require separate accounting.
\end{proof}

\subsection{Development before pricing}
\label{app:provider-entry}

Consider a finite set $\mathcal J$ of at least two potential providers, each able to develop a copy of the fixed technology $\mathbf d$ at cost $K=K(\mathbf d)\geq0$. Providers first simultaneously decide whether to incur this cost. After observing entry, the entrants simultaneously set their tariffs. Senders then subscribe and signal, and the receiver assesses them. Development cost is incurred on entry, regardless of subsequent sales. Let $G=G(\mathbf d,\hardQ)$ be the maximum monopoly operating profit defined in \Cref{eq:monopoly-operating-value}.

\begin{proposition}
\label{prop:provider-entry}
If $K>0$, the sequential development and pricing game admits subgame-perfect equilibria with pure provider strategies, finite-support D1 signaling equilibria, and the following entry outcomes,
\begin{enumerate}[label=\textup{(\roman*)}]
\item If $G<K$, no provider enters;
\item If $G>K$, exactly one provider enters, charges an optimal monopoly tariff, and earns $G-K$;
\item If $G=K$, both no entry and entry by one provider can be supported.
\end{enumerate}
In these constructions, a sole entrant charges a tariff attaining $G$ after every one-entrant history. After every history with two or more entrants, all entered providers set $F_j=0$ and $p_j=c$.
\end{proposition}

\begin{proof}[Proof of \Cref{prop:provider-entry}]
Fix the technology, its development cost $K>0$, and the finite set of potential providers. Specify strategies for the subsequent pricing and signaling stages after every realized entry set. With no entrants, signaling proceeds without paid AI. A sole entrant posts a tariff attaining $G$. \Cref{app:market-proofs} establishes attainment, including subscription at indifference and equality in the conditions supporting shared signaling. With two or more entrants, every provider posts $F=0$ and $p=c$. After every possible tariff menu, use the signaling equilibrium in \Cref{lem:menu-signaling-continuation}. For a single offer, this agrees with \Cref{lem:signaling-continuation-selection}.

These strategies are optimal in each pricing subgame. A sole entrant maximizes operating profit because development cost is sunk. With multiple entrants, the prescribed tariffs form a Nash equilibrium by \Cref{prop:provider-competition}. The signaling equilibria satisfy sender optimality, Bayes' rule, receiver optimality, and D1, including after tariff deviations.

If $G<K$, prescribe no entry. A unilateral entrant would earn $G-K<0$, compared with zero from remaining out. If $G>K$, prescribe entry by one designated provider. It earns $G-K>0$ and prefers entry to withdrawal. Any additional entrant would induce competitive pricing and earn $-K<0$. If $G=K$, entry into an empty market yields zero, and a sole entrant is indifferent to withdrawal. Both no entry and entry by one provider are therefore supported.

Under these pricing equilibria, two or more entrants each lose $K$ and can obtain zero by remaining out. Multiple entry cannot be an equilibrium when $K>0$. This conclusion is conditional on the specified pricing equilibria. The proof does not characterize all equilibria of the pricing subgames.
\end{proof}

Potential competition can coexist with pooling in these equilibria. In the usage-subsidy example in \Cref{sec:market}, $G=0.01323$. With $K=0.001$, a sole entrant earns $0.01223$ and the signaling outcome pools. An additional entrant would earn zero operating profit under the specified competitive pricing equilibrium and lose its development cost.

If $K=0$, the same strategies for the subsequent pricing and signaling stages support a sole entrant for every $G\geq0$ and any entry set with at least two providers. An additional entrant earns zero, so remaining out is a weak best response. At a multiple-entry outcome, every incumbent earns zero and would also obtain zero by withdrawing. No entry is supported when $G=0$. If $G>0$, then entry into an empty market is profitable. Sender subscription at indifference follows the same subscription rule used in monopoly pricing.

The construction allows tariffs to respond to observed development decisions. With simultaneous entry and binding tariff posting, an entrant instead faces its rivals' existing offers. The competitive pricing equilibrium cannot then be used to evaluate its entry deviation. \Cref{supp:simultaneous-entry} gives the corresponding undercutting calculation. Throughout the sequential construction, providers choose tariffs and anticipate the signaling equilibrium selected under the rule in \Cref{sec:model}. They choose neither receiver beliefs nor commit senders to an outcome.


\section{Proofs of design competition}
\label{app:ed-proofs}

I first specify signaling after arbitrary tariff menus, then verify prices, design deviations, and the welfare comparison. Improvement expenditure includes a tool that makes no sales.

\subsection{Signaling equilibria with heterogeneous tools}

\begin{lemma}
\label{lem:ed-menu-continuation}
Under \cref{eq:ed-feasible,eq:ed-separating-restrictions}, every feasible design pair and pair of nonnegative tariffs has a finite-support D1 signaling equilibrium satisfying the selection rule in \Cref{sec:model}. It can be chosen as pooling, partial pooling, or least-cost separation.
\end{lemma}

\begin{proof}[Proof of \Cref{lem:ed-menu-continuation}]
For a fixed set of subscriptions, all artifact input can be assigned to the subscribed plan with the smallest type-specific unit cost. Dropping other positive-fee subscriptions weakly reduces expenditure. Mixing manual and assisted production also cannot improve on the cheaper unit method when coverage is unrestricted. This gives \cref{eq:ed-menu-cost}, including at production ties.

The envelopes are continuous, strictly increasing, and piecewise affine, with $C_i(0)=0$. Every high-type slope is at least $\epsilon_h^0-\bar u>0$, so the costs are coercive. For $q>0$, we have $C_h(q)<C_\ell(q)$: the high type can use the low type's minimizing method, pay the same tariff, and incur less effort. An active low-type slope cannot exceed $\kappa_\ell$, because its nonnegative fee would otherwise make the plan more costly than manual production. As a result, for $q_2>q_1$, we get
\begin{equation*}
C_h(q_2)-C_h(q_1)
\geq \left(\epsilon_h^0-\bar u\right)(q_2-q_1)
>\rhoHard\bigl[C_\ell(q_2)-C_\ell(q_1)\bigr].
\end{equation*}
Replacing artifact deterrence with hard evidence lowers high-type expenditure until useful capacity is exhausted. Therefore, the least-cost separating signal is unique, with $\hardsig_R=\min\{\hardQ,\DeltaT\}$ and $C_\ell(q_R)=(\DeltaT-\hardQ)_+$.

For these costs, define $V$ by \cref{eq:menu-advantage-frontier} and put $b_i=U_i-\thL$. The threshold comparison in the proof of \Cref{lem:menu-signaling-continuation} gives the global D1 condition $V(\DeltaT-b_h)\leq b_h-b_\ell$.

The existence and selection construction in that proof uses the cost properties verified above, together with $0<\rhoHard<1$ and finite $\hardQ$. Its fixed point in \cref{eq:menu-payoff-fixed-point} and three maximizing-signal cases apply to these menus. The same support reduction and payoff comparisons yield pooling, partial pooling, or least-cost separation satisfying the selection rule.
\end{proof}

\subsection{Pricing after every design history}

\begin{proof}[Proof of \Cref{prop:ed-pricing}]
Let $r=\DeltaT-\hardQ>0$. At the proposed prices, both assisted high-type unit costs equal $a=m_h^0-u_2$. The low-type envelope has slope $a+\delta_*$. Manual production is dominated for both types. Their artifact cost advantage is $\delta_*q$, so a small quality increase defeats any shared signal under D1. The hard-evidence ranking in \Cref{lem:ed-menu-continuation} gives \cref{eq:ed-signals}. The low type uses the origin $(0,0)$. A more costly separating signal fails D1 by deviation to the least-cost signal. The least-cost signal itself passes D1: if a cheaper high-type signal had greater advantage, then increasing its quality until low-type cost reached $\DeltaT$ would preserve the greater advantage and contradict minimality. Hence, the finite-support observable outcome is unique.

\textit{Unilateral tariff deviations.} Hold a rival at an offer with zero membership fee, high-type slope $a\leq m_h^0$, and gap $\delta_r$. The reference is productive for both types: $a<\kappa_h$ and $a+\delta_r<\kappa_\ell$. For the deviator, write $a_j=c+\epsilon_{hj}$ and $\delta_j=\epsilon_{\ell j}-\epsilon_{hj}$. After any $F\geq0$, $p\geq0$, we have
\begin{align*}
C_h(q)&=\min\{aq,F+(p+\epsilon_{hj})q\},\\
C_\ell(q)&=\min\{(a+\delta_r)q,F+(p+\epsilon_{\ell j})q\}.
\end{align*}
The difference $C_\ell-C_h$ increases on every segment. If both types use one plan, its slope is that plan's positive gap. If only the high type uses the deviator, attracting it requires $p+\epsilon_{hj}\leq a$, so the slope is $a+\delta_r-p-\epsilon_{hj}>0$. If only the low type uses the deviator, the slope is $p+\epsilon_{\ell j}-a>0$ by \cref{eq:ed-separating-restrictions}. Continuity covers switching points. Thus, the signaling equilibrium separates after every unilateral tariff deviation and satisfies $C_\ell(q)=r$ and $\hardsig=\hardQ$.

If the deviator sells to high types at that quality, its margin per subscriber is $M=F+(p-c)q$. Private cost minimization requires $M\leq(a-a_j)q$. When $a\leq a_j$, positive profit is impossible. Suppose $a>a_j$. If the reference minimizes low-type cost, then $q=r/(a+\delta_r)$ and
\begin{equation*}
M\leq\frac{r(a-a_j)}{a+\delta_r}
\leq\frac{r(a-a_j)}{a+\min\{\delta_r,\delta_j\}}.
\end{equation*}
If the deviator minimizes low-type cost, $M=r-(a_j+\delta_j)q$. Its high-type bill is at most $aq$, so the low-type envelope is at most $[a+\min\{\delta_r,\delta_j\}]q$. Therefore,
\begin{equation*}
q\geq\frac{r}{a+\min\{\delta_r,\delta_j\}},
\end{equation*}
and
\begin{equation*}
M\leq\frac{r[a-a_j+\min\{\delta_r,\delta_j\}-\delta_j]}{a+\min\{\delta_r,\delta_j\}}
\leq\frac{r(a-a_j)}{a+\min\{\delta_r,\delta_j\}}.
\end{equation*}
The inequalities include ties. Receiving only a share of tied purchases cannot yield a larger positive profit. With no sales, operating profit is zero. Thus every tariff deviation obeys
\begin{equation*}
\profit_j^{\mathrm{op}}\leq
\frac{\prior r(a-a_j)_+}{a+\min\{\delta_r,\delta_j\}}.
\end{equation*}

Against provider 2's proposed tariff, provider 1 attains this bound with $F_1=0$, $p_1=c+u_1-u_2$. Its lower high-type editing requirement wins the purchase tie when $u_1>u_2$. Against provider 1's tariff, provider 2 faces $a=a_2$ and cannot earn positive profit. When $u_1=u_2$, the same bound gives zero for both firms. Development expenditure is sunk and does not affect these pricing best responses. \Cref{lem:ed-menu-continuation} completes strategies at tariff menus involving deviations by both providers.

\textit{Resources conditional on the menu.} Let $\beta=a+\delta_*$. The cheapest high-type real artifact cost has slope $m_h^0-u_1$, attained by provider 1. Since $\beta<\kappa_\ell$ and $m_h^0-u_1>\epsilon_h^0-\bar u$, \cref{eq:ed-separating-restrictions} gives
\begin{equation*}
\rhoHard<\frac{m_h^0-u_1}{\beta}.
\end{equation*}
Any separating allocation must impose low-type imitation cost at least $\DeltaT$, so $\beta q_h+\hardsig_h\geq\DeltaT$. Its high-type real cost is at least $(m_h^0-u_1)q_h+\rhoHard\hardsig_h$, and low-type real expenditure is nonnegative. Minimizing that lower bound uses all useful hard capacity and $q_h=r/\beta$. The proposed allocation attains it. Both incurred improvements are fixed in this comparison.
\end{proof}

For $\hardQ\geq\DeltaT$, the uniform marginal ranking makes $(q_R,\hardsig_R)=(0,\DeltaT)$ the unique least-cost separating signal. A shared signal must use $\hardsig=\hardQ$, imposing low-type cost at least $\DeltaT$, which precludes pooling and partial pooling. Hard separation passes D1 and purchases no positive AI input. Any nonzero improvement therefore loses its incremental cost; baseline designs and $F_j=0$, $p_j=c$ form a pricing equilibrium and induce hard-signal separation at these capacities.

\subsection{Global design deviations}

The preceding prices give an operating return after every feasible design history. Against a rival $(s,w)$, it is
\begin{equation}
\label{eq:ed-design-return}
R(u,v\mid s,w)=
\begin{cases}
\displaystyle\frac{\prior r(u-s)}{m_h^0-s+\min\{\delta_0+u-v,\delta_0+s-w\}},&u>s,\\[6pt]
0,&u\leq s.
\end{cases}
\end{equation}
Net profit is $R(u,v\mid s,w)-K(u,v)$. I specify the pricing equilibria in \Cref{prop:ed-pricing} after every design pair. Then, senders play a signaling equilibrium under the selection rule in \Cref{sec:model}. Providers anticipate these choices without controlling sender behavior or receiver beliefs.

\begin{proposition}
\label{prop:ed-pure-design}
Suppose \cref{eq:ed-feasible,eq:ed-separating-restrictions} hold, $\hardQ<\DeltaT$, and
\begin{equation}
\label{eq:ed-pure-conditions}
\begin{aligned}
u^*=\frac{\prior r}{k_hm_\ell^0}<\bar u,\\ k_\ell\geq\frac{\prior r}{(m_\ell^0-\bar v)^2}.
\end{aligned}
\end{equation}
With the pricing equilibria in \Cref{prop:ed-pricing}, the sequential game has pure provider equilibria with design pairs $((u^*,0),(0,0))$ and its permutation. The improving provider earns net profit $k_h(u^*)^2/2$, and its rival earns zero. Under these pricing equilibria, no pure symmetric design equilibrium exists.
\end{proposition}

\begin{proof}[Proof of \Cref{prop:ed-pure-design}]
Fix a rival $(s,0)$. If $u\leq s$, operating return is zero for every $v$, so a positive-cost choice cannot outperform the baseline. If $u>s$ and $v\leq u-s$, the denominator in \cref{eq:ed-design-return} is $m_\ell^0$, and increasing $v$ only adds cost. For $v>u-s$, the derivative of net profit in $v$ is
\begin{equation*}
\frac{\prior r(u-s)}{(m_\ell^0+u-s-v)^2}-k_\ell v
\leq\left[\frac{\prior r}{(m_\ell^0-\bar v)^2}-k_\ell\right]v\leq0.
\end{equation*}
The first inequality uses $u-s<v$ and $m_\ell^0+u-s-v\geq m_\ell^0-\bar v>0$. The payoff is continuous at $v=u-s$. Thus, $v=0$ is a best response to every $(s,0)$ over the full feasible rectangle.

Against $(0,0)$, net profit is $\prior ru/m_\ell^0-k_hu^2/2$ and has the unique interior maximizer $u^*$, earning $k_h(u^*)^2/2$. Against $(u^*,0)$, a choice $u\leq u^*$ earns no operating return. For $u>u^*$, after setting $v=0$ its net profit equals
\begin{equation*}
\frac{\prior r(u-u^*)}{m_\ell^0}-\frac{k_hu^2}{2}
=-\frac{k_h(u-u^*)^2}{2}-\frac{k_h(u^*)^2}{2}<0.
\end{equation*}
The baseline earns zero and is a best response. Both pure design profiles withstand every unilateral feasible design deviation. Pricing and signaling are optimal at every later history by \Cref{prop:ed-pricing,lem:ed-menu-continuation}.

At identical nonzero designs, both operating profits under these pricing equilibria are zero. Either firm can choose the baseline, pay no incremental cost, and obtain nonnegative operating return. At identical baseline designs, a sufficiently small $u>0$, $v=0$ earns positive net profit. No pure symmetric design profile is an equilibrium under the prescribed pricing equilibria.
\end{proof}

\subsection{Profitable improvements and welfare}\label{app:ed-welfare-example}

Take $(\kappa_\ell,\kappa_h)=(10,8)$, $\DeltaT=1$, $\prior=.8$, $\hardQ=.45$, $\rhoHard=.02$, $(\epsilon_h^0,\epsilon_\ell^0)=(.5,1)$, $c=.2$, $(\bar u,\bar v)=(.1,.25)$, and $(k_h,k_\ell)=(5,.005)$.

\begin{proof}[Proof of \Cref{prop:ed-harmful-improvement}]
These parameters give $\delta_0=.5$, $m_h^0=.7$, $m_\ell^0=1.2$, $\prior r=.44$, and $b=m_\ell^0-\bar v=.95$. All restrictions hold strictly, as $m_h^0=.7<.75=\epsilon_\ell^0-\bar v$, and $\rhoHard=.02<.04=(\epsilon_h^0-\bar u)/\kappa_\ell$.

Against the baseline, net profit at $(u,v)$ is
\begin{equation*}
f(u,v)=\frac{.44u}{1.2-\max\{v-u,0\}}-\frac{5u^2}{2}-\frac{.005v^2}{2}.
\end{equation*}
If $v\leq u$, its value is at most $.44^2/[2(5)(1.2)^2]=.01344444\ldots$. If $u\leq .005$, its value is at most $.44(.005)/.95=.00231579\ldots$. For $u\geq .005$ and $v>u$,
\begin{equation*}
\frac{\partial f}{\partial v}
=\frac{.44u}{(1.2+u-v)^2}-.005v
\geq\frac{.44(.005)}{1.2^2}-.005(.25)
=.00027777\ldots>0.
\end{equation*}
An optimum in this remaining region has $v=.25$. At $v=.25$, the objective is strictly concave in $u$,
\begin{equation*}
f(u,.25)=\frac{.44u}{.95+u}-\frac{5u^2}{2}-\frac{.005(.25)^2}{2}.
\end{equation*}
Its derivative vanishes at the unique solution of
\begin{equation}
\label{eq:ed-example-root}
5u^*(.95+u^*)^2=.44(.95).
\end{equation}
This solution lies between $.07896033$ and $.07896035$, and hence in $(.005,.1)$. Its net profit is $.01802162\ldots$, strictly exceeding both preceding upper bounds. Thus, $(u^*,.25)$ is a global best response to the baseline.

Against $(u^*,.25)$, any $u\leq u^*$ earns zero operating return. If $u>u^*$, even $v=.25$ leaves the deviator's gap larger than the rival's. The denominator in \cref{eq:ed-design-return} is $.95$, and its net profit is at most
\begin{equation*}
\frac{.44(u-u^*)}{.95}-\frac{5u^2}{2}
\leq\frac{.44^2}{2(5)(.95)^2}-\frac{.44u^*}{.95}
=-.01511958\ldots<0.
\end{equation*}
Choosing the baseline gives zero. This verifies both global design best responses. Combining them with the pricing equilibria at every design history and D1 signaling at every tariff history gives a sequential equilibrium with pure provider strategies. The permutation gives a second such equilibrium.

\textit{Welfare.} High types use the improved provider. The private low-type slope is $b+u^*$, so
\[
q_R=\frac{.55}{.95+u^*}=.53452012\ldots,
\]
\[
C^{\mathrm{real}}=.8\left[(.7-u^*)q_R+.02(.45)\right]=.27276655\ldots.
\]
Total improvement expenditure is $K_{\mathrm{tot}}=5(u^*)^2/2+.005(.25)^2/2=.01574309\ldots$. Operating profit is $.44u^*/(.95+u^*)=.03376471\ldots$, and net profit is positive as above.

With both designs fixed at the baseline, cost pricing gives $q_{\mathrm{base}}=.55/1.2=.45833333\ldots$, zero incremental improvement expenditure, and
\begin{equation*}
C_{\mathrm{base}}^{\mathrm{real}}=.8\left[.7q_{\mathrm{base}}+.02(.45)\right]=.26386667\ldots.
\end{equation*}
Both outcomes have zero assessment error. Fees and usage payments cancel, and establishment and capacity costs are common. Therefore,
\begin{equation*}
\welfare_{\mathrm{design}}-\welfare_{\mathrm{base}}
=C_{\mathrm{base}}^{\mathrm{real}}-C^{\mathrm{real}}-K_{\mathrm{tot}}
=-.02464297\ldots<0
\end{equation*}
for every $\receiverweight>0$. The root interval above bounds the rounded quantities away from each asserted inequality. Exact rational parameters and \cref{eq:ed-example-root} define the example. Manual separation instead costs $.8[8(.55)/10+.02(.45)]=.3592$, exceeding $.28850964\ldots=C^{\mathrm{real}}+K_{\mathrm{tot}}$.
\end{proof}

The global deviation bounds and welfare inequality are strict. The example persists under sufficiently small perturbations of the primitives. This robustness concerns the constructed equilibria and comparison with the fixed-design baseline.

\section{Proofs of welfare and capacity results}
\label{app:welfare-proofs}

This appendix proves the welfare results in \Cref{sec:welfare} and the verification results in \Cref{sec:capacity}. The first three subsections retain the fixed-technology welfare comparisons. The verification analysis then allows the installed provider to choose improvements and prices, with capacity chosen either before or simultaneously with its offer. All prospective comparisons begin from the same baseline and include newly chosen improvement expenditure. \Cref{app:ed-proofs} separately treats competing established providers.

Retain $\delta=\epsilon_\ell-\epsilon_h$, $g=\kappa_\ell-\kappa_h$, and $s=g-\delta>0$ from \Cref{app:market-proofs}. Write $b\coloneqq\kappa_h+\delta=\kappa_\ell-s$ for the imitation cost per assisted unit under an optimal separating tariff. Also write $r=(\DeltaT-\hardQ)_+$ for the artifact's remaining deterrence requirement and $T=\min\{\hardQ,\DeltaT\}$ for hard input in Riley separation. The resource saving per assisted unit is $w_i=\kappa_i-\epsilon_i-c$, as in \cref{eq:production-savings}.

\subsection{Information and resource accounting}

\begin{proof}[Proof of \Cref{prop:welfare-identity}]
Bayes consistency and the posterior-mean best response imply
\[
\E[\action]=\E\bigl[\E[\theta\mid\sigvec]\bigr]=\thbar.
\]
At every realized production allocation, sender costs consist of effort and payments to providers. Provider profits equal those payments less compute expenditure and incurred development and improvement costs. Adding sender utilities and provider profits cancels membership and usage payments, including when usage is priced below compute cost. Taking expectations gives $\thbar-C^{\mathrm{real}}-K_{\mathrm{tot}}$. Adding receiver utility $-\receiverweight\mathrm{MSE}/2$ and subtracting capacity cost proves \cref{eq:welfare-master}. The argument applies to mixed signals, multiple subscriptions, and technologies developed but unused.

Conditional on posterior $\belief$, binary types have variance $\belief(1-\belief)\DeltaT^2$. In the canonical pooling and partial-pooling outcomes, all high types belong to the shared group, whose population mass is $\prior/\belief$. Its contribution to assessment error is
\[
\frac{\prior}{\belief}\belief(1-\belief)\DeltaT^2
=\prior(1-\belief)\DeltaT^2.
\]
Any low types at the origin are identified and contribute zero error. This establishes \cref{eq:shared-assessment-error}. Separation has zero error.

The production choices in \cref{eq:monopoly-separating-output,eq:shared-payments} give the resource costs in \cref{eq:separating-resource-cost,eq:shared-resource-cost}. Without paid incremental AI, the high type produces quality $r/\kappa_\ell$ manually and supplies hard input $T$, yielding \cref{eq:baseline-resource-cost}.
\end{proof}

\subsection{Separating products and private Riley costs}

For an optimal separating offer with positive operating profit, $z_*=\min\{L,r/b\}$ and the manual remainder is $(r-bz_*)/\kappa_\ell$. Hence,
\[
q_R=z_*+\frac{r-bz_*}{\kappa_\ell}
=\frac{r}{\kappa_\ell}+\frac{s}{\kappa_\ell}z_*.
\]
The attaining tariff makes the high type indifferent about adoption, so its private artifact cost is $\kappa_hq_R$. The low type remains at the origin. Relative to manual separation, aggregate sender payoff falls by $\prior\kappa_hsz_*/\kappa_\ell$, proving \cref{eq:separating-sender-surplus-change}.

Capacity is the same in both economies, so hard-signal effort cancels from the resource comparison. Subtracting assisted artifact cost from the manual baseline gives
\begin{align*}
\frac{\kappa_hr}{\kappa_\ell}
-\left[(\epsilon_h+c)z_*+\frac{\kappa_h(r-bz_*)}{\kappa_\ell}\right]
&=\left(w_h-\frac{\kappa_hs}{\kappa_\ell}\right)z_*.
\end{align*}
Both outcomes reveal type perfectly. Multiplying this difference by $\prior$ and subtracting fresh development cost proves \cref{eq:social-value-separation-product}. Applying the welfare identity to a shared product and the no-AI economy similarly gives \cref{eq:social-value-pooling-product}.

The proof below uses the parameters
\begin{equation}
\label{eq:welfare-example-primitives}
(\kappa_\ell,\kappa_h,\epsilon_\ell,\epsilon_h,L,\prior,\DeltaT,\rhoHard,\hardQ)
=(10,8,0.01,0.002,0.5,0.8,1,0.02,0.45).
\end{equation}

\begin{proof}[Proof of \Cref{prop:riley-resource-inefficiency}]
First consider a local change under a fixed tariff. Suppose $0<\hardQ<\DeltaT$ and that private artifact costs and actual artifact resource cost are affine for qualities immediately above $q_R$. Assume resource cost is continuous from the right at the chosen Riley allocation. Let $\nu_h$ denote the high type's marginal real artifact cost there. All derivatives in this local comparison are taken from the right. If
\begin{equation}
\label{eq:riley-local-resource-condition}
\frac{\nu_h}{C_\ell'(q_R)}<\rhoHard
<\frac{C_h'(q_R)}{C_\ell'(q_R)},
\end{equation}
then a small increase in artifact quality can save resources while preserving deterrence.

For sufficiently small $t>0$, set
\[
q'=q_R+t \; \text{ and } \;
\hardsig'=\hardQ-C_\ell'(q_R)t.
\]
The hard signal remains nonnegative, and low-type imitation cost remains $\DeltaT$. High-type private cost rises by $[C_h'(q_R)-\rhoHard C_\ell'(q_R)]t>0$, while its real cost changes by $[\nu_h-\rhoHard C_\ell'(q_R)]t<0$. For small $t$, the perturbed private cost remains below $\DeltaT$.

Place the low type at the origin, assign posterior one at the new high signal, and assign posterior zero at every unused signal. Both on-path incentive constraints and all off-path PBE constraints hold. The resulting allocation is a pure separating PBE using fewer resources. It fails D1 because its high-type private cost exceeds the minimum separating cost, as established in \Cref{app:equilibrium-proofs}.

To show that this can occur under a globally optimal monopoly tariff, use \cref{eq:welfare-example-primitives} with $c=0.01$. The tariff
\[
p=7.994 \;\text{ and }\; F=0.004\frac{0.55}{8.008}
\]
attains $R=0.4389010989\ldots$, whereas $H\leq0.06986$. Thus, its selected outcome separates. At $q_R=0.55/8.008$, the high type subscribes at indifference and uses maximal input, making actual resource cost continuous from the right. The right-hand slopes are $C_\ell'=8.004$, $C_h'=7.996$, and $\nu_h=0.012$, which satisfy \cref{eq:riley-local-resource-condition}.

Taking $t=0.01$ gives $q'=0.0786813187\ldots$ and $\hardsig'=0.36996$. Low-type imitation cost remains one, and high-type private cost rises by $0.0783592$ but remains below one. The change in expected resource cost is
\[
0.8[0.012-0.02(8.004)](0.01)=-0.00118464,
\]
reducing it from $0.0078593407\ldots$ to $0.0066747007\ldots$. This proves the claim under the same globally optimal tariff while leaving its selected D1 signaling outcome unchanged.
\end{proof}


\subsection{Competition and signaling resources}

\begin{proof}[Proof of \Cref{prop:competition-resource-cost}]
Let $a\coloneqq\epsilon_h+c$ and $m\coloneqq\epsilon_\ell+c=a+\delta$ denote the per-unit resource costs of assisted production for the high and low types, respectively. Since $w_h>0$, $a<\kappa_h$ and $m<b<\kappa_\ell$, with $b-m=w_h>0$.

Under competitive pricing, separating quality is $r/m$ when $r\leq mL$ and $L+(r-mL)/\kappa_\ell$ otherwise. Under optimal monopoly pricing, it is $r/b$ when $r\leq bL$ and $L+(r-bL)/\kappa_\ell$ otherwise. The thresholds $mL$ and $bL$ thus mark when AI coverage is exhausted under competition and monopoly, respectively. Both outcomes use the same hard capacity, so hard-signal effort cancels from the comparison.

Subtracting monopoly resource cost from competitive resource cost gives
\begin{equation*}
\frac{C_C^{\mathrm{real}}-C_R^{\mathrm{real}}}{\prior}
=\begin{cases}
\displaystyle ar\left(\frac1m-\frac1b\right),
&0<r\leq mL,\\[5pt]
\displaystyle aL+\frac{\kappa_h(r-mL)}{\kappa_\ell}-\frac{ar}{b},
&mL\leq r\leq bL,\\[5pt]
\displaystyle\frac{\kappa_hw_hL}{\kappa_\ell},
&r\geq bL.
\end{cases}
\end{equation*}
The first and third expressions are strictly positive. The middle expression is affine in $r$ and agrees with those positive values at its endpoints, so it is also strictly positive. This proves \cref{eq:competition-resource-order}, including cases in which coverage binds.
\end{proof}


\subsection{Optimal improvements and tariffs}
\label{app:verification-reduction}

The capacity analysis in \Cref{sec:capacity} starts from the same installed baseline at every prospective choice. Write $\delta=\delta_0+u-v$ and $g=\kappa_\ell-\kappa_h$, and retain the improvement rectangle and costs in \Cref{sec:capacity}. All maximizations below include the option $(u,v)=(0,0)$. Common sunk development expenditure is suppressed.

\textit{Coverage and revealing deviations.} A shared quality satisfies $0<x<(\DeltaT-\hardQ)/\kappa_\ell$. Its recovery quality is $q_E=gx/\delta$ when recovery occurs within coverage. The sufficient coverage bound in \Cref{sec:capacity} gives
\[
q_E<\frac{g\DeltaT}{\kappa_\ell\underline\delta}<L.
\]
By \Cref{lem:maximum-recovery-expenditure}, the greatest recovery expenditure available at this quality is $\beta x$, where $\beta=\kappa_h(g-\delta)/\delta$. The uniform gap bound is equivalent to $\beta\geq\kappa_\ell/\prior$.

At the unconstrained pooling optimum, $x=(\prior\DeltaT-\hardQ)/\kappa_\ell$ with $\hardQ\leq\prior^2\DeltaT$. Its smallest recovery expenditure occurs at $\hardQ=\prior^2\DeltaT$, where
\[
\beta x\geq\frac{\beta\prior(1-\prior)\DeltaT}{\kappa_\ell}\geq(1-\prior)\DeltaT.
\]
At the unconstrained partial pool, set $\belief=\sqrt{\hardQ/\DeltaT}>\prior$. Its quality is $x=\belief(1-\belief)\DeltaT/\kappa_\ell$, and the revealing-deviation constraint becomes
\[
\beta x\geq(1-\belief)\DeltaT
\quad\Longleftrightarrow\quad
\beta\belief\geq\kappa_\ell.
\]
This follows from $\belief>\prior$. Both unconstrained shared optima are feasible at every design. The same coverage bound makes optimal separating input smaller than $L$, since $g/(\kappa_\ell\delta)>1/(\kappa_h+\delta)$ for $0<\delta<g$.

\textit{Profit maximization.} The shared qualities in \cref{eq:capacity-shared-output} uniquely maximize aggregate subscribing low-type input. Substitution into \cref{eq:low-type-demand} gives
\begin{equation*}
A(\hardQ)=
\begin{cases}
\displaystyle\frac{(1-\prior)(\prior\DeltaT-\hardQ)}{\kappa_\ell},
&0\leq\hardQ\leq\prior^2\DeltaT,\\[5pt]
\displaystyle\frac{\prior}{\kappa_\ell}(\sqrt{\DeltaT}-\sqrt{\hardQ})^2,
&\prior^2\DeltaT<\hardQ<\DeltaT,\\[5pt]
0,&\hardQ\geq\DeltaT.
\end{cases}
\end{equation*}
Baseline low-type resource savings are positive. Shared operating profit at a design is, thus, $(w_\ell^0+v)A(\hardQ)$. High-type improvement adds cost without increasing that profit. Maximizing the strictly concave low-type objective gives
\[
u_H=0,
\]
\[
v_H=\min\{\bar v,A(\hardQ)/k_\ell\},
\]
\[
\widehat H=\left(w_\ell^0+v_H\right)A(\hardQ)-k_\ell v_H^2/2.
\]
When $v_H<\bar v$, the value is $w_\ell^0A+A^2/(2k_\ell)$. Every price $0\leq p\leq\kappa_h-\epsilon_\ell$ satisfying the required recovery inequality implements the optimum with $F=(\kappa_\ell-p-\epsilon_\ell)x_H$. Participating low types adopt at production indifference, as in \Cref{app:implementation}.

The separating value follows from \Cref{prop:monopoly-separation}, since coverage is slack. Maximizing the larger fixed-design value over the rectangle is equivalent to comparing the separately maximized net values in \Cref{sec:capacity}. These maxima exist by continuity and compactness. This argument establishes global values without assuming concavity of the separating problem.

For later constructions, put $r=\DeltaT-\hardQ$, $w=\kappa_h-\epsilon_h^0+u-c$, and $b=\kappa_h+\delta_0+u-v$. At a positive interior separating optimum with $\hardQ<\DeltaT$, differentiation gives
\begin{equation}
\begin{aligned}
\label{eq:verification-separating-foc}
k_hu&=\frac{\prior r(\epsilon_\ell^0+c-v)}{b^2},\\
k_\ell v &=\frac{\prior r w}{b^2}.
\end{aligned}
\end{equation}
Global optimality of solutions to these equations is checked separately in each construction. A positive-saving design is implemented by
\[
p=\kappa_h-\frac{\epsilon_\ell+\epsilon_h}{2},
\]
\[
F=(\kappa_h-p-\epsilon_h)\frac{\DeltaT-\hardQ}{b}.
\]
This tariff satisfies $p+\epsilon_\ell>\kappa_h$ and has the unique observable separating outcome. At $\hardQ\geq\DeltaT$, all offers separate, no positive input is purchased, and costly improvements are strictly dominated by retaining the baseline.

\subsection{Higher required capacity and information}
\label{app:verification-information-loss}

\begin{proof}[Proof of \Cref{prop:verification-information-loss}]
Take
\[
(\kappa_\ell,\kappa_h,\DeltaT,\prior)=(10,8,1,0.95),
\]
\[
(\epsilon_h^0,\epsilon_\ell^0,c,L)=(0.5,1,7.6,1),
\]
\[
(\bar u,\bar v,k_h,k_\ell)=(0.25,0.05,0.05,10),
\]
\[
\rhoHard=0.02,\quad\receiverweight=0.1,\quad\capcost(\hardQ)=0.2\hardQ.
\]
Every design satisfies \Cref{ass:comparative-advantage}. The gap lies in $[0.45,0.75]$, below the sufficient upper bound $15.2/17.6$. The required coverage is $2/4.5<1$. Baseline low-type savings equal $1.4$.

\textit{Minimum capacity zero.} At zero capacity, the separating design $(u,v)=(0.25,0)$ earns $0.014723214\ldots$, exceeding the maximal shared value $0.006651128125$. Separation is globally optimal. To identify its design, positive high-type savings require $u>0.1$. On that region,
\[
\frac{\partial R}{\partial u}=\frac{0.95(8.6-v)}{(8.5+u-v)^2}
\geq\frac{0.95(8.55)}{8.75^2}>0.05(0.25).
\]
Net profit increases with $u$ throughout that region. Designs with $u\leq0.1$ yield nonpositive net separating profit. Thus, $u=0.25$ is globally optimal. At this value, net profit is strictly concave in $v$, since its second derivative is at most $2(0.95)(0.15)/8.7^3-10<0$. Its derivative is positive at zero and negative at $\bar v$. The unique optimum solves
\[
10v=\frac{0.95(0.15)}{(8.75-v)^2},
\]
\[
v=0.00018615\ldots.
\]
Use the separating tariff in \Cref{app:verification-reduction}. At zero capacity it yields complete information. Every positive capacity costs more and cannot improve accuracy. Zero is a global receiver best response under either timing, independently of subsequent provider choices in the sequential game.

\textit{Minimum capacity $0.9$.} At $\hardQ=0.9$, the shared optimum has $(u,v)=(0,0.000025)$, $x_H=0.005$, and net profit $0.000350003125$. Its posterior is $0.95$, giving assessment error $0.0475$. An implementing interior tariff is
\[
p=3.5000125,
\]
\[
F=0.0275000625.
\]
Its recovery expenditure exceeds $0.05$.

Shared provision is globally optimal throughout $[0.9,1)$. Put $r=1-\hardQ\in(0,0.1]$. Positive separating savings require $u>0.1$, and $b=8.5+u-v>8.55>8.5$. Dropping low-type improvement cost and maximizing over an unrestricted $u$ bounds the separating value by
\[
\widehat R\leq\max\left\{0,\frac{(0.95/8.5)^2}{0.1}r^2-\frac{0.095}{8.5}r\right\}.
\]
Baseline shared profit is at least $0.03325r^2$. On the partial-pooling interval this follows from $(1-\sqrt{\hardQ})^2=r^2/(1+\sqrt{\hardQ})^2\geq r^2/4$. On the pooling interval, $r\in[0.0975,0.1]$, it follows from $0.007(r-0.05)\geq0.03325r^2$. The difference is concave in $r$ and positive at both endpoints. Finally,
\[
\left[\frac{(0.95/8.5)^2}{0.1}-0.03325\right]0.1<\frac{0.095}{8.5}.
\]
The shared lower bound exceeds the separating upper bound for every $0<r\leq0.1$.

Sequentially, assessment error remains constant until $\hardQ=0.9025$. Above that threshold the marginal accuracy gain is $0.02375/\sqrt{\hardQ}<0.2$. At and above one, accuracy is complete and further capacity adds cost. The receiver's global optimum is $0.9$.

Simultaneously, hold the displayed design and tariff fixed. Recovery remains sufficient at every higher capacity. The outcome is partially pooled above $0.9$ until $\hardQ=0.95$, and then separates. The marginal accuracy gain is $0.0475<0.2$ throughout the partial pool and zero after separation. Thus, $0.9$ is again the receiver's global best response. The provider's optimal offer and capacity form a pure equilibrium under both timings. Raising the external minimum from zero to $0.9$ therefore raises equilibrium assessment error from zero to $0.0475$.
\end{proof}

The comparison concerns equilibria in two games with different capacity requirements. It does not assert uniqueness of every equilibrium at the zero minimum. In any sequential equilibrium, the receiver's option to retain its minimum implies
\[
\frac{\receiverweight}{2}\bigl[\mathrm{MSE}(\hardQ_0)-\mathrm{MSE}(\hardQ^*)\bigr]
\geq\capcost(\hardQ^*)-\capcost(\hardQ_0).
\]
A voluntary costly expansion must improve accuracy relative to that feasible option.

\subsection{Sequential and simultaneous capacity choices}
\label{app:verification-timing}

Suppose $\kappa_h-\epsilon_h^0+\bar u-c\leq0$, so shared provision is globally optimal at every $\hardQ<\DeltaT$. For a fixed optimal shared design and tariff, let $x$ be its adoption threshold and $E$ its recovery expenditure. The selected outcome at a deviating capacity is pooling if $\kappa_\ell x+\hardQ\leq\prior\DeltaT$ and $E\geq(1-\prior)\DeltaT$. It is partial pooling if
\[
\prior\DeltaT<\kappa_\ell x+\hardQ<\DeltaT,
\]
\[
E\geq\DeltaT-\kappa_\ell x-\hardQ.
\]
Otherwise, it separates. \Cref{thm:pure-d1,lem:signaling-continuation-selection} establish these conditions, including recovery equalities under our selection rule.

Within partial pooling, $\belief=(\kappa_\ell x+\hardQ)/\DeltaT$ and $\mathrm{MSE}=\prior(1-\belief)\DeltaT^2$. The receiver's fixed-offer marginal accuracy gain is
\[
m\coloneqq\frac{\receiverweight\prior\DeltaT}{2}.
\]
When the provider reoptimizes its offer as capacity changes, the marginal accuracy gain is $m\sqrt{\DeltaT/\hardQ}/2$, proving \cref{eq:verification-marginal-information}. The design affects profit and resources, but the restrictions in \Cref{app:verification-reduction} make the optimal shared posterior independent of design.

\textit{Global test for an interior simultaneous equilibrium.} Assume $\capcost$ is continuously differentiable and strictly convex. Consider a candidate $\overline Q\in(\max\{\hardQ_0,\prior^2\DeltaT\},\DeltaT)$ satisfying $\capcost'(\overline Q)=m$. Evaluate the optimal shared design and $x=x_H(\overline Q)$ there. Define
\[
\begin{aligned}
E_{\max}&=\frac{\kappa_h(g-\delta)}{\delta}x,\\
T_0 & =\min\{(1-\prior)\DeltaT,\DeltaT-\kappa_\ell x-\hardQ_0\},\\
\belief_0&=\max\{\prior,(\kappa_\ell x+\hardQ_0)/\DeltaT\},\\
\overline\belief&=\sqrt{\overline Q/\DeltaT}.
\end{aligned}
\]
Some provider-optimal implementing tariff supports a pure equilibrium at $\overline Q$ if and only if
\begin{equation}
\label{eq:verification-simultaneous-global-test}
\begin{aligned}
E_{\max}&\geq T_0,\\
\frac{\receiverweight\prior\DeltaT^2}{2}(\overline\belief-\belief_0)
&\geq\capcost(\overline Q)-\capcost(\hardQ_0).
\end{aligned}
\end{equation}

To prove the test, first note that recovery is hardest at the minimum capacity: its requirement is constant in pooling and decreases in partial pooling. If the first inequality fails, every optimal implementing tariff induces separation at some lower feasible capacity. Then, the receiver obtains complete information at a lower cost and strictly prefers that deviation to the proposed partial pool. If it holds, choose the recovery-maximizing price $p=\kappa_h-\epsilon_\ell$ and fee $F=(\kappa_\ell-\kappa_h)x$. Sharing persists at every feasible lower capacity and at every higher capacity until full information is reached. The receiver's payoff decreases in the pooling interval, is strictly concave in the partial-pooling interval with maximizer $\overline Q$, and decreases after full information. Its only global candidates are $\hardQ_0$ and $\overline Q$. Their payoff comparison is the second inequality. All tariffs preserving sharing at lower capacities produce the same comparison. This also proves necessity. Strict recovery slack permits an interior usage price.

\textit{Capacity ordering in the main text.} Suppose additionally $\prior>1/2$ and $\hardQ_0\geq\prior^2\DeltaT$. Throughout the sequential choice interval below $\DeltaT$, $\sqrt{\DeltaT/\hardQ}/2<1$. The receiver's sequential marginal return is strictly smaller than its fixed-offer return. If a pure interior simultaneous equilibrium exists, strict convexity gives $\capcost'(\overline Q)=m$, whereas the sequential payoff derivative at $\overline Q$ is negative. The sequential payoff derivative is strictly decreasing, because its accuracy return decreases and marginal capacity cost increases. At $\DeltaT$, information is reached continuously, and capacities above it add costs. The sequential global optimum consequently lies strictly below $\overline Q$, possibly at $\hardQ_0$.

\textit{The comparison in \Cref{tab:verification-timing}.} Take
\begin{equation}
\label{eq:verification-shared-example}
\begin{gathered}
(\kappa_\ell,\kappa_h,\DeltaT,\prior)=(10,8,1,0.8),\\
(\epsilon_h^0,\epsilon_\ell^0,c,L)=(0.002,0.01,8.1,100),\\
(\bar u,\bar v,k_h,k_\ell)=(0.001,0.004,10,2),\\
\rhoHard=0.02, \; \text{ and } \;\receiverweight=0.4.
\end{gathered}
\end{equation}
Let $\hardQ_0=0.72$ and $\capcost(\hardQ)=0.1\hardQ^2$. The gap lies in $[0.004,0.009]$, the sufficient coverage bound is $50<100$, and every design satisfies \Cref{ass:comparative-advantage}. High-type savings are at most $-0.101$, while baseline low-type savings equal $1.89$. Shared provision is optimal below one, with $u=0$ and $v=A(\hardQ)/2$ on the whole feasible capacity interval.

Sequentially, the marginal accuracy gain $0.08/\sqrt{\hardQ}$ is below marginal capacity cost $0.2\hardQ$ for every $\hardQ\geq0.72$. The receiver retains $0.72$. Simultaneously, the candidate is $\overline Q=0.8$. Its optimal design and quality are
\[
v=0.00044582472\ldots,
\]
\[
x=0.00944271910\ldots.
\]
An interior optimal tariff uses $p=0.1$ and $F=0.09339270170\ldots$. At the minimum, $\kappa_\ell x+0.72=0.81442719\ldots>0.8$, so the fixed offer remains partially pooled throughout all lower feasible deviations. Its recovery expenditure exceeds the largest required amount, $0.18557281\ldots$. The receiver's partial-pooling objective has derivative $0.16-0.2\hardQ$, uniquely maximizing it at $0.8$. After complete information is reached, capacity only adds cost. This verifies both inequalities in \cref{eq:verification-simultaneous-global-test}. Substitution into the quality, error, and improvement formulas gives \Cref{tab:verification-timing}.

\begin{table}[!htbp]
\centering
\begin{tabular}{lrr}
\toprule
& Sequential & Simultaneous\\
\midrule
Capacity & $0.7200$ & $0.8000$\\
Assessment error & $0.1212$ & $0.0845$\\
Shared artifact quality & $0.0129$ & $0.0094$\\
Low-type improvement & $0.000918$ & $0.000446$\\
\bottomrule
\end{tabular}
\caption{\textbf{Verification and provider adjustment.} Both columns report pure equilibria for the same primitives in \cref{eq:verification-shared-example}, with $\hardQ_0=0.72$ and $\capcost(\hardQ)=0.1\hardQ^2$. High-type improvement is zero. Assessment error is expected squared error.}
\label{tab:verification-timing}
\end{table}

\subsection{Receiver investment and welfare}
\label{app:verification-welfare}

The accounting identity in \Cref{prop:welfare-identity} applies to the newly chosen improvements. For any specified provider offer at each capacity, let $B(\hardQ)=C^{\mathrm{real}}(\hardQ)+K_{\mathrm{tot}}(\hardQ)$ denote total signaling and incremental improvement resources. Subtracting welfare at two prospective capacities proves \cref{eq:social-restoration-value}. The receiver omits the change in $B$ from its objective.

For an optimal separating design with positive savings and $\hardQ<\DeltaT$,
\[
B_R(\hardQ)=\prior\left[\frac{(\epsilon_h^0-u+c)(\DeltaT-\hardQ)}{\kappa_h+\delta_0+u-v}+\rhoHard\hardQ\right]+K(u,v).
\]
At an interior shared optimum, define $B_H^0$ as shared signaling resources under the baseline editing requirements, evaluated at the optimal shared quality and posterior. Improvement reduces those resources by $vA$ and costs $k_\ell v^2/2$, giving
\[
B_H=B_H^0-\frac{A^2}{2k_\ell}.
\]
At a bounded optimum use $B_H=B_H^0-v_HA+k_\ell v_H^2/2$. In particular, the baseline resource bills needed below are
\[
\begin{aligned}
B_P^0&=[(1-\prior)(\epsilon_\ell^0+c)+\prior\kappa_h]\frac{\prior\DeltaT-\hardQ}{\kappa_\ell}
+[(1-\prior)+\prior\rhoHard]\hardQ,\\
B_M^0&=\prior\DeltaT\left[\frac{\epsilon_\ell^0+c}{\kappa_\ell}(1-\belief)^2
+\left(1+\frac{\kappa_h}{\kappa_\ell}\right)\belief(1-\belief)+\rhoHard\belief^2\right],
\end{aligned}
\]
where $\belief=\sqrt{\hardQ/\DeltaT}$ in the second line.

\begin{proof}[Proof of \Cref{prop:receiver-restoration-distortion}]
\textit{Excessive investment.} Take
\begin{equation*}
\begin{gathered}
(\kappa_\ell,\kappa_h,\DeltaT,\prior)=(10,8,1,0.8),\\
(\epsilon_h^0,\epsilon_\ell^0,c,L)=(0.99,1,6.65,40),\\
(\bar u,\bar v,k_h,k_\ell)=(0.02,0.004,10,10),\\
\rhoHard=0.7,\quad\receiverweight=1,\quad\hardQ_0=0,\; \text{ and }\;\capcost(\hardQ)=0.2\hardQ.
\end{gathered}
\end{equation*}
The gap lies in $[0.006,0.03]$, sufficient coverage is $33.333\ldots<40$, and the gap bound is satisfied. Every feasible design satisfies \Cref{ass:comparative-advantage}. In particular, $\rhoHard<\min\{8/10,0.97/1\}$. Baseline low-type savings are $2.35$.

\textit{Global design choices.} For separation, $w=0.36+u>0$ and $b=8.01+u-v\in[8.006,8.03]$. The operating-profit Hessian has $R_{uu}<0$,
\[
|R_{uv}|\leq\frac{0.8(8.03)}{8.006^3}<0.01252,
\]
\[
R_{vv}\leq\frac{2(0.8)(0.38)}{8.006^3}<0.001185.
\]
Subtracting the cost Hessian $10I$ makes net profit strictly concave. Its derivative is positive at each zero improvement, and the first-order conditions imply $u<0.00955<\bar u$ and $v<0.000475<\bar v$. The unique global separating design solves \cref{eq:verification-separating-foc} for every $\hardQ<1$. The shared optimum is $u=0$, $v=A/10$, also interior.

\textit{Provider choices.} At zero, $\widehat H=0.0376128$. Set $M_u=0.8(7.65)/8.006^2$ and $M_v=0.8(0.38)/8.006^2$. Bounding the operating-profit gradient uniformly and then maximizing the quadratic cost bound gives
\[
\widehat R(0)\leq\frac{0.8(0.36)}{8.01}+\frac{M_u^2+M_v^2}{20}<0.036413<\widehat H(0).
\]
At $\hardQ=0.2$, baseline separating profit is $0.028764045\ldots>\widehat H(0.2)=0.0282072$. On $[0,0.64]$, the envelope derivative satisfies
\[
\frac{d(\widehat R-\widehat H)}{d\hardQ}\geq0.047-\frac{0.8(0.38)}{8.006}>0.
\]
There is exactly one profit crossing $\hardQ_*\in(0,0.2)$. On $[0.64,1)$, $\widehat H/(1-\hardQ)$ decreases: both $A/(1-\hardQ)$ and $A^2/(1-\hardQ)$ decrease when $A=0.08(1-\sqrt{\hardQ})^2$. At $0.64$, this ratio is already below baseline separating profit per unit of $1-\hardQ$. Separation remains optimal above the crossing. Specify that the provider chooses a separating offer at the tie.

Solving the strictly concave separating problem and the scalar profit crossing gives
\[
\hardQ_*=0.11770469544\ldots,
\]
\[
(u_*,v_*)=(0.00839867839\ldots,0.00040447394\ldots).
\]
These numerical values locate the analytically established crossing. Of course, the existence and global choices do not depend on numerical searches.

\textit{Global receiver and welfare choices.} Below $\hardQ_*$, the outcome pools and receiver investment gain is $-0.2\hardQ$. At and above it, assessment error is zero and gain is $0.08-0.2\hardQ$. Its unique optimum is $\hardQ_*$. Initial total resources are $B_H(0)=0.6343872$. During pooling, we have $B_H'(\hardQ)=-0.033+0.02v_H$. Adding marginal capacity cost $0.2$ makes total social expenditure increase. Every positive pooling investment lowers welfare.

For any separating design with positive savings and $0\leq\hardQ\leq1$, total resources and capacity expenditure satisfy
\[
B_R(\hardQ)+0.2\hardQ
\geq\frac{0.8(7.62)}{8.03}(1-\hardQ)+(0.8\cdot0.7+0.2)\hardQ
\geq0.759153175\ldots.
\]
The accuracy gain is at most $0.08$, so every separating target has welfare gain at most $0.08+0.6343872-0.759153175\ldots<0$. At and above one, improvements vanish, the high type supplies hard evidence one, and further capacity only adds cost. Welfare is globally maximized at zero. At the receiver's optimum, its gain is $0.05645906\ldots$ and welfare gain is $-0.04724337\ldots$.

\textit{Insufficient investment.} Use \cref{eq:verification-shared-example}, now with $\hardQ_0=0.45$ and $\capcost(\hardQ)=0.05\hardQ^2$. All feasible designs satisfy the extension's restrictions. Separation has zero maximal value, and shared profit is positive below one. For all feasible capacities, $u=0$ and $v=A/2<0.004$. Pooling continues up to $0.64$, with assessment error $0.16$, followed by partial pooling.

The receiver's best target in the partial-pooling interval solves $0.08/\sqrt{\hardQ}=0.1\hardQ$, giving $\hardQ^\dagger=0.8^{2/3}=0.861773876\ldots$. Its gain over retaining $0.45$ is
\[
0.2[0.16-0.8(1-\sqrt{\hardQ^\dagger})]-0.05[(\hardQ^\dagger)^2-0.45^2]
=-0.006476868\ldots.
\]
Higher pooling capacities bring no accuracy gain, and capacities above one add cost without improving information. Retaining $0.45$ is the receiver's unique global optimum.

Welfare increases throughout $[0.45,1]$. In pooling, its derivative is
\[
0.5862-0.02v_H-0.1\hardQ\geq0.5862-0.02(0.0035)-0.1(0.64)>0.
\]
In partial pooling, put $\belief=\sqrt{\hardQ}\in[0.8,1]$. Differentiating the baseline resource bill and adding accuracy and capacity costs gives $0.8[0.969+0.011/\belief]-0.1\hardQ$. The improvement adjustment adds $v_H A'(\hardQ)$. Here, $v_H\leq0.0016$ and $A'\geq-0.02$, so the complete welfare derivative is at least $0.8(0.98)-0.1-0.0016(0.02)>0$. Resources and error reach their full-information values continuously at one, and higher capacities only add cost. The unique welfare optimum is one.

At $0.45$, total resources are $0.37795775$. At one they are $0.016$. Thus, the welfare gain at full capacity is $0.2(0.16)+0.37795775-0.016-0.05(1-0.45^2)=0.35408275$. Even the expansion to $0.46$ raises welfare by $0.00540631$, with unchanged accuracy and a reduction in low-type improvement from $0.0035$ to $0.0034$. Both constructions use global design, tariff, and capacity choices.
\end{proof}

These constructions use finite-support signaling equilibria, one installed provider, and prospective choices from a common baseline. They retain the equilibrium selection and production rules in \Cref{sec:model} when comparing all offers and capacity deviations. The excessive-investment construction specifies a separating provider offer at its profit tie. This choice is separate from sender equilibrium selection. \Cref{app:verification-selection} discusses alternative provider choices and selection assumptions.


\clearpage
\etocdepthtag.toc{main}
\phantomsection
\hypersetup{bookmarksdepth=0}
\addcontentsline{toc}{section}{Supplement}
\hypersetup{bookmarksdepth=}
\pdfbookmark[0]{Supplement}{supplement-overview}
\etocdepthtag.toc{supplement}

\setcounter{section}{0}
\renewcommand{\thesection}{S\arabic{section}}
\renewcommand{\theHsection}{supplement.\arabic{section}}

\crefalias{section}{supplement}
\crefalias{subsection}{supplement}
\crefalias{subsubsection}{supplement}
\crefalias{appendix}{supplement}
\crefalias{subappendix}{supplement}
\crefalias{subsubappendix}{supplement}

\begin{center}
{\LARGE\bfseries Supplement\par}
\medskip
{\large The Market for AI-Assisted Signaling\par}
\end{center}

\bigskip
\begingroup
\singlespacing
\small
\hypersetup{linkcolor=black}
\renewcommand{\contentsname}{Contents of the supplement}
\etocsettagdepth{main}{none}
\etocsettagdepth{appendix}{none}
\etocsettagdepth{supplement}{subsection}
\tableofcontents
\endgroup
\clearpage

\section{Robustness and scope}
\label{supp:robustness-scope}

This section considers production choices at subscription indifference, alternative equilibrium selection rules, perturbations of assistance costs, unconditional cash rebates, differentiated technologies, and simultaneous entry and tariff posting.

\subsection{Production indifference and provider returns}
\label{app:production-indifference}

The subscription rule selects a production method among private-cost minimizers. Its role differs between separating and shared provision. The following comparison changes how subscription indifference is resolved while retaining the signaling equilibrium in \Cref{sec:model}.

\begin{lemma}
\label{lem:production-indifference}
Fix a tariff satisfying \cref{eq:shared-signal-regime} and a selected shared outcome at $(x,\hardQ)$. Every participating low type can minimize private cost either by producing manually without subscribing or by subscribing and using input $x$. These are its only minimizing pure production choices. The high type produces manually. Let $D_\ell(x,\hardQ)$ be the participating low-type mass in \cref{eq:low-type-demand}. If a fraction $\beta\in[0,1]$ of this mass uses AI, operating profit and expected signaling resource cost are
\begin{align}
\Pi_{P,M}^{\mathrm{op}}(\beta)
&=\beta D_\ell(x,\hardQ)w_\ell x,
\label{eq:shared-profit-production-tie}\\
C_{P,M}^{\mathrm{real}}(\beta)
&=D_\ell(x,\hardQ)(\kappa_\ell x+\hardQ)
+\prior(\kappa_hx+\rhoHard\hardQ)
-\beta D_\ell(x,\hardQ)w_\ell x.
\label{eq:shared-resources-production-tie}
\end{align}
Private signaling costs, sender payoffs, and assessment error are independent of $\beta$. Holding development and capacity costs fixed,
\begin{equation}
\label{eq:shared-welfare-production-tie}
\welfare(\beta)-\welfare(1)
=-(1-\beta)D_\ell(x,\hardQ)w_\ell x.
\end{equation}
If manual production is selected at every subscription tie, the best selected shared operating value is zero. When senders choose manual production at subscription indifference, the separating value $R$ in \cref{eq:profit-separation-general} remains the supremum of separating operating profits and can be approached with strictly preferred high-type adoption.
\end{lemma}

\begin{proof}[Proof of \Cref{lem:production-indifference}]
Since $p+\epsilon_\ell\leq\kappa_h<\kappa_\ell$ and $x<L$, a subscribing low type uniquely minimizes production cost by using input $x$. Its assisted cost is $F+(p+\epsilon_\ell)x=\kappa_\ell x$, equal to manual production without subscription. Positive $F$ excludes subscription without input. Any intermediate input uses more costly manual production after paying the fee and is strictly worse. The high type's per-unit saving from adoption is smaller by $(\kappa_\ell-\kappa_h)-(\epsilon_\ell-\epsilon_h)>0$, so its saving at $x$ cannot cover the fee. It uniquely produces manually.

Each adopting low type generates operating profit $w_\ell x$ and saves the same amount in resources relative to manual production. Multiplication by the adopting mass $\beta D_\ell$ gives \cref{eq:shared-profit-production-tie,eq:shared-resources-production-tie}. Both minimizing methods have the same private cost. Changing their probabilities leaves the cost envelopes, observable signaling incentives, D1 tests, and selected posteriors unchanged. In particular, assessment error remains $\prior(1-\belief)\DeltaT^2$. The welfare identity in \cref{eq:welfare-master} gives \cref{eq:shared-welfare-production-tie}. Manual production at subscription ties sets $\beta=0$ at every selected shared outcome, yielding zero revenue and compute expenditure.

The separating profit bound in \Cref{prop:monopoly-separation} uses the weak adoption constraint and remains valid when senders choose manual production at subscription indifference. If $R>0$, choose an attaining tariff in \cref{eq:separating-tariff} and reduce its fee from $F_*$ to $F_*-t$, with $t>0$ sufficiently small. Keep $p$ fixed. Write $q_*$ and $z_*$ for its original separating quantities. If $q_*<L$, the binding low-type imitation constraint gives $q_t=q_*+t/(p+\epsilon_\ell)<L$. The high type strictly prefers adoption, and its contribution to operating profit is
\begin{equation*}
F_*-t+(p-c)q_t
=w_hz_*-\frac{\epsilon_\ell+c}{p+\epsilon_\ell}t.
\end{equation*}
If $q_*\geq L$, that constraint instead gives $q_t=q_*+t/\kappa_\ell$ and input $L$, with contribution $w_hL-t$. High-type adoption is again strict. In both cases $p+\epsilon_\ell>\kappa_h$, so Riley separation remains the unique finite-support D1 observable outcome. Multiplying by $\prior$ and taking $t$ to zero approaches $R$. If $R=0$, an inactive separating offer attains zero.
\end{proof}

Parameter perturbations that preserve a selected canonical shared outcome preserve this production indifference. With $w_\ell>0$, subscription at indifference raises both profit and welfare relative to manual production at the same shared signal. It does not determine the welfare ranking against separation or against no paid AI, which also changes signaling and information. For the usage-subsidy primitives in \Cref{sec:market}, $R=0$. Manual production at subscription ties removes every positive operating return under our equilibrium selection criterion, and rules out a profitable fresh launch with positive development cost. When $R>0$, profitable provision can instead survive through separating offers. These statements concern the selected canonical shared outcomes. At the price boundary, other D1 shared qualities on the flat interval can lie above the adoption threshold. The equilibrium selection rule selects the least-quality outcome used here.

\subsection{Alternative equilibrium selection rules}
\label{supp:alternative-selection}

Selection matters when several D1 outcomes coexist, including some offers at the price boundary. For example, when pooling is feasible and $E\leq\DeltaT-v$, Riley separation can also survive. An alternative equilibrium selection rule may select it whenever it passes D1. Such a rule must still respect D1: under a strict price inequality, it cannot select separation when $x<q_R<q_E$, where Riley fails D1. The admissible alternatives are those characterized in \Cref{app:equilibrium-proofs}, including partial pooling and mixed equilibria.

The separating tariff family in \cref{eq:separating-tariff} has a unique D1 separating observable outcome and continues to attain separating operating profit $R$ under alternative equilibrium selection rules. Shared operating profits and product choices must be recomputed if the equilibrium selection rule changes. The competitive deviation bound in \cref{eq:competitive-deviation-bound} holds for every signaling outcome with cost-minimizing private production, and the cost-price menu itself has a unique D1 observable outcome. The welfare identity in \cref{eq:welfare-master} requires Bayes consistency and posterior-mean assessments and holds independently of the equilibrium selection rule. Particular monopoly and verification comparisons depend on the specified provider choices and signaling outcomes.

In the design extension, the constructed pricing menus and menus reached by unilateral tariff deviations uniquely separate under \cref{eq:ed-separating-restrictions}. At other tariff menus, \Cref{lem:ed-menu-continuation} supplies a signaling equilibrium consistent with the equilibrium selection rule. The design equilibria also specify which pricing equilibrium follows each design pair, but pricing uniqueness is not established.


\subsection{Imperfect substitution and perturbations}
\label{supp:perturbations}

Under the technology, $\epsilon_h<\epsilon_\ell$ and assistance narrows the effort difference between types. Hence, pooling occurs with distinct assisted effort costs. Its existence depends on the participation and revealing-deviation inequalities in \cref{eq:pool-existence}.

Fix an offer with $p+\epsilon_\ell<\kappa_h$, and suppose $x\in(0,L)$, $v<\prior\DeltaT$, and $E>\dpool$. The adoption threshold, cost envelopes, and recovery expenditure vary continuously with the primitives within the technology class. Small perturbations preserve these strict inequalities and hence a D1 pool, whose quality adjusts to the new adoption threshold. When the provider can choose the tariff, \cref{eq:pooling-minimum-coverage,eq:pooling-feasible-quality-interval} describe the corresponding pooling feasibility region.

For example, take the technology and population in \cref{eq:welfare-example-primitives}, with $c=8.1$, $p=0.1$, and $F=0.2967$. Then, $x=0.03$, $v=0.75<0.8$, and $E=0.3317>0.2$. Both assisted effort costs are positive and distinct. Moreover, $E>\DeltaT-v=0.25$, so \cref{eq:unique-pooling-region} applies. The relevant inequalities are strict, and sufficiently small perturbations within \Cref{ass:comparative-advantage} preserve unique pooling at this offer. This establishes an open pooling region. Whether this particular tariff maximizes profit after a perturbation is a separate question.

At participation, recovery, or price equalities, the selected signaling outcome or optimal offer can change discontinuously. Subscription at indifference and the equilibrium selection rule remain relevant to attainment of monopoly profits. Pooling can be robust on an open region while profit attainment still depends on how subscription and signaling indifference are resolved. Changes that reverse the technology restrictions, make adoption observable, or give artifact quality a direct productive benefit require separate analyses.


\subsection{Unconditional cash rebates}
\label{supp:cash-rebates}

Allow $F<0$ while retaining $p\geq0$. A sender can collect the membership rebate without using the tool, subscriptions remain nonexclusive, and there is no minimum spending requirement. The rebate is available even at the origin $(0,0)$.\footnote{Promotional compute credits have a different contract structure. Google Cloud's \href{https://docs.cloud.google.com/free/docs/free-cloud-features}{free trial}, for example, provides credits toward eligible service usage, including certain AI services (accessed October 13, 2026). Such credits offset expenditure on usage, but they do not represent the unrestricted cash rebate collectible without usage considered here.}

\begin{proposition}
\label{prop:unconditional-cash-rebates}
Under the equilibrium selection rule and private-production rules, a negative membership fee is strictly dominated in monopoly. In a finite market with nonexclusive subscriptions, no pure provider pricing equilibrium has a negative unconditional membership fee.
\end{proposition}

\begin{proof}[Proof of \Cref{prop:unconditional-cash-rebates}]
For a single offer with $F<0$, every sender subscribes at every signal. Define variable artifact cost as
\[
M_i(q)=\min\{p+\epsilon_i,\kappa_i\}\min\{q,L\}
+\kappa_i(q-L)_+.
\]
Minimized artifact cost is $F+M_i(q)$. Replacing $F$ by zero increases every signal's cost by $-F>0$ and reduces each sender's equilibrium payoff by the same amount. Signal comparisons and the D1 thresholds $U_i+c_i(\sigvec)$ are unchanged. Bayes consistency, sender-Pareto rankings, and the rules for resolving ties in shared mass and artifact quality are also preserved. Input choices remain unchanged under the production rules. The provider collects $-F$ more per sender while incurring the same compute and development costs, so profit rises by $-F$.

For a finite menu, every sender collects every negative membership fee. Their sum,
\[
b_F\coloneqq\sum_j\min\{F_j,0\},
\]
enters the cost of every signal as a common additive constant for both types. Removing this constant leaves the same private-production problem and signaling incentives. A provider with $F_j<0$ can raise its fee by $t\in(0,-F_j)$ while keeping it negative. Every sender still subscribes, and costs after removing the common constant, input opportunities, and cost-minimizing allocations remain unchanged. The allocation rule resolves the same production ties, while the equilibrium selection rule preserves observable signaling outcomes under the common payoff shift. Operating profit rises by $t$ on the unit sender mass. hence, a negative unconditional fee cannot be a pure pricing best response.
\end{proof}

The competitive deviation argument also permits unconditional rebates. Replacing a customer's production with a cost-priced rival's method bounds the deviator's total bill by compute expenditure. An unused negative-fee subscription loses money. Thus, the pricing profile in \Cref{prop:provider-competition} remains an equilibrium when these rebates are allowed.

Unconditional rebates and below-cost usage pricing have different effects. A rebate available without usage shifts every signal's cost by the same amount. A usage price $p<c$ changes marginal signaling costs and can be financed by positive membership revenue, as \Cref{supp:usage-subsidy} shows. Regular two-part tariffs separate access payments from usage charges while allowing both to affect adoption and deviations. Exclusivity, minimum spending requirements, usage-contingent rebates, allowances, and nonlinear usage prices alter these choices and lie outside the stated contract class. This comparison also retains the restriction $p\geq0$.


\subsection{Differentiated technologies}
\label{supp:differentiated-technologies}

The competitive bill comparison requires a rival that can reproduce the same production method at the same resource cost. Differentiation can invalidate that comparison. Consider $(\kappa_\ell,\kappa_h)=(10,8)$,
$\DeltaT=1$, $\prior=0.8$, $\hardQ=0.45$, and $\rhoHard=0.1$. Two providers have coverage $L=1$ and compute cost $c=1$. Their editing coefficients $(\epsilon_\ell,\epsilon_h)$ are $(1,0.5)$ and $(2,1)$, respectively. Both technologies satisfy \Cref{ass:comparative-advantage}. Initially, both providers set $F=0$ and $p=1$. Senders use the better tool at positive artifact quality, with private marginal costs $2$ for the low type and $1.5$ for the high type within coverage. The unique D1 outcome separates at $(q_R,\hardsig_R)=(0.275,0.45)$, and both providers earn zero operating profit.

The better provider can profitably raise its membership fee to $0.01$ while retaining $p=1$. For $0\leq q\leq1$, minimized artifact costs become
\[
C_\ell(q)=\min\{3q,0.01+2q\},
\]
\[
C_h(q)=\min\{2q,0.01+1.5q\}.
\]
Both tools cover the same units. Once the better tool's fee is paid, its marginal cost is lower for either type, so combining the tools cannot reduce these costs. Low types switch to the better tool at $q=0.01$, and high types at $q=0.02$. The artifact cost advantage $C_\ell(q)-C_h(q)$ increases up to $0.01$, is flat on $[0.01,0.02]$, and increases thereafter.

A shared D1 signal must exhaust hard capacity and lie on this flat interval. No shared outcome survives there. If the origin is used, low-type indifference implies that every shared posterior is at most $0.50$. If the origin is unused, Bayes consistency requires at least one shared posterior at or below $\prior=0.8$, since any high-only signal has posterior one. At such a shared signal, full recognition raises the high type's assessment by at least $0.2$. Moving just beyond $q=0.02$ gives it a greater cost advantage while raising its artifact cost by at most $0.02$ plus an arbitrarily small amount. Thus, D1 forces a profitable revealing deviation.

The least-cost separating signal is now $(0.27,0.45)$, as $C_\ell(0.27)=0.55$, so low-type imitation costs one. The nondecreasing cost advantage satisfies the separating D1 test. High types subscribe to the better tool, paying artifact cost $0.415$, compared with $0.54$ under the inferior tool. Low types remain at the origin and pay no positive membership fee. The better provider earns $0.8(0.01)=0.008>0$. Free membership and cost-priced usage fail to form a Nash equilibrium in this differentiated market.

A superior tool can retain paying subscribers even when an inferior alternative offers usage at resource cost. The example establishes the scope of the identical-technology result. \Cref{sec:design-competition} endogenizes editing requirements for two established providers with unrestricted overlapping coverage and bounded improvements. Its restrictions give pricing equilibria after every design pair and permit global checks of design deviations.\footnote{A general analysis combining entry, endogenous coverage, and differentiated features remains open.}

\subsection{Simultaneous entry and tariff posting}
\label{supp:simultaneous-entry}

The entry construction in \Cref{prop:provider-entry} allows providers to choose tariffs after observing development decisions. If entry and binding tariffs are chosen simultaneously, a unilateral entry deviation instead holds the incumbent's tariff fixed. The same sole-provider profile need not remain an equilibrium.

Consider the monopoly subsidy example in \Cref{supp:usage-subsidy}, with $F=0.34965$, $p=0$, $x=0.035$, $c=8.1$, and development cost $K=0.001$. An entrant can offer the same technology and usage price with a lower membership fee, $F'=F-t$, for small $t>0$. Its offer is cheaper for assisted production, so the effective menu has shared quality and associated costs
\[
x'=0.035-\frac{t}{9.99},
\]
\[
v'=0.8-\frac{10t}{9.99},
\]
\[
E'=0.33465-t.
\]
For sufficiently small $t$, the participation and recovery conditions remain satisfied, and the selected signaling outcome pools. Low types subscribe to the entrant under the assumption of subscription at indifference, while high types produce manually. The entrant's profit net of development cost is
\[
(1-\prior)w_\ell x'-K
=0.01223-\frac{0.378}{9.99}t>0.
\]
For example, $t=0.0001$ yields $0.0122262162\ldots$. Entry is profitable against the fixed monopoly tariff, so this sole-provider profile is not an equilibrium when entry and tariff posting are simultaneous.

In the sequential game, the same entry decision leads to a pricing subgame with two developed copies. Free membership and cost-priced usage form a Nash pricing equilibrium in that subgame, leaving the additional entrant with zero operating profit and a net loss of $K$. thus, decision timing changes the entrant's return. The example rules out the displayed sole-provider profile under simultaneous entry and tariff posting, and it leaves open the existence of other equilibria under that timing. Likewise, the sequential construction establishes an equilibrium without claiming uniqueness across pricing equilibria.

\clearpage
\section{Monopoly calculations and examples}
\label{supp:monopoly-calculations}

\subsection{Constrained shared qualities}
\label{supp:shared-quality-calculations}

Fix a technology and capacity with $w_\ell>0$ and $\hardQ<\DeltaT$. Let $A_\ell(x,\hardQ)=xD_\ell(x,\hardQ)$ and $B_Q=\min\{L,(\DeltaT-\hardQ)/\kappa_\ell\}$. Retain $g$, $\delta$, and $s$ from \Cref{app:market-proofs}, the thresholds $x_1,x_2$ from \cref{eq:entry-maximization-thresholds}, and the maximum recovery expenditure $\overline E(x)$ from \cref{eq:maximum-recovery-expenditure}.

For $w_\ell>0$, pure-pooling profit increases with $x$. The recovery condition $\overline E(x)\geq\dpool$ can be satisfied if and only if
\begin{equation}
\label{eq:pooling-minimum-coverage}
L\geq L_{\min}^P\coloneqq\frac{g\dpool}{\kappa_hs}.
\end{equation}
When this condition holds, define
\[
x_{\min}^P\coloneqq\frac{\delta\dpool}{\kappa_hs},
\]
\[
x_{\max}^P\coloneqq L-\frac{g\dpool}{\kappa_h(\kappa_\ell-\epsilon_\ell)}.
\]
The rising and falling parts of \cref{eq:maximum-recovery-expenditure}, together with low-type participation, give the exact pure-pooling interval
\begin{equation}
\label{eq:pooling-feasible-quality-interval}
\left[x_{\min}^P,\ \min\left\{x_{\max}^P,
\frac{\prior\DeltaT-\hardQ}{\kappa_\ell}\right\}\right],
\end{equation}
provided its upper endpoint is at least its lower endpoint. Equivalently, pure pooling is possible if and only if \cref{eq:pooling-minimum-coverage} holds and $\hardQ\leq\prior\DeltaT-\kappa_\ell x_{\min}^P$. Its best positive-profit quality is the upper endpoint.

The partial-pooling objective is concave, with stationary quality given by \cref{eq:partial-pooling-optimal-quality} when $\hardQ>0$, and decreases when $\hardQ=0$. The constrained maximum can be found from finitely many candidate qualities. Partition $[0,B_Q]$ at $x_1$, $x_2$, and $(\prior\DeltaT-\hardQ)/\kappa_\ell$ whenever these points lie inside the domain. On each resulting interval, the recovery inequality in \cref{eq:closed-shared-quality-set} is linear. Intersect it with that interval and inspect the endpoints of each nonempty feasible interval. Include the inactive quality $x=0$, and, when $\hardQ>0$, inspect $x_M$ if it lies in the feasible partial-pooling region. The largest value of $(w_\ell)_+A_\ell$ at these candidates is $H$. Every candidate with positive value implements an actual shared outcome.


\subsection{A globally optimal usage subsidy}
\label{supp:usage-subsidy}

The example in \Cref{sec:market} uses $(\kappa_\ell,\kappa_h)=(10,8)$, $(\epsilon_\ell,\epsilon_h)=(0.01,0.002)$, $L=0.5$, $c=8.1$, $\prior=0.8$, $\DeltaT=1$, $\hardQ=0.45$, and $\rhoHard=0.02$. The per-unit resource savings are $w_h=-0.102$ and $w_\ell=1.89$, so the best separating operating value is zero.

Since $\hardQ<\prior^2\DeltaT=0.64$, the unconstrained shared optimum is $x_P=(0.8-0.45)/10=0.035$. The tariff $p=0$, $F=(10-0.01)x_P=0.34965$ gives this adoption threshold. \Cref{eq:recovery-expenditure-piecewise} yields
\begin{equation*}
E=9.99(0.035)+\left(0.002-\frac{8(0.008)}{2}\right)0.5
=0.33465>\dpool=0.2.
\end{equation*}
The quality lies below coverage, and low-type participation binds at $v=10(0.035)+0.45=0.8=\prior\DeltaT$. The selected outcome is pure pooling, with operating profit
\begin{equation*}
H=(1-0.8)(1.89)(0.035)=0.01323.
\end{equation*}

This tariff attains the unconstrained shared-profit upper bound and earns more than any separating offer. Hence, it is globally optimal for the fixed technology. Membership revenue of $0.06993$ covers compute expenditure of $0.05670$, leaving the stated positive operating profit despite free usage.

Every tariff supporting a shared signal for this technology satisfies $p\leq\kappa_h-\epsilon_\ell=7.99<c$. Subsidized usage is thus compatible with profit-maximizing provision through a two-part tariff. A fresh launch is profitable whenever its development cost $K(\mathbf d)$ is below $0.01323$.


\subsection{A bounded monopoly improvement problem}
\label{supp:monopoly-improvements}

Retain the primitives in \Cref{supp:usage-subsidy}. Let the provider choose independent improvements $u_h\in[0,0.001]$ and $u_\ell\in[0,0.004]$. Set $\epsilon_h=0.002-u_h$,
$\epsilon_\ell=0.01-u_\ell$, 
$L=0.5$, and
\begin{equation*}
K(u_h,u_\ell)=\frac{k_h}{2}u_h^2+u_\ell^2.
\end{equation*}
where $k_h>0$
The remaining primitives are 
\[
(\kappa_\ell,\kappa_h,c,\prior,\DeltaT,\hardQ,\rhoHard)=(10,8,8.1,0.8,1,0.45,0.02).
\]
Every design in this compact rectangle satisfies \Cref{ass:comparative-advantage}. In particular, $\epsilon_h\geq0.001$, $\epsilon_\ell\geq0.006$, and $\epsilon_h/\epsilon_\ell\geq0.1>\rhoHard$.

Throughout the rectangle, $w_h=-0.102+u_h<0$, so $R=0$. The unconstrained shared-profit upper bound is attained at $x=0.035$ by $p=0$ and $F=(10-\epsilon_\ell)x$. This quality lies in the middle case of \cref{eq:recovery-expenditure-piecewise} for every design. Recovery expenditure satisfies
\begin{equation*}
E=(10-\epsilon_\ell)(0.035)
+\left[\epsilon_h-4(\epsilon_\ell-\epsilon_h)\right](0.5)
\geq0.33215>0.2=\DeltaT-\kappa_\ell x-\hardQ.
\end{equation*}
The price inequality is strict and low-type participation binds at $\kappa_\ell x+\hardQ=0.8$. Thus, the tariff supports a unique finite-support D1 pool. It attains the shared upper bound, giving
\begin{equation*}
G-K
=0.007(1.89+u_\ell)-\frac{k_h}{2}u_h^2-u_\ell^2.
\end{equation*}
This expression is strictly concave on the rectangle. Its unique maximizing design is $u_h^*=0$ and $u_\ell^*=0.0035$, with operating profit $0.0132545$ and net profit $0.01324225>0$. A globally optimal tariff has $p=0$ and $F=0.3497725$. The provider improves low-type productivity alone because its customers are participating low types, and the optimal shared quality is already feasible without a high-type improvement. Positive net profit also makes launch optimal when the provider can decline development.

\section{Additional results on design competition}
\label{supp:ed-additional-results}

\subsection{Symmetric mixed design equilibrium}
\label{supp:ed-mixed-design}

A symmetric strategy profile need not impose identical realized designs. 

\begin{corollary}
\label{cor:ed-mixed-design}
Under the conditions of \Cref{prop:ed-pure-design}, each provider choosing $v=0$ and independently drawing $u$ uniformly on $[0,u^*]$ is a symmetric mixed design equilibrium with the pricing equilibria in \Cref{prop:ed-pricing}. Each provider earns zero expected net profit.
\end{corollary}

\begin{proof}[Proof of \Cref{cor:ed-mixed-design}]
The low-type improvement bound in the proof of \Cref{prop:ed-pure-design} rules out $v>0$ pointwise against every rival $(s,0)$. For $0\leq u\leq u^*$, expected operating return is
\begin{equation*}
\frac{1}{u^*}\int_0^u\frac{\prior r(u-s)}{m_\ell^0}\,ds
=\frac{k_hu^2}{2},
\end{equation*}
which covers incremental expenditure. For $u>u^*$, subtracting expenditure from expected return gives $-k_h(u-u^*)^2/2\leq0$. Staying at the baseline gives zero. These bounds cover all designs in the rectangle, including asymmetric deviations. The pricing equilibria after each design pair and signaling equilibria after each tariff menu are already verified.
\end{proof}

\subsection{Resources and verification}
\label{supp:ed-resources-verification}

With $u_1>u_2$ and the installed menu fixed at the prices in \Cref{prop:ed-pricing}, expected signaling resources are
\begin{equation*}
C^{\mathrm{real}}(\hardQ)=\prior\left[(m_h^0-u_1)\frac{\DeltaT-\hardQ}{m_h^0-u_2+\delta_*}+\rhoHard\hardQ\right].
\end{equation*}
Differentiating gives the installed-menu derivative discussed in the main text. It is negative by the resource ratio proved in \Cref{prop:ed-pricing}. Improvement expenditure is already sunk here. An accuracy-motivated receiver gains no further information from capacity along these separating outcomes, although capacity saves resources.

For prospective improvements under \Cref{prop:ed-pure-design}, $q_R=(\DeltaT-\hardQ)/m_\ell^0$ and the leader chooses $u^*(\hardQ)=\prior(\DeltaT-\hardQ)/(k_hm_\ell^0)$. Total signaling and improvement resources are
\begin{equation*}
C_{\mathrm{tot}}(\hardQ)
=\prior\left[\frac{m_h^0(\DeltaT-\hardQ)}{m_\ell^0}+\rhoHard\hardQ\right]
-\frac{[\prior(\DeltaT-\hardQ)]^2}{2k_h(m_\ell^0)^2}.
\end{equation*}
The last term subtracts the net resource saving from the optimal high-type improvement: its production saving is twice its improvement expenditure. Differentiation gives
\begin{equation}
\label{eq:ed-capacity-derivative}
\frac{dC_{\mathrm{tot}}}{d\hardQ}
=\prior\left[\rhoHard-\frac{m_h^0-u^*(\hardQ)}{m_\ell^0}\right].
\end{equation}

\begin{corollary}
\label{cor:ed-verification}
There is a capacity game satisfying the extension's restrictions in which the receiver chooses capacity before improvements, the selected signaling outcome separates after every capacity choice, and receiver-optimal capacity is strictly below welfare-maximizing capacity.
\end{corollary}

\begin{proof}[Proof of \Cref{cor:ed-verification}]
Retain the baseline, feasible rectangle, and signaling parameters of \Cref{app:ed-welfare-example}, but set $k_\ell=1$. Let the receiver choose $\hardQ\in[.45,1]$ before improvements, with $\capcost(\hardQ)-\capcost(.45)=.1(\hardQ-.45)$. For every $\hardQ<1$ in that interval, \cref{eq:ed-pure-conditions} holds, as $u^*(\hardQ)\leq .07333333\ldots<.1$ and
\begin{equation*}
\frac{\prior(1-\hardQ)}{(m_\ell^0-\bar v)^2}
\leq\frac{.44}{.95^2}<1=k_\ell.
\end{equation*}
Specify either pure asymmetric equilibrium in \Cref{prop:ed-pure-design} at each such capacity. At $\hardQ=1$, both providers retain their baseline, charge $F=0$, $p=c$, and hard evidence alone separates. The prescribed provider strategies and signaling equilibria satisfy optimality after every design and tariff history, as established in the proofs of \Cref{prop:ed-pricing,prop:ed-pure-design}.

Assessment error is zero at every capacity, so the receiver's payoff differs only through its strictly increasing capacity expenditure. Its unique optimum is $.45$. At that capacity, \cref{eq:ed-capacity-derivative} equals $-.40177777\ldots$. The derivative decreases as capacity rises because $u^*(\hardQ)$ decreases. Adding marginal capacity expenditure $.1$ leaves total social cost decreasing throughout $[.45,1]$. The continuous welfare objective is uniquely maximized at $\hardQ=1$. This specifies a complete verification game rather than a local comparison of installed technologies.
\end{proof}

The equilibrium in \Cref{app:ed-welfare-example} also has a local capacity response. Put $b=m_\ell^0-\bar v=.95$. Near $\hardQ=.45$, its strict deviation bounds persist, $v^*=\bar v$, and $k_hu(b+u)^2=\prior(\DeltaT-\hardQ)b$. Hence,
\begin{equation*}
u'(\hardQ)=-\frac{\prior b}{k_h(b+u)(b+3u)}.
\end{equation*}
Its total signaling and improvement cost is $\prior[(m_h^0-u)(\DeltaT-\hardQ)/(b+u)+\rhoHard\hardQ]+k_hu^2/2+k_\ell\bar v^2/2$, whose derivative is
\begin{equation*}
\prior\left[\rhoHard-\frac{m_h^0-u}{b+u}\right]
-\frac{\prior(\DeltaT-\hardQ)m_h^0}{(b+u)^2}u'(\hardQ).
\end{equation*}
At $.45$ it equals $-.43064139\ldots$. This calculation includes changing improvement expenditure. The global verification construction in \Cref{cor:ed-verification} uses the high-type-only equilibrium, which is specified at every admissible capacity.

\section{Additional welfare and verification results}
\label{supp:welfare-verification-additions}

\subsection{Numerical welfare and resource comparisons}
\label{supp:welfare-examples}

\textit{A profitable separating launch that lowers welfare.} Use the primitives in \cref{eq:welfare-example-primitives}.

With $c=7$ and $K=0.001$, $b=8.008$, $s=1.992$, and $q_R=z_*=0.55/8.008$. Separating operating profit is $R=0.0548351648\ldots$. The unconstrained shared bound in \cref{eq:unconstrained-shared-value} gives $H\leq0.02093$, so separation is strictly optimal over all tariffs for this technology. Net provider profit is $0.0538351648\ldots$. Expected signaling resource cost is
\[
C_R^{\mathrm{real}}=0.8\left[7.002\frac{0.55}{8.008}+0.02(0.45)\right]
=0.3919252747\ldots,
\]
compared with the manual baseline
\[
C_0^{\mathrm{real}}=0.8[0.8(0.55)+0.02(0.45)]=0.3592.
\]
Including development cost, welfare changes by $-0.0337252747\ldots$ for every $\receiverweight>0$.

\textit{Monopoly and competition.} For the parameters in \cref{eq:welfare-example-primitives} with $c=0.01$, competitive quality is $0.554$ and expected signaling resource cost is $0.3576$. Under optimal monopoly pricing, the corresponding resource cost is $0.0078593407\ldots$.

\subsection{Failure of a pure simultaneous equilibrium}
\label{app:verification-no-pure}

Retain the primitives in \cref{eq:verification-shared-example}, but set $\hardQ_0=0.45$ and $\capcost(\hardQ)=0.05\hardQ^2$. At every capacity below one, the provider's optimal outcome is shared, with positive quality and profit. At a pooling optimum, the low-type participation condition binds: $\kappa_\ell x_H+\hardQ=0.8$. A sufficiently small upward capacity deviation holding the offer fixed enters partial pooling. At a partial-pooling optimum, such a deviation remains partially pooled. Recovery feasibility is preserved because the required expenditure falls. Under either outcome the right-hand marginal receiver gain is $0.16-0.1\hardQ>0$ for $\hardQ<1$. No such capacity is a best response.

At $\hardQ=1$, every provider best response has zero improvement, since operating profit is zero for every offer. Hold any such finite tariff fixed. If it admits a positive shared threshold $x$, then sharing is impossible for all capacities in a left neighborhood of one because $10x+\hardQ\geq1$. If it has no positive shared threshold, it admits no selected shared signaling outcome at any capacity. The receiver can reduce capacity slightly, preserve separation, and save capacity cost. Capacities above one similarly admit a reduction to one. No pure simultaneous equilibrium exists. The argument covers every optimal tariff, including those that are unused at full capacity.

The sequential outcome in this specification retains $0.45$, as shown in the insufficient-investment construction in \Cref{app:verification-welfare}. It fails the simultaneous best-response condition. Its quality is $x=0.035$, so holding its optimal offer fixed and raising capacity to $0.65$ yields full information. The receiver's gain is
\[
\frac{0.4}{2}(0.16)-0.05(0.65^2-0.45^2)=0.021>0.
\]
The distinction between commitment and a mutual best-response outcome is substantive in this example. No claim about mixed-strategy existence or convergence of repeated adjustments is needed.

\subsection{Provider offer ties and equilibrium selection}
\label{app:verification-selection}

The excessive-investment construction specifies separation at $\hardQ_*$. If the provider instead chooses sharing there, receiver gain is $-0.2\hardQ_*$ at the tie and $0.08-0.2\hardQ$ just above it. The positive supremum $0.08-0.2\hardQ_*$ is approached from above but is unattained. No pure sequential receiver best response exists when the provider chooses sharing at the profit tie. Selecting separation at the tie gives the pure equilibrium constructed in \Cref{app:verification-welfare}. This issue concerns the provider's choice among optimal offers. The equilibrium selection rule applies after an offer has been chosen.

The examples do not require choosing sharing over an admissible separating observable outcome at their displayed shared offers. In the excessive construction one can use $p=0$ throughout its optimal shared interval. Its recovery expenditure is $E=(1-v_H)(2-\delta)x_H/\delta>\DeltaT-\kappa_\ell x_H-\hardQ$. For example, uniformly on $[0,\hardQ_*]$, $x_H>0.068$, $1-v_H>0.9984$, and $\delta\leq0.01$, giving $E>13$. The other shared constructions use the interior tariffs in \Cref{app:verification-information-loss}, or $p=0.1$ for \cref{eq:verification-shared-example}; their recovery inequalities are likewise strict on the capacity intervals where sharing is optimal. With a strict shared price and this stronger recovery inequality, Riley separation fails D1 and the canonical shared observable outcome is unique. The separating implementations have $p+\epsilon_\ell>\kappa_h$ and a unique observable separating outcome.

These facts do not establish the same optimal policies under every alternative equilibrium selection rule. The equilibrium selection rule is used to bound profits over all offers and to evaluate every capacity deviation. Exact profit attainment also uses subscription at production indifference. Allowing different adoption at such ties can change sales even when private costs and observable signaling choices coincide. Off-path beliefs can remain multiple without changing the selected outcome. The constructions concern finite-support signaling equilibria, one installed provider, and prospective choices from a common baseline. 

\section{Parameters used in the figures}
\label{app:figure-parameters}
The figures evaluate the paper's analytical expressions at illustrative parameter values. Artifact quality is measured in physical units, hard capacity in low-type effort-cost units, and costs and profits in payoff units.

For \Cref{fig:ai-signaling-costs}, $(\kappa_\ell,\kappa_h)=(4,3)$, $(\epsilon_\ell,\epsilon_h)=(1.5,1)$, $L=4$, $F=2$, and $p=0.5$. These values give $x=1$, $q_E=2$, and $E=2$. The displayed quality range lies below AI coverage.

For \Cref{fig:provider-profit-envelope}, $(\kappa_\ell,\kappa_h)=(10,8)$, $(\epsilon_\ell,\epsilon_h,L)=(0.01,0.002,0.5)$, $c=7.9$, $\prior=0.8$, $\DeltaT=1$, $\rhoHard=0.02$, and $K=0.001$. Pooling gives way to partial pooling at $Q=0.640$. Shared and separating operating values cross at $Q\simeq0.791$, and net launch profit reaches zero at $Q\simeq0.898$. At the latter two thresholds, both adjacent choices are optimal. The no-launch threshold concerns fresh development. Installed providers do not subtract sunk development cost again.

For \Cref{fig:welfare-regions}, $(\kappa_\ell,\kappa_h)=(10,8)$, $(\epsilon_\ell,\epsilon_h,L)=(0.01,0.002,0.5)$, $c=7.4$, $\prior=0.5$, $\DeltaT=1$, $\rhoHard=0.02$, and $K=0.003$. Panel (a) uses $\receiverweight=0.05$. Its transitions occur at $Q=0.250$, $Q\simeq0.305$, and $Q\simeq0.920$. At the latter two thresholds, both adjacent choices maximize profit. In panel (b), the curve defined by $\welfare=\welfare_0$ intersects the axes at $\receiverweight=0.094$ and $Q\simeq0.091$. Development cost is incurred only upon launch, and capacity cost cancels in comparisons at the same capacity.

In \Cref{fig:provider-profit-envelope,fig:welfare-regions}, physical technology is fixed, both tariff components are reoptimized at each capacity, and the optimal shared outputs satisfy the deviation constraints wherever shared outcomes are displayed. The outcome sequences are specific to these parameter values.

For \Cref{fig:verification-design-timing}, both panels use $(\kappa_\ell,\kappa_h)=(10,8)$, $\DeltaT=1$, and $\rhoHard=0.02$. The panels use different parameter sets.

Panel (a) uses $\prior=0.95$, $\receiverweight=0.1$, and $\capcost(\hardQ)=0.2\hardQ$, with
\[
\begin{aligned}
(\epsilon_h^0,\epsilon_\ell^0,c,L)&=(0.5,1,7.6,1),\\
(\bar u,\bar v,k_h,k_\ell)&=(0.25,0.05,0.05,10).
\end{aligned}
\] Improvements and both tariff components are reoptimized at each capacity. Separating and shared net profits coincide at $\hardQ\simeq0.869423$, where the provider chooses separation. Above that threshold, the selected outcome pools until $\hardQ=\prior^2\DeltaT=0.9025$, then partially pools until full information at $\hardQ=1$. Pooling assessment error is $0.0475$. Squares mark equilibria at minimum requirements $\hardQ_0=0$ and $\hardQ_0=0.9$ under both timings. The curve elsewhere records the provider's optimal response to a given capacity. \Cref{app:verification-information-loss} gives the offers and global best-response checks.

Panel (b) uses $\prior=0.8$, $\receiverweight=0.4$, $\hardQ_0=0.72$, and $\capcost(\hardQ)=0.1\hardQ^2$, with
\[
\begin{aligned}
(\epsilon_h^0,\epsilon_\ell^0,c,L)&=(0.002,0.01,8.1,100),\\
(\bar u,\bar v,k_h,k_\ell)&=(0.001,0.004,10,2).
\end{aligned}
\] The marginal accuracy gains are $0.16$ with the offer held fixed and $0.08/\sqrt{\hardQ}$ with provider adjustment; marginal capacity cost is $0.2\hardQ$. Each accuracy return is evaluated at the optimal offer for that capacity, with open endpoints marking left-hand limits at $\hardQ=1$. Sequential capacity is $0.72$, and simultaneous capacity is $0.80$. \Cref{app:verification-timing} verifies both equilibria.


\end{document}